\documentclass[11pt,a4paper]{article}
\usepackage{geometry}
\usepackage[english]{babel}
\usepackage{mathrsfs}
\usepackage{bbm}
\usepackage{wrapfig}
\usepackage{amsmath}
\usepackage{amssymb}
\usepackage{mathtools}
\usepackage{amsfonts}
\usepackage{setspace}
\usepackage{jcap}

\usepackage{color,soul}
\usepackage{xcolor}
\usepackage{tikz}

\allowdisplaybreaks

\newcommand{\V}{V}
\newcommand{\DD}{\mathcal{D}}
\newcommand{\RR}{\mathcal{R}}

\newcommand{\Oo}[1]{
	\ifnum#1=1
		\mathcal{O}\left(\frac{1}{c}\right)
	\else
		\mathcal{O}\left(\frac{1}{c^{#1}}\right)
	\fi
}

\newcommand{\Ss}{\mathcal{S}}

\newcommand{\dchi}{\frac{d}{d\chi}}
\newcommand{\dd}[2]{\frac{d#1}{d#2}}
\newcommand{\ddchi}{\frac{d^2}{d\chi^2}}
\newcommand{\HH}{\mathcal{H}}
\newcommand{\tDD}{\tilde{\mathcal{D}}}

\newcommand{\DP}{\mathcal{D}^{(P)}}
\newcommand{\DVW}{\mathcal{D}^{(VW)}}
\newcommand{\DWW}{\mathcal{D}^{(WW)}}
\newcommand{\Dh}{\mathcal{D}^{(h)}}
\newcommand{\DVV}{\mathcal{D}^{(VV)}}
\newcommand{\DUW}{\mathcal{D}^{(UW)}}

\newcommand{\tDm}[1]{{}_{0}\tilde{\mathcal{D}}^{(#1)}_{\rm R}}
\newcommand{\kap}[1]{\kappa^{(#1)}}
\newcommand{\dx}{\delta x}

\newcommand{\vp}{v_{S\parallel}}
\newcommand{\vs}{v_S}
\newcommand{\vm}{{}_{-1}v_S}
\newcommand{\vpl}{{}_{1}v_S}

\newcommand{\cc}[1]{\frac{1}{c^{#1}}}
\newcommand{\cgen}[2]{\frac{#1}{c^{#2}}}
\newcommand{\bk}[1]{\bar{k}^{#1}}

\newcommand{\eplus}[1]{e_{+}^{#1}}
\newcommand{\eminus}[1]{e_{-}^{#1}}

\newcommand{\pt}{\partial_{\theta}}
\newcommand{\pp}{\partial_{\phi}}
\newcommand{\fkk}[1]{\frac{\bar{k}^{#1}}{\bar{k}^{0}}}

\newcommand{\slpa}{\ensuremath{\mathrlap{\!\not{\phantom{\partial}}}\partial}}

\newcommand{\slpae}{\partial_{\theta}+\frac{i}{\sin\theta}\partial_{\phi}}
\newcommand{\slpab}{\partial_{\theta}-\frac{i}{\sin\theta}\partial_{\phi}}
\newcommand{\slpaee}{\partial_{\theta}+\frac{i}{\sin\theta}\partial_{\phi}-\cot\theta}
\newcommand{\slpabb}{\partial_{\theta}-\frac{i}{\sin\theta}\partial_{\phi}+\cot\theta}
\newcommand{\intchi}{\int_{0}^{\chi_{S}}d\chi (\chi_S -\chi)}

\newcommand{\ichi}{\int_{0}^{\chi_{S}}d\chi}
\newcommand{\bound}{_{0}^{\chi}}
\newcommand{\dc}{d\chi}
\newcommand{\aaa}{\delta}
\newcommand{\bbb}{\beta}
\newcommand{\ccc}{\alpha}
\newcommand{\ddd}{\gamma}

\title{Full-sky weak lensing: a nonlinear post-Friedmann treatment}
\author[a]{Hedda A. Gressel} 
\author[b]{Camille Bonvin}
\author[a]{Marco Bruni}
\author[a]{David Bacon}
\affiliation[a]{Institute of Cosmology and Gravitation, University of
  Portsmouth\\
   Dennis Sciama Building, Portsmouth PO1 3FX, United Kingdom}
 \affiliation[b]{D\'epartement de Physique Th\'eorique and Center for Astroparticle Physics, University of Geneva, CH-1211 Geneva, Switzerland}
\emailAdd{hedda.gressel@port.ac.uk}
\emailAdd{camille.bonvin@unige.ch}
\emailAdd{marco.bruni@port.ac.uk}
\emailAdd{david.bacon@port.ac.uk}
\abstract{We present a full-sky derivation of weak lensing observables in the Post-Friedmann (PF) formalism. Weak lensing has the characteristic of mixing small scales and large scales since it is affected by inhomogeneities integrated along the photon trajectory.
With the PF formalism, we develop a modelling of lensing observables which encompasses both leading order relativistic effects and effects that are due to the fully non-linear matter distribution at small scales. 
We derive the reduced shear, convergence and rotation up to order $1/c^4$ in the PF approximation, accounting for scalar, vector and tensor perturbations, as well as galaxies' peculiar velocities. We discuss the various contributions that break the Kaiser-Squires relation between the shear and the convergence at different orders. We pay particular attention to the impact of the frame-dragging vector potential on lensing observables and we discuss potential ways to measure this effect in future lensing surveys.}
\arxivnumber{XXX}
\begin{document}
\maketitle
\flushbottom
\section{Introduction}
\label{sec:intro}
Weak gravitational lensing (WL) --the statistical analysis of distorted galaxy images-- is a rich source of information about the evolution of the large-scale structure of the Universe. It follows from the equivalence principle that light is bent by gravitational masses, which consequently distorts the galaxy images along the line of sight. When distortions are small enough such that no caustics are generated, we enter the regime of weak gravitational lensing, where distortions can only be detected statistically.  The distortions can be split into a convergence (which changes the apparent size of galaxies), a shear and a rotation~\cite{1991MNRAS.251..600B,1992ApJ...388..272K,1991ApJ...380....1M}. Cosmic shear has first been detected in the early 2000s~\cite{Bacon:2000sy,Kaiser:2000if,vanWaerbeke:2000rm,Wittman:2000tc}. Estimators for the convergence, combining measurement of galaxies sizes and luminosities, have been constructed recently~\cite{Schmidt:2011qj,Casaponsa:2012tq,Heavens:2013gol,Alsing:2014fya}, and a first detection at small scales was achieved in 2012~\cite{Schmidt:2011qj}.  
Until recently, weak lensing surveys covered only parts of the sky, see for example DES \cite{Abbott:2017wau} with a coverage of around 5000 square degrees. But future surveys such as Euclid~\cite{EuclidWeb} and LSST~\cite{LSSTWeb} will deliver high precision data on more than a third of the sky. With this vast amount of high precision data, weak lensing is becoming a promising tool to map the Universe. 

Yet, the WL analysis is challenging: while we use different approximation schemes for large and small scales, weak lensing effects cover all scales. By integrating along the light path, large and small scales are mixed. 
For example, the correlation between two galaxies far apart is affected by relativistic effects, which distort the photon trajectory beyond the Newtonian treatment and lead to relativistic corrections to the convergence and the shear~\cite{Bonvin:2008ni,2010PhRvD..81h3002B,Schmidt:2012ne,Yoo:2018qba}. A full-sky relativistic treatment is therefore necessary to capture these effects. If in addition, these galaxies are aligned with respect to the line of sight, their correlation will be strongly affected by non-linear effects, since their photons' trajectories traverse the same non-linear structures. Similarly, since all trajectories end up at the observer, any non-linear structure close to the observer will generate non-linear correlations between pairs of galaxies even if those are widely separated in the sky. Analyses of weak lensing data at large scales do therefore require modelling relativistic effects and non-linear effects at the same time.

Furthermore, an important part of the high precision data from future weak lensing surveys will be data from small, non-linear scales. It is standard to assume that the Newtonian approximation is sufficient to model structure formation on those scales, e.g.\ in N-body simulations. However, with Euclid aiming at 1\% accuracy, it is not clear if the Newtonian treatment is still sufficient. For example, the ``frame-dragging" vector potential \cite{Bardeen:1980kt}, which is a purely relativistic effect, may affect weak lensing observables at small scales. In~\cite{Bruni:2013mua,Thomas:2015kua}, the vector frame dragging gravito-magnetic potential was computed, showing that its magnitude is small but not entirely negligible, with its power spectrum of order $10^{-5}$ that of the non-linear scalar potential on non-linear scales\footnote{By non-linear scales we mean here those scales where the non-linear power spectrum of the scalar potential differs from the linear power spectrum.}. This is a robust result, independently confirmed in~\cite{Adamek:2013wja,Adamek:2015eda} by using an N-body weak-field code based on General Relativity~\cite{Adamek:2016zes}. A similar result, with a stronger gravito-magnetic effect, was found in~\cite{Thomas:2015dfa} for the Hu-Sawicki $f(R)$ gravity model~\cite{Hu:2007nk}. Analyses of weak lensing data at small scales may therefore also require a non-linear relativistic treatment. 

In this paper, we use the post-Friedmann (PF) formalism developed in~\cite{Milillo:2015cva}, to bridge the gap between the small scale and large scale descriptions of weak gravitational lensing. 
The PF formalism is a post-Newtonian-type approximation scheme in a cosmological setting that combines both the fully non-linear Newtonian treatment of the dynamics on small, fully non-linear scales and the relativistic perturbative analysis on large scales. Therefore, it seems to be the ideal approximation for a thorough weak lensing analysis. Furthermore, the PF formalism provides an apt framework for N-body simulations with relativistic corrections~\cite{Bruni:2013mua,Thomas:2015kua,Thomas:2015dfa}, with a first attempt to consider WL in this formalism in~ \cite{Thomas:2014aga}. 
Analogously to post-Newtonian approximations, in the PF scheme the relevant variables are expanded in inverse powers of the speed of light $c$, once the Robertson-Walker background has been subtracted. 
In this sense, it is a weak-field approximation (aka post-Minkowskian expansion)  on a Robertson-Walker background, where variables are expanded in inverse power of $c$ like in a post-Newtonian expansion. In addition, it includes both vector and tensor potentials, non-linearly sourced by the matter distribution. In particular, the frame-dragging vector potential is sourced in cosmology by the momentum-density vector field. An example of an effect arising from this vector potential  in another  context is the frame-dragging Lense-Thirring effect, which has been measured in the Solar system by Gravity Probe B \cite{Everitt:2011hp}. This exemplifies that this purely relativistic effect is small, but not unmeasurable, and it can be sourced in a  weak-field regime such as that of the Earth.

In this paper we provide a detailed analysis of the convergence, shear, and rotation up to order $\Oo{4}$ in the PF approximation scheme. We include scalar, vector, and tensor potentials and put a specific focus on how the vector potential affects the distorted images of galaxies. We also consistently include the effect of galaxies' peculiar velocities, which contribute to our observables through redshift perturbations. 
Our results are expressed in terms of the spherical spin operator and are valid in the full sky. We show how the vector potential contributes to the shear and convergence, but not to the rotation. Furthermore, we argue that the terms involving the vector potential violate the Kaiser-Squires relation. 

The rest of the paper is organised as follows: in section~\ref{sec:PF} we introduce the PF formalism. We discuss the metric and its features such as the validity on all scales. In section~\ref{sec:sec3}, we derive the magnification matrix. Starting from the geodesic deviation equation, we derive the Jacobi mapping up to order $\Oo{4}$ with the PF formalism, including also redshift perturbations. In section~\ref{sec:sec4}, we introduce spherical coordinates and spin operators. This allows us to extract the convergence, shear and rotation, without relying on the flat-sky approximation. We conclude in section~\ref{sec:con}.

\section{Post-Friedmann Approximation Scheme}
\label{sec:PF}
The post-Friedmann formalism is a generalisation of the post-Minkowskian (weak field)  approximation to cosmology, where the Minkowski background is replaced by a Robertson-Walker one. It was developed in \cite{Milillo:2015cva} to consider the approximate non-linear general relativistic dynamics  in $\Lambda$CDM cosmology. The aim of this formalism is to unite different approximations on different  cosmological scales, from small scales, where - at leading order -  the dynamics should be sufficiently well described by  the Newtonian approximation, to the largest scales, at which standard relativistic perturbation theory should be applicable. In the  PF scheme   the  metric and the 4-velocity of CDM, described as a dust fluid, are first written as perturbations of a flat Robertson-Walker  background,   then they are  expanded in inverse powers of the speed of light $c$. In doing so, the crucial difference with a standard post-Newtonian scheme is that the Hubble (background) flow is separated from the velocity perturbation, and therefore the scheme is also valid on  Hubble scales and beyond\footnote{In a  traditional post-Newtonian scheme,  the whole 4-velocity field is expanded in inverse powers of $c$;  if such a scheme is applied in cosmology,  it can  therefore only deal with sub-horizon scales.}. The matter density contrast field $\delta=(\rho-\bar{\rho})/\bar{\rho}$ (where $\bar{\rho}$  denotes the background matter density)  and the  velocity perturbation $\beta=v/c$  are fundamental exact dimensionless variables 
 that are not expanded into contributions at different orders; rather other quantities, e.g.\ the energy momentum tensor, contain contributions of different orders in $\beta$.

The nonlinear dynamical equations are consistently derived at different orders in inverse powers of $c$ by expanding the  Einstein equations. 
When the resulting $c^{-4}$ equations are linearised, this scheme recovers standard first-order general relativistic perturbation theory and can therefore be used to describe structure formation on the largest scales. 

At leading $c^{-2}$ order, however, the PF formalism yields the fully non-linear Newtonian dynamics of CDM in a flat $\Lambda$CDM background. However, in this framework the Newtonian dynamics are just an approximation, and the spacetime is inhomogeneous. In this Newtonian regime for the dynamics, additionally to the Newtonian scalar potential and consistently derived  from the Einstein field equations at $c^{-3}$ order, one recovers a metric gravito-magnetic ``frame-dragging" vector potential \cite{Bardeen:1980kt} as the leading-order  contribution to the  $g_{0i}$ metric components in Poisson gauge, see below. This vector potential is sourced by the  Newtonian momentum density  that  can be extracted from  standard Newtonian N-body simulation, and as such  it has been computed   in \cite{Bruni:2013mua,Thomas:2015kua} in $\Lambda$CDM, and in \cite{Thomas:2015dfa} for the Hu-Sawicki \cite{Hu:2007nk} $f(R)$ gravity model. 

 Thus, in  seeking whether the gravito-magnetic potential could also be measurable on cosmological scales, one main motivation  in this paper is to look for gravito-magnetic  effects in weak lensing. We emphasise however that the formalism developed  here is purely geometrical, i.e.\ we do not make specific assumptions about the dynamics, and as such it could be applicable not only in $\Lambda$CDM, but also in different cosmologies, e.g.\ some dark energy or modified gravity model (see~\cite{Clifton:2011jh} for a review).

\subsection{The Metric}
The metric of the PF formalism in the Poisson gauge reads \cite{Milillo:2015cva}
\begin{align}
g_{00}=&-e^{-\frac{2U_{N}}{c^{2}}-\frac{4U_{P}}{c^4}}+\mathcal{O}\left(\frac{1}{c^{6}}\right)\\
=&-\left[1-\frac{2U_{N}}{c^{2}}+\frac{1}{c^{4}}\left(2U_{N}^{2}-4U_{P}\right)\right]+\mathcal{O}\left(\frac{1}{c^{6}}\right)\, ,\label{eq:g00}\\
g_{0i}=&-\frac{a}{c^{3}}B_{N i}-\frac{a}{c^{5}}B_{Pi}+\mathcal{O}\left(\frac{1}{c^{7}}\right)\,,\\
g_{ij}=&\,e^{\frac{2V_{N}}{c^{2}}+\frac{4V_{P}}{c^4}}\delta_{ij}+\frac{1}{c^{4}}h_{ij}+\mathcal{O}\left(\frac{1}{c^{6}}\right)\\
=&a^{2}\left[\left(1+\frac{2V_{N}}{c^{2}}+\frac{1}{c^{4}}\left(2V_{N}^{2}+4V_{P}\right)\right)\delta_{ij}+\frac{1}{c^{4}}h_{ij}\right]+\mathcal{O}\left(\frac{1}{c^{6}}\right)\,.\label{eq:gij}
\end{align}
The subscripts $N$ and $P$ of the metric potentials refer to Newtonian and post-Friedmann contributions, respectively. In the Poisson gauge, the vector fields $B_{Ni}$ and $B_{Pi}$ are divergenceless and the tensor field $h_{ij}$ is transverse and trace-free. A Lagrangian gauge (a generalisation of the synchronous-comoving gauge of SPT) version of the PF formalism was derived in \cite{Rampf:2016wom}.

\subsubsection*{Validity on All Scales}
The PF formalism is an approximation scheme that is valid on both small and large scales. It differs from traditional post-Newtonian (PN) approximations in the following way:  the PN formalism is derived from the post-Minkowski approximation with the assumption that velocities are much smaller than the speed of light $c$~ \cite{1972gcpa.book.....W,poisson2014gravity}. The PF approximation has a FLRW background instead of a Minkowskian and only \emph{peculiar} velocities $v$ are assumed to be small $v/c\ll 1$. The latter assumption does not restrict the validity of the approximation to small scales:
let us assume that $x^{i}$ are comoving, spatial coordinates, then the physical coordinate of a fluid element is $r^{i}=a x^{i}$. The time derivative of $r^{i}$ yields $\dot{r}^{i}=Hr^{i}+v^{i}$, which is the sum of the Hubble flow and the deviation from it, i.e.\ the peculiar velocity. If we assume that $|\dot{r}^{i}|\ll c$, our approach would only be valid on scales much smaller than the Hubble horizon. However, if we only assume that $v\ll c$, we are not restricted to specific scales.

\noindent\emph{Newtonian regime:} at leading order, the Einstein Field Equations yield the standard equations of Newtonian cosmology \cite{Milillo:2015cva}. One obtains the Poisson equation as well as constraint equations demanding $V_{N}=U_{N}$. However, the Einstein Field Equation involving $G^{0}_{\, i}$ also has leading order contributions, which determine the frame-dragging potential $B_{Ni}$. It follows that $B_{Ni}$ is sourced by the purely Newtonian quantities $\bar{\rho}v_{i}$.
For this reason, this leading order contribution carries the subscript $N$, even if frame-dragging is a purely relativistic effect.

\noindent\emph{Relativistic regime:} we define ``resummed variables'' $\phi=-\left(U_N+\frac{2}{c^2}U_P\right)$ and $\psi=-\big(V_N+\frac{2}{c^2}V_P\big)$. When one linearises the Einstein Field Equations substituting the resummed variables, we recover the first order of standard relativistic perturbation theory (see~\cite{Milillo:2015cva} for more detail).

 The validity on all scales is especially beneficial for the analysis of weak gravitational lensing, since as discussed in the introduction, the integral along the line of sight mixes small scale and large scale effects.


\section{Derivation of the Magnification Matrix}
\label{sec:sec3}
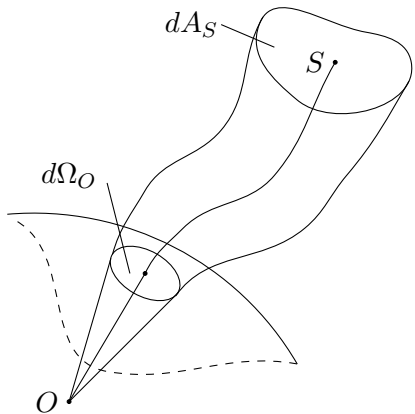
\begin{wrapfigure}{l}{7cm}
\begin{tikzpicture}
\draw (4,0.5) arc [radius=4, start angle=30, end angle= 95];
\draw [dashed] (4,0.5) to [out=170,in=-10] (1.5,0.4) to [out=170,in=-30] (0.3,2.4);
\draw[fill] (1,0) circle [radius=0.025];
\node [left] at (1,0) {$O$};
\draw (1.55,1.9) -- (1,0) -- (2.44,1.45);
\draw [rotate around={-30:(2,1.7)},black](2,1.7) ellipse (5mm and 3mm); 
\draw (1,0) -- (2,1.7);
\draw[fill] (2,1.7) circle [radius=0.025];
\draw (1.8,1.7) -- (1.5,2.9);
\node [left] at (1.5,3) {$d\Omega_O$};
\draw[fill] (4.5,4.5) circle [radius=0.025];
\node [left] at (4.5,4.5) {$S$};
\draw (3.5,5) to [out=70,in=170] (4.9,4.9) to [out=-10,in=100](5.5,4.5) to [out=-80,in=-40](4,4) to [out=140,in=-110](3.5,5);
\draw (3.55,5.1) to [out=-120,in=50](3,3.6) to [out=-130,in=60](2,2.8) to [out=-120,in=75](1.57,1.95);
\draw (4.5,4.5) to [out=-120,in=50](3.7,3) to [out=-130,in=45](2.5,2.3) to [out=-135,in=65](2,1.7);
\draw (5.47,4.25) to [out=-120,in=50](4.65,3) to [out=-130,in=40](4.1,2.3) to [out=-140,in=55](2.4,1.4);
\draw (3.7,4.7) -- (3,5);
\node [left] at (3.1,5) {$dA_S$};
\end{tikzpicture}
\caption{The surface $dA_S$ is related to the solid angel $d\Omega_O$ at the observer $O$.}
\end{wrapfigure}
In weak lensing we study the distortion of images due to inhomogeneities between the source and the observer. We follow the propagation of a light bundle (i.e.\ a collection of nearby light rays) in a perturbed geometry \cite{1992grle.book.....S,Wald:1984rg,1961RSPSA.264..309S,1994CQGra..11.2345S,2006PhR...429....1L,2001PhRvD..63b3004U,Yoo:2018qba,Grimm:2018nto}. We consider two neighbouring geodesics $x^{\mu}(\lambda)$ and $y^{\mu}(\lambda)=x^{\mu}(\lambda)+\xi^{\mu}(\lambda)$, which start at $\lambda=0$ at the observer $O$ ($\lambda$ is an affine parameter). For an infinitesimal light bundle, the connection vector $\xi^{\mu}$ lies on the null surface, $\xi^{\mu}k_\mu=0$, where $k^\mu=dx^\mu/d\lambda$ denotes the tangent vector to the congruence of light rays.  The evolution of the connection vector along the photon geodesics is given by the Sachs equation~\cite{1961RSPSA.264..309S}
\begin{align}
\frac{D^{2}\xi^{\mu}}{D\lambda^{2}}=R^{\mu}_{\  \nu\alpha\beta}\xi^{\beta}k^{\nu}k^{\alpha}\,,\label{eq:gde}
\end{align}
where $D/D\lambda\equiv k^\mu \nabla_\mu$ denotes the covariant derivative along the geodesics.

Let us assume that a light beam is emitted at the source $S$ and measured at the observer $O$. Furthermore, $u_{O}^{\mu}$ denotes the 4-velocity of the observer. We define an orthonormal spacelike basis $n^{\mu}_{a}$ with $a=1,\,2$, which is orthogonal to $k^{\mu}$ and to $u_{O}^{\mu}$ at the observer, and which obeys $g_{\mu\nu} n_a^\mu n_b^\mu=\delta_{ab}$. At the observer $\{n_{1}^{\mu},n_{2}^{\mu},k^{\mu},u_{O}^{\mu}\}$ form a basis, which can be parallel transported along the geodesics
\begin{align}
\frac{Dn_{a}^{\mu}}{D\lambda}=0\quad \text{and}\quad\frac{Du_{O}^{\mu}}{D\lambda}=0. \label{eq:dn}
\end{align}
It is standard to refer to the two dimensional space spanned by $n^{\mu}_{a}(\lambda)$ as the \emph{screen space}.
The deviation vector expressed in this basis reads
\begin{align}
\xi^{\mu}=\xi^{a}_{\textbf{n}}n_{a}^{\mu}+\xi_{\textbf{k}}k^{\mu}+\xi_{\textbf{u}}u^{\mu}_{O},\label{eq:xixi}
\end{align}
with $\xi^{a}_{\textbf{n}}(0)=0$,  $\xi_{\textbf{k}} (0)=0$, and $\xi_{\textbf{u}}(0)=0$. 
From $\xi^{\mu}k_{\mu}=0$ it follows that $\xi_{\textbf{u}}(\lambda)=0$ for all $\lambda$.
Substituting~\eqref{eq:xixi} into~\eqref{eq:gde} we obtain an evolution equation for the two components $\xi^{a}_{\textbf{n}}$ with $a=1,2$~\cite{2010PhRvD..81h3002B,1961RSPSA.264..309S}:
\begin{align}
\frac{D^{2}\xi^{c}_{\textbf{n}}}{D\lambda^{2}}=\mathcal{R}^{c}_{\enspace a}\xi^{a}_{\textbf{n}},\enspace \text{with}\enspace\mathcal{R}^{c}_{\enspace a}= R^{\mu}_{\enspace \nu\alpha \beta}k^{\nu}k^{\alpha}n^{c}_{\mu}n_{a}^{\beta}\label{eq:gde2}.
\end{align}
Let $\theta^{a}_{O}$ be the vectorial angle between two neighbouring geodesics at the observer $O$
\begin{equation}
\left.\theta^{a}_{O}\equiv\frac{d\xi^{a}_{\textbf{n}}}{d\lambda}\right |_{\lambda=0}\, .
\end{equation} 
Since eq.~\eqref{eq:gde2} is a linear second-order differential equation, with initial conditions $\xi^{a}_{\textbf{n}}(0)=0$, the solution can be written as
\begin{align}
\xi^{a}_{\textbf{n}}(\lambda)=\DD_{ab}(\lambda)\theta^{b}_{O}\, .\label{eq:xiD}
\end{align}
The matrix $\DD_{ab}$ denotes the linear Jacobi mapping, which relates the observed angle $\theta^{b}_{O}$ between two neighbouring geodesics at the observer $O$ to the distance $\xi^{a}_{\textbf{n}}$ between the geodesics at the source $S$. $\DD_{ab}$ is called the magnification matrix: it gives a measure of the distortion between the shape of the observed image and the shape of the source. Note that the indices $a$ and $b$ are raised and lowered with $\delta_{ab}$.

Substituting \eqref{eq:xiD} into~\eqref{eq:gde2} we obtain an evolution equation for $\DD_{ab}$~\cite{2010PhRvD..81h3002B,1994CQGra..11.2345S,2006PhR...429....1L,2001PhRvD..63b3004U}
\begin{align}
\frac{d^{2}}{d\lambda^{2}}\DD_{ab}=\RR_{ac}\DD_{c b}\, ,\label{eq:dd}
\end{align}
with initial conditions $\DD_{ab}(0)=0$ and $\left.\frac{d\DD_{ab}}{d\lambda}\right |_{\lambda=0}=\delta_{ab}$. The affine parameter $\lambda$ is a perturbative quantity. In order to take these perturbations into account, we rewrite the evolution equation~\eqref{eq:dd} in terms of the unperturbed parameter $\chi$, defined as $\chi\equiv c\left(\eta_{O}-\eta\right)$, where $\eta$ denotes conformal time. Note that we choose our time coordinate as $x^{0}=c\left(\eta_{O}-\eta\right)=\chi$ (see section~\ref{sec:31} for a more detailed discussion). As a consequence $d\chi/d\lambda=dx^0/d\lambda=k^0$.
The total derivative with respect to $\lambda$ transforms then into
\begin{align}
\frac{d}{d\lambda}=\frac{d\chi}{d\lambda}\frac{d}{d\chi}=k^{0}\frac{d}{d\chi}\label{eq:derlc}\, .
\end{align}
Substituting~\eqref{eq:derlc} into~\eqref{eq:dd} we obtain
\begin{align}
\frac{d^{2}}{d\chi^{2}}\DD_{ab}+\frac{1}{k^{0}}\frac{dk^{0}}{d\chi}\frac{d}{d\chi}\DD_{ab}=\frac{1}{(k^{0})^{2}}\RR_{ac}\DD_{cb}\, .\label{eq:ddd}
\end{align} 

To solve eq.~\eqref{eq:ddd}, we need to calculate $k^0$ and $\RR_{ab}$ and solve the equation order by order in powers of $1/c$. The calculation can be simplified by using the fact that null geodesics are not affected by conformal transformations. As a consequence, the calculation can be performed without the Friedmann expansion, i.e.\ for the metric $ds^2$ defined through 
\begin{equation}
d\tilde{s}^2=a^2 ds^2\, ,  
\end{equation}
where $d\tilde{s}^2$ is the line element associated to the metric~\eqref{eq:g00}-\eqref{eq:gij}. 
The effect of the expansion can then simply be taken into account at the end by rescaling the mapping by the conformal factor~\cite{2010PhRvD..81h3002B} 
\begin{equation}
\tilde{\DD}_{ab}(\chi_S)=a(\chi_S)\DD_{ab}(\chi_S)\, ,   \label{eq:tildeD} 
\end{equation}
where $\DD_{ab}$ denotes the Jacobi mapping for the metric $ds^2$, and $\tilde{\DD}_{ab}$ is the expression for the metric $d\tilde s^2$.

The matrix $\tilde{\DD}_{ab}(\chi_S)$ represents the Jacobi mapping for sources situated at constant conformal time $\chi_S$. However, observationally we select sources at constant redshift $z_S$. Since the observed redshift is itself affected by perturbations, $z_S=\bar{z}_S+\delta z_S$, where $1+\bar z_S=1/a_S$, this will modify the expression of the Jacobi mapping~\footnote{Note that we normalise the scale factor to 1 today: $a_O=1$.}. In particular, we can write
\begin{align}
\tilde{\DD}_{ab}\left(\chi_{S}\right)&=\tilde{\DD}_{ab}\left(\chi_{S}\left(\bar{z}_{S}\right)\right)= \tilde{\DD}_{ab}\left(\bar{z}_{S}\right)=
\tilde{\DD}_{ab}\left(z_{S}-\delta z_S\right)\notag\\
&=\tilde{\DD}_{ab}\left(z_{S}\right)-\frac{d}{dz_{S}}\tilde{\DD}_{ab}\left(z_{S}\right)\delta z_{S}+\frac{1}{2}\frac{d^2}{dz_{S}^2}\tilde{\DD}_{ab}\left(z_{S}\right)\delta z_{S}^2\notag\\
&-\frac{1}{3!}\frac{d^3}{dz_{S}^3}\tilde{\DD}_{ab}\left(z_{S}\right)\delta z_{S}^3+\frac{1}{4!}\frac{d^4}{dz_{S}^4}\tilde{\DD}_{ab}\left(z_{S}\right)\delta z_{S}^4
+\Oo{5}\, ,\label{eq:Dz}
\end{align}
where in the second and third lines the matrix $\tilde{\DD}_{ab}\left(z_{S}\right)$ and its derivatives~\footnote{The derivatives in eq.~\eqref{eq:Dz} are formally given by $d^n\tilde D_{ab}(\bar z_S)/d \bar z_S^n\Big|_{\bar z_S=z_S}$} are formally given by eqs.~\eqref{eq:ddd} and~\eqref{eq:tildeD} where $\chi_S$ can now be interpreted as $\chi(z_S)$ and $1+\bar z_S$ can be replaced by $1+z_S$.

The Jacobi mapping $\tilde \DD_{ab}(z_S)$ is usually decomposed into a convergence $\kappa$, a shear $\gamma=\gamma_1+i\gamma_2$ and a rotation $\omega$
\begin{align}
\tilde\DD_{ab}=\frac{\chi(z_S)}{1+z_S}\begin{pmatrix} 1-\kappa - \gamma_1 & -\gamma_2-\omega\\
-\gamma_2 +\omega & 1-\kappa +\gamma_1 \end{pmatrix}\, .\label{eq:oJM}
\end{align}
The prefactor $\chi(z_S)/(1+z_S)$ represents the magnification of images due to the background expansion of the Universe, for sources situated at the observed redshift $z_S$. The convergence $\kappa$ denotes the magnification or demagnification of images due to perturbations. The shear $\gamma$ is the trace-free, symmetric part of $\tilde\DD_{ab}$ and refers to the change in the shape. The rotation $\omega$, the antisymmetric part of $\tilde\DD_{ab}$, represents a rotation without any change in the shape. Note that what we observe when we measure the ellipticity of galaxies is not directly the shear $\gamma$ but rather the reduced shear $g$ which is the ratio of the anisotropic and isotropic deformations~\cite{Bartelmann:1999yn, 2010PhRvD..81h3002B,Kilbinger:2014cea}
\begin{equation}
g\equiv \frac{\gamma}{1-\kappa}\, .    
\end{equation}
The rotation $\omega$ does in principle contribute to ellipticity orientation (see~\cite{2010PhRvD..81h3002B}). However, we will see that the rotation is of order $\Oo{4}$ and contributes consequently to the ellipticity at the order $\Oo{6}$. Therefore our ellipticity measurement is dominated by the reduced shear.

Following~\cite{2010PhRvD..81h3002B}, the convergence, shear and rotation can be expressed in terms of the spin-0 and spin-2 components of the Jacobi mapping $\tilde\DD_{ab}$
\begin{align}
{}_{0}\tilde\DD\equiv\, &\tilde\DD_{11}+\tilde\DD_{22}+i\left(\tilde\DD_{12}-\tilde\DD_{21}\right)\, , \label{eq:0D}  \\
{}_{2}\tilde\DD\equiv\, & \tilde\DD_{11}-\tilde\DD_{22}+i\left(\tilde\DD_{12}+\tilde\DD_{21}\right)\,.\label{eq:2D}
\end{align}
The spin-0 field contains the contribution from the magnification and the rotation, whereas the spin-2 field is related to the shear distortion. Comparing eqs.~\eqref{eq:0D} and~\eqref{eq:2D} with~\eqref{eq:oJM} we obtain 
\begin{align}
\kappa=&1-\frac{1+z_S}{2\chi_{S}}\textrm{Re}\big[{}_{0}\tilde\DD\big]\,,        &\omega=&-\frac{1+z_S}{2\chi_{S}}\textrm{Im}\big[{}_{0}\tilde\DD\big]\,, \label{eq:kappa}\\
\gamma=&-\frac{1+z_S}{2\chi_{S}}{}_{2}\tilde\DD\,,\label{eq:gamma} 
\end{align}
where Re and Im denote the real and imaginary parts of the spin-0 component. The reduced shear then becomes
\begin{equation}
g=- \frac{{}_{2}\tilde\DD}{\textrm{Re}\big[{}_{0}\tilde\DD\big]}\, .\label{eq:rs}
\end{equation}
The advantage of expressing the shear in terms of the spin-2 component of the magnification matrix is that this allows us to expand it onto spin-weighted spherical harmonics. We can then uniquely decompose it into an E-component (or scalar gradient) and a B-component (or curl)~\cite{Newman:1966ub}. Contrary to the $\gamma_1$ and $\gamma_2$ components, the E and B components are invariant under a rotation of the coordinate system around the line of sight. As a consequence, this decomposition is particularly well adapted to a full-sky survey where the line of sight direction varies from patch to patch of the sky.

%
%
\subsection{Resolution of $\DD_{ab}(\chi_S)$ up to order $\frac{1}{c^4}$}
\label{sec:31}
In this section, we compute the Jacobi mapping for the orders $\Oo{2}$, $\Oo{3}$, and $\Oo{4}$. We denote by a bar, $\bar f$, quantities associated with the background metric, and by, $f^{(n)}$, quantities of order $\mathcal{O}\left(\frac{1}{c^n}\right)$. We start by computing $k^\mu$ and $n_a^\mu$ in section~\ref{sec:311}, and $\RR_{ab}$ in section~\ref{sec:312}. Then, in section~\ref{sec:313} we use these results to solve for the Jacobi mapping $\DD_{ab}$ up to order $\Oo{4}$.
From eq.~\eqref{eq:ddd} we see that we need expressions for $k^{0}$ and $\RR_{ab}$ up to order $\Oo{4}$. Since $\RR_{ab}=R_{\mu\nu\alpha\beta}k^\nu k^\alpha n_a^\mu n_b^\beta$, this requires us to calculate $k^i$ and $n_a^\mu$ up to order $\Oo{2}$ (since $R_{\mu\nu\alpha\beta}$ is at least of order $\Oo{2}$).

We choose the coordinate system $\big(x^0=c(\eta_0-\eta),x^i\big)$, so that $x^\mu$ has dimension of length. The derivative of any function $f(x^0,x^i)$ with respect to $x^0$ and $x^i$ (denoted respectively by $f_{,0}$ and $f_{,i}$) does not change its order in the expansion $1/c$. The photon wave vector $k^\mu$ is defined as
\begin{equation}
k^\mu=\frac{dx^{\mu}}{d\lambda}\, ,
\end{equation}
where $\lambda$ is an affine parameter with $\lambda_O=0$. We choose $\lambda$ with dimension of length so that $k^0$ and $k^i$ are both dimensionless. The background $\bar k^0$ and $\bar k^i$ are of order zero in the expansion $1/c$. We have furthermore the freedom to choose $\bar\lambda$ such that $\bar k^0=\delta_{ij}\bar k^i\bar k^j=1$. In the following we keep track however of $\bar k^0$ and $\bar k^i$ as a consistency check. We know indeed that the Jacobi mapping has to be independent of the normalisation of $\bar k^0$ and $\bar k^i$. As such it can only depend on $\bar k^\mu$ through the ratio $\bar k^i/\bar k^0$.

Note that in the calculation of the Jacobi mapping, we do not consider the perturbations at the observer, contrary to what is done (at linear order in standard PT) in~\cite{Yoo:2018qba}. The reason is the following: the scalar potentials at the observer do not contribute to the shear or rotation, because they affect all light rays around the observer in the same way. As such they do not distort an isotropic bundle of light and are therefore irrelevant for the shear and rotation. The scalar potentials at the observer do however contribute to the isotropic part of the Jacobi mapping. However, when one constructs estimators to measure the convergence, one always subtracts the average size of galaxies at redshift $z$. Since the scalar potentials at the observer do contribute to this average size, they are by construction {\it removed} from the estimator. This procedure reflects the well-known fact that the monopole part of the perturbations cannot be observationally separated from the background contribution. Hence to consistently compare our theoretical modelling with observations, we need to remove the scalar contributions at the observer.

The situation with the vector and tensor contributions at the observer is different: these contributions do affect the Jacobi mapping (including the anisotropic part), and contrary to the scalar contributions, they are not removed when one subtracts the angular average. However, similarly to what happens with the CMB temperature, the vector contribution at the observer is degenerate with the effect of the observer peculiar velocity, which strongly dominates the signal. This effect generates a dipolar modulation around the observer, that can be removed from the Jacobi mapping by fitting for a dipole. In the following we neglect therefore both the vector contribution at the observer and the observer peculiar velocity. 

Finally, the tensor contribution at the observer does contribute to the Jacobi mapping. This contribution has been calculated in detail in~\cite{Schmidt:2012nw}. Here we do not re-derive these terms at the observer, but we emphasize that they should be added to the final expression for consistency.

\subsubsection{Calculation of the wave vector $k^{\mu}$ and the screen-space basis $n^{\mu}_{a}$}
\label{sec:311}
We start by calculating the wave vector $k^{\mu}$. We use the fact that it is parallel transported along the null geodesic
\begin{equation}
\frac{Dk^{\mu}}{D\lambda}=0\, . \label{eq:gek}
\end{equation}
Analogously to the previous section, we rewrite the derivative with respect to $\lambda$ as a derivative with respect to $\chi=c(\eta_{0}-\eta)$. The geodesic equation~\eqref{eq:gek} becomes
\begin{equation}
\frac{dk^{\mu}}{d\chi}=-\Gamma^{\mu}_{\, \nu\alpha}\frac{k^{\nu}k^{\alpha}}{k^0}\label{eq:gege}.
\end{equation}
The solution for $k^i$ up to order $\Oo{3}$ reads 
\begin{align}
k^{i}=&\bk{i}\left(1-\frac{2}{c^2}V_{N}\right)+\frac{2}{c^2}\int_{0}^{\chi}d\chi'W_{N}^{,i}\bk{0}  
+\cc{3}B^{i}_N\bk{0}- \cc{3}\int \bound d\chi'B_{Nm}^{\hspace{2mm} ,i}\bk{m}  
\,, \label{eq:ki}
\end{align}
where we have defined the Weyl potential $W_{A}=\frac{1}{2}\left(U_{A}+V_{A}\right)$ for $A=N,P$. 

To solve for $k^0$ up to order $\Oo{4}$ we need to go beyond the so-called Born approximation, and integrate eq.~\eqref{eq:gege} along the perturbed geodesic. We have
\begin{equation}
k^0(\chi)=\bar k^0+\int_0^\chi d\chi' G(\chi')\, ,  \label{eq:k0gen} \end{equation}
where we have defined
\begin{equation}
G=-\Gamma^{0}_{\, \nu\alpha}\frac{k^{\nu}k^{\alpha}}{k^0}\, .
\end{equation}
Since $G$ is at least of order $\Oo{2}$, the solution at order $\Oo{2}$ and $\Oo{3}$ can be found by integrating~\eqref{eq:k0gen} along the background geodesic. We obtain 
\begin{align}
k^{0(2)}&= \frac{2}{c^2}U_N\bk{0}-\frac{2}{c^2}\int\bound d\chi' W_{N,0}\bk{0}  \, ,\label{eq:k02}  \\
k^{0(3)}&=-\frac{1}{c^3}\int\bound d\chi' B_{Ni,j}\frac{\bk{i}\bk{j}}{\bk{0}}  \, . 
\end{align}

At order $\Oo{4}$ on the other hand, we need to integrate along the perturbed trajectory $x^\mu(\chi)=\bar{x}^\mu(\chi)+\delta x^\mu(\chi)$. We have
\begin{equation}
G\big(x^\mu(\chi')\big)= G\big(\bar x^\mu(\chi') +\delta x^\mu(\chi')\big)
=G\big(\bar x^\mu(\chi')\big)+\delta x^\mu(\chi')\partial_\mu G\big(\bar x^\mu(\chi')\big)+\Oo{6}\, . \label{eq:Fexpand}
\end{equation}
Since $G$ is at least of order $1/c^2$, it is enough to consider only the first term of the Taylor expansion in eq.~\eqref{eq:Fexpand}. We need to calculate $\delta x^\mu(\chi)$ at order $1/c^2$. We have
\begin{equation}
\frac{d x^\mu}{d\chi}=\frac{dx^\mu}{d\lambda}\frac{d\lambda}{d\chi}=\frac{k^\mu}{k^0}\, .    
\end{equation}
Using eq.~\eqref{eq:ki} and~\eqref{eq:k02} we obtain
\begin{align}
\dx^{i}=&-\cgen{2}{2}\int^{\chi}_{0}d\chi'W_{N}\frac{\bk{i}}{\bk{0}}+\cgen{2}{2}\int_{0}^{\chi}d\chi'\int^{\chi'}_{0}d\chi''\left(W_{N}^{,i}-W_{N,j}\frac{\bk{i}\bk{j}}{\left(\bk{0}\right)^2}\right)\,,\label{eq:deltaxi}\\
\delta x^{0}=&0\,. \label{eq:deltax0}
\end{align}
Inserting this into eqs.~\eqref{eq:Fexpand} and~\eqref{eq:k0gen} we obtain at order $\Oo{4}$
\begin{align}
k^{0(4)}=& \,\frac{4}{c^4}U_P\bk{0}-\frac{4}{c^4}\int\bound d\chi' W_{P,0}\bk{0}+\frac{2}{c^4}\bk{0}\left(  U_N-\int\bound d\chi'W_{N,0}\right)^2\notag\\
&-\cc{4}\frac{1}{2\bk{0}} h_{ij}\bk{i}\bk{j}+\cc{4}\frac{1}{2\left(\bk{0}\right)^2}\int\bound d\chi' h_{ij,m}\bk{i}\bk{j}\bk{m} +\notag\\
&  -\cgen{4}{4} U_{N,i}\bk{i}  \int^{\chi}_{0}d\chi'\left[ W_{N} -\left(\chi-\chi'\right)\left(\bk{0} W_{N}^{,i}-W_{N,j}\frac{\bk{i}\bk{j}}{\bk{0}}\right)    \right]+\notag\\
&+\cgen{4}{4}\int \bound d\chi'  U_{N,i}\bk{i}\left[ W_{N} - \int_{0}^{\chi'}d\chi''\left(\bk{0} W_{N}^{,i}-W_{N,j}\frac{\bk{i}\bk{j}}{\bk{0}}\right)          \right]+\notag\\
 &+\cgen{4}{4}\int \bound d\chi' W_{N,0i} \bk{i} \int^{\chi'}_{0}d\chi''\left[    W_{N}   -  \left( \chi' - \chi''   \right)\left(\bk{0} W_{N}^{,i}-W_{N,j}\frac{\bk{i}\bk{j}}{\bk{0}}\right)     \right] \, . \label{eq:k0}
\end{align}

We then calculate the screen vectors $n^{\mu}_{a}$. They are constructed at the observer to be orthogonal to $u_O^\mu(0)$ and $k^\mu(0)$, and then parallel transported along the geodesics  
\begin{equation}
\frac{Dn^{\mu}_{a}}{D\lambda}=0\, . \label{eq:genn}
\end{equation}
The solutions up to order $\Oo{2}$ are
\begin{align}
n^{0}_{a}=&\cc{2}\int^{\chi}_{0}d\chi'U_{N,i}\bar{n}^{i}_{a}\,,\label{eq:n0}\\
n^{i}_{a}=&\left(1-\frac{V_{N}}{c^{2}}\right)\bar{n}^{i}_{a}-\frac{1}{c^{2}}\frac{\bk{i}}{\bk{0}}\int^{\chi}_{0}d\chi'V_{N,j}\bar{n}^{j}_{a}\, .\label{eq:ni}
\end{align}
\subsubsection{Calculation of $\RR_{ab}$}
\label{sec:312}
We now compute the contracted Riemann tensor  $\mathcal{R}_{ab}$ up to order $\Oo{4}$. The Riemann tensor is at least of order $\Oo{2}$. As a consequence, $\mathcal{R}_{ab}$ at order $\Oo{2}$ and $\Oo{3}$ will be obtained by contracting the Riemann tensor at the order $\Oo{2}$ and $\Oo{3}$ respectively, with the background $k^\mu$ and $n_a^\mu$. At the order $\Oo{4}$ on the other hand, we will also have contributions from $k^\mu$ and $n_a^\mu$ at order $\Oo{2}$.

To simplify the calcualtion we perform the following conformal transformation
\begin{align}
ds^{2}=&-e^{-\frac{2}{c^2}U_{N}-\frac{4}{c^4}U_{P}}c^2 d\eta^{2}-\frac{2}{c^3}B^{N}_{i}dx^{i}cd\eta+\left(e^{\frac{2}{c^2}V_{N}+\frac{4}{c^4}V_{P}}\delta_{ij}+\cc{4}h_{ij}\right)dx^{i}dx^{j}\label{eq:gtilde}\\
=&e^{\frac{2}{c^2}V_{N}+\frac{4}{c^4}V_{P}}\left[-e^{-\frac{4}{c^2}W_{N}-\frac{8}{c^4}W_{P}}c^2d\eta^{2}-\frac{2}{c^3}B^{N}_{i}dx^{i}cd\eta+\left(\delta_{ij}+\cc{4}h_{ij}\right)dx^{i}dx^{j}\right]+\Oo{5}\, .\notag
\end{align}
We first compute the Riemann tensor in the metric $\hat{g}_{\aaa\bbb}$ defined through $g_{\aaa\bbb}=\Omega^{2}\hat{g}_{\aaa\bbb}$, with $\Omega=e^{\frac{1}{c^2} V_{N}+\frac{2}{c^4}V_{P}}$. Then we use the relation between conformally transformed Riemann tensors~\cite{Wald:1984rg} to obtain the result for the metric $g_{\ccc\bbb}$
\begin{align}
R_{\ccc\bbb\ddd \aaa}=&\hat{R}_{\aaa\bbb\ccc\ddd}-2g_{\ccc\left[\ddd\right.}\nabla_{\left.\aaa\right]}\nabla_{\bbb}\ln \Omega+2g_{\bbb\left[\ddd\right.}\nabla_{\left.\aaa\right]}\nabla_{\ccc}\ln \Omega-2\left(\nabla_{\left[\ddd\right.}\ln \Omega\right)g_{\,\left. \aaa\right]\ccc}\nabla_{\bbb}\ln \Omega\notag\\
&+2\left(\nabla_{\left[\ddd\right.}\ln \Omega\right)g_{\left. \aaa\right]\bbb}\nabla_{\ccc}\ln \Omega+2g_{\bbb\left[\ddd\right.}g_{\left. \aaa\right]\ccc}g^{\epsilon \zeta}\left(\nabla_{\epsilon}\ln \Omega\right)\nabla_{\zeta}\ln \Omega\, . \label{eq:RgenW}
\end{align}
We obtain up to order $\Oo{4}$
\begin{align}
\RR_{ab}^{\left(2\right)}
=&\frac{\left(\bk{0}\right)^{2}}{c^2}\left[2\bar{n}^{i}_{a}\bar{n}^{j}_{b} W_{N,ij}+\delta_{ab}\frac{d^2 V_{N}}{d\chi^2}\right]\, ,\label{eq:R2}\\
\RR^{(3)}_{ab}=&\frac{\left(\bk{0}\right)^2}{c^3}\bar{n}_a^{i}\bar{n}_{b}^{j}\left(\frac{dB^{N}_{\left(i,j\right)}}{d\chi}-\frac{\bk{k}}{\bk{0}}B_{k,ij}^N\right),\label{eq:R3}\\
\RR_{ab}^{\left(4\right)}=&\frac{\left(\bk{0}\right)^2}{c^4}\delta_{ab}\left[2\frac{d^2}{d\chi^2}V_{P}-\frac{d}{d\chi}V_{N}\frac{d}{d\chi}V_{N}+2\left(\frac{dU_N}{d\chi}-W_{N,0}\right)\frac{dV_N}{d\chi}\right]\notag\\
&+\frac{\left(\bk{0}\right)^2}{c^4}\bar{n}^{i}_{a}\bar{n}^{j}_{b}
\Bigg[4W_{P,ij}-4 W_{N,ij}^2 +4W_{N,i} W_{N,j}-4W_{N,ij}V_N \notag\\
&\hspace{2.2cm}-4W_{N,ij}\bk{0}\left(\bk{i}\bar{n}^{m}_{a}\bar{n}^{j}_{b} +\bk{j}\bar{n}^{m}_{b}\bar{n}^{i}_{a}\right) \int_{0}^{\chi} W_{N,m} d\chi'\Bigg]\notag\\
&+\frac{\left(\bk{0}\right)^2}{c^4}\bar{n}^{i}_{a}\bar{n}^{j}_{b}\frac{1}{2}\left[\frac{d^2}{d\chi^2}h_{ij}-\frac{1}{\bk{0}}\frac{d}{d\chi}\left(h_{jp, i}+h_{i p,j}\right)\bk{p}+ h_{mp,i j}\frac{\bar k^{p}\bar k^{m}}{\left(\bk{0}\right)^2}\right]\, .\label{eq:R4}
\end{align}

\subsubsection{Calculation of the Jacobi mapping $\DD_{ab}(\chi_S)$} 
\label{sec:313}
In subsections~\ref{sec:311} and~\ref{sec:312} we have derived expressions for $k^\mu$ and  $n^{\mu}_a$ in eqs.~\eqref{eq:k0} to~\eqref{eq:ni}, and for $\RR_{ab}$ up to order $\Oo{4}$ in eqs.~\eqref{eq:R2}, \eqref{eq:R3}, and~\eqref{eq:R4}. We now use these results and substitute them into the evolution equation~\eqref{eq:ddd}, which we solve order by order in the expansion $1/c$.
\subsubsection*{Order $\Oo{2}$}
At this order, the evolution equation~\eqref{eq:ddd} for $\DD_{ab}^{(2)}$  reduces to
 \begin{align}
\frac{d^{2}}{d\chi^{2}}\DD_{ab}^{(2)}=&-\frac{1}{\bk{0}}\frac{dk^{0(2)}}{d\chi}\delta_{ab}+\frac{\chi}{\left(\bk{0}\right)^2}\RR_{ab}^{(2)},\label{eq:ddddd}
\end{align}
 where we have used that $\bar{\DD}_{ab}=\chi\delta_{ab}$. We integrate \eqref{eq:ddddd} two times
\begin{equation}
\DD_{ab}^{(2)}(\chi_S)=\frac{1}{\bk{0}}\int^{\chi_{S}}_{0}d\chi \left(2-k^{0(2)}\right)\delta_{ab}+\frac{1}{\left(\bk{0}\right)^2}\int_{0}^{\chi_{S}}d\chi\left(\chi_{S}-\chi\right)\chi\RR_{ab}^{(2)}\, ,\label{eq:d2sol}
\end{equation}
and substitute \eqref{eq:R2} and \eqref{eq:k0} up to order $\Oo{2}$ into \eqref{eq:d2sol} to obtain
\begin{align}
\DD_{ab}^{(2)}\left(\chi_{S}\right)=&\frac{1}{c^2}\chi_{S}V_{NS}\delta_{ab} - \frac{2}{c^2} \int^{\chi_{S}}_{0}d\chi \left[W_{N}+(\chi_S-\chi) W_{N,i}\fkk{i}\right]\delta_{ab}\notag\\
&+\frac{2}{c^2}\int_{0}^{\chi_{S}}d\chi\left(\chi_{S}-\chi\right)\chi \bar{n}^{i}_{a}\bar{n}^{j}_{b}W_{N,ij}\,.\label{eq:DD2}
\end{align}
The solution~\eqref{eq:DD2} coincides with the first order solution for $\DD_{ab}$ using standard perturbation theory (SPT) \cite{2010PhRvD..81h3002B,Bonvin:2005ps,Bonvin:2008ni}. This follows from the fact that the metric at order $\Oo{2}$ in the PF formalism is mathematically identical to the metric at first order in SPT. At this order, the difference between SPT and the PF framework becomes apparent when one uses field equations, to relate the metric to the matter content in the Universe. While in SPT the density and velocity are perturbative quantities, in PF these quantities are unperturbed and e.g.\ the density contrast can become larger than 1.

In \eqref{eq:DD2}, the Jacobi mapping $\DD_{ab}$ involves the Weyl potential $W_{N}$ as well as the scalar potential $V_{N}$\footnote{At this order, the Einstein field equations yield that $V_{N}=U_{N}$ and therefore $W_{N}=V_{N}=U_{N}$. We decided to keep $U_{N}$ and $V_{N}$ throughout the calculations in order to keep the result as general as possible. For example in modified gravity theories the scalar potentials may differ.}. From the metric~\eqref{eq:gtilde}, using the fact that null geodesics are invariant under conformal transformations, we would expect that $\DD_{ab}$ depends only on the Weyl potential $W_{N}$. However, as previously noted in~\cite{2010PhRvD..81h3002B}, the term involving $V_{N}$ is generated by the parallel transport of the basis $n^{\mu}_{a}$, which is not conformally invariant.

Finally, let us note that, as expected, the Jacobi mapping does not depend on the choice of normalisation for $\bar k^\mu$ since it depends only on the ratio $\bar k^i/\bar k^0$.

\subsubsection*{Order $\Oo{3}$}
At order $\Oo{3}$, \eqref{eq:ddd} yields
\begin{align}
\frac{d^{2}}{d\chi^{2}}\DD^{(3)}_{ab}+\frac{1}{\bk{0}}\frac{dk^{0(3)}}{d\chi}\frac{d}{d\chi}\bar{\DD}_{ab}=\frac{1}{(\bk{0})^{2}}\RR_{ac}^{(3)}\bar{\DD}_{cb}\,.\label{eq:d3}
\end{align} 
Analogously to the previous order, we integrate along the background geodesic and substitute~\eqref{eq:k0} and~\eqref{eq:R3} into~\eqref{eq:d3} to obtain
\begin{align}
\DD_{ab}^{(3)}\left(\chi_{S}\right)=&-\frac{1}{\bk{0}}\int_{0}^{\chi_{S}}d\chi k^{0(3)}\delta_{ab}+\frac{1}{\left(\bk{0}\right)^2}\int_{0}^{\chi_{S}}d\chi \left(\chi_{S}-\chi\right)\chi\RR_{ab}^{(3)}\notag\\
=&\cc{3}\int_{0}^{\chi_{S}}d\chi\left(\chi_{S}-\chi\right) B_{Ni,j}\frac{\bk{j} \bk{i}}{\left(\bk{0}\right)^2}\delta_{ab}\notag\\
&+\cc{3}\int_{0}^{\chi_{S}}d\chi \left(\chi_{S}-\chi\right)\chi\bar{n}^{i}_{a}\bar{n}_{b}^{j}\left[\frac{dB^{N}_{(i,j)}}{d\chi}-\frac{\bk{m}}{\bk{0}}B_{m,ij}^N\right]\, .\label{eq:DD3}
\end{align} 
The last line in~\eqref{eq:DD3} coincides with the vector part in \cite{2010PhRvD..81h3002B}, in which the shear has been computed up to second order in SPT. The second line in~\eqref{eq:DD3} will contribute only to the convergence, since it is proportional to $\delta_{ab}$. 


\subsubsection*{Order $\Oo{4}$}
At order $\Oo{4}$, we have to go beyond Born approximation. Following~\cite{2010PhRvD..81h3002B} we rewrite the evolution equation~\eqref{eq:ddd} as 
\begin{align}
\frac{d^{2}}{d\chi^{2}}\DD_{ab}=&\frac{1}{\chi}\Ss_{ab} (\chi)\, ,\\
\text{with}\quad\Ss_{ab}\equiv& \chi\left( -\frac{1}{k^{0}}\frac{dk^{0}}{d\chi}\frac{d}{d\chi}\DD_{ab}+\frac{1}{(k^{0})^{2}}\RR_{ac}\DD_{cb}\right)\label{eq:S}.
\end{align}
Integrating eq.~\eqref{eq:S} by parts we obtain
\begin{equation}
\DD_{ab}(\chi_S)=\chi_S \delta_{ab}+\int_0^{\chi_S}d\chi\frac{\chi_S-\chi}{\chi}\Ss_{ab}(\chi)\, . 
\label{eq:DintS}
\end{equation}
The integral in eq.~\eqref{eq:DintS} is along the perturbed geodesic, i.e.\ $\Ss_{ab}(\chi)=\Ss_{ab}\big(x^\mu(\chi)\big)$, where $x^\mu(\chi)$ denotes the perturbed trajectory of the photon. Similarly to the calculation of $k^0$ in section~\ref{sec:311}, we write
\begin{equation}
\Ss_{ab}\big(x^\mu(\chi)\big)
=\Ss_{ab}\big(\bar x^\mu(\chi)\big)+\delta x^\mu(\chi)\partial_\mu \Ss_{ab}\big(\bar x^\mu(\chi)\big)+\Oo{6}\, . \label{eq:Sexpand}
\end{equation}
Since $\Ss_{ab}$ is at least of order $1/c^2$, it is enough to consider only the first term of the Taylor expansion in eq.~\eqref{eq:Sexpand}. 

Inserting eqs.~\eqref{eq:deltaxi} and~\eqref{eq:deltax0} into~\eqref{eq:Sexpand} we obtain 
\begin{align}
\Ss_{ab}(x^\mu)= \Ss_{ab}(\bar x^\mu)+
\frac{\chi}{\left(\bk{0}\right)^2}\delta x^{j}\left(\RR^{(2)}_{ac}\bar{\DD}_{c b}\right){}_{,j} -\frac{\chi}{\bk{0}}\delta x^{j}\left(\frac{d k^{0(2)}}{d\chi}\right){}_{,j}\delta_{ab}\label{eq:tt}\, ,
\end{align}
which is computed in appendix~\ref{App:B}, eqs.~\eqref{eq:twoterms} to \eqref{eq:dS2}.

We can now insert~\eqref{eq:tt} into~\eqref{eq:DintS} and integrate along the background geodesic to obtain $\DD_{ab}^{(4)}$. The Jacobi map can be divided into a part proportional to $\delta_{ab}$ and a part proportional to $n^{i}_{a}n^{j}_{b}$. Only the latter contributes to the shear and rotation, whereas both parts contribute to the convergence. 

At order $\Oo{4}$, we group the terms of $\DD_{ab}$ according to the potentials or their couplings, i.e.\ we write
\begin{align}
\DD^{(4)}_{ab}=\DP_{ab}+\DVV_{ab}+\DWW_{ab}+\DUW_{ab}+\DVW_{ab}+\Dh_{ab},\label{eq:D4split}
\end{align}
where the subscripts refer to the couplings of the potentials. The detailed derivation is given in appendix~\ref{App:B}, here we write only the results.
The contribution $\DP_{ab}$ is a purely relativistic contribution generated by the relativistic potentials $U_P$, $V_P$, and $W_P$:
\begin{align}
\DP_{ab}=&\frac{2}{c^4}\chi_{S}V_{PS}\delta_{ab} - \frac{4}{c^4} \int^{\chi_{S}}_{0}d\chi \left[W_{P}+(\chi_S-\chi) W_{P,i}\fkk{i}\right]\delta_{ab}\notag\\
&+\frac{4}{c^4}\int_{0}^{\chi_{S}}d\chi\left(\chi_{S}-\chi\right)\chi \bar{n}^{i}_{a}\bar{n}^{j}_{b}W_{P,ij}\,.\label{eq:DD41}
\end{align}
$\DP_{ab}$ take the same form as the Jacobi mapping $\DD^{(2)}_{ab}$ in~\eqref{eq:DD2}, which is due to the form of the metric \eqref{eq:g00} - \eqref{eq:gij}.
The $\DVV_{ab}$ contribution contains all the terms quadratic in $V_N$:
\begin{align}
    \DVV_{ab}=\cc{4}\left[\frac{\chi_S}{2}V_{NS}^2-2\intchi \chi\left(\frac{dV_N}{d\chi}\right)^2 \right]\delta_{ab}\,.
\end{align}
This terms contributes only to the convergence, since they are proportional to $\delta_{ab}$. The $\DWW_{ab}$ contribution contains all the terms quadratic in $W_N$. It can be split into a part proportional to $\delta_{ab}$ and a part proportional to $n_a n_b$, which contributes also to the shear and rotation: 
\begin{align}
\DWW_{ab}=&\delta_{ab}\cc{4}\left\{\ichi\left[4 W_{N,m}\int\bound d\chi' W_{N}\frac{\bk{m}}{\bk{0}}+2W_N^2+2\left(\int\bound d\chi' W_{N,l}\fkk{l}\right)^{2}\right.\right.+\notag\\
&\left.-4 W_{N,m}\int\bound d\chi'\left(\chi -\chi'\right)\left(W_{N}^{,m}-W_{N,j}\frac{\bk{m}\bk{j}}{\left(\bk{0}\right)^2}\right)   +4W_N\int\bound d\chi'W_{N,l}\fkk{l}\right]+ \notag \\
&  +4  \intchi \left[W_{N,m}\left(\int\bound \dc' \left(W_{N}^{,m}-W_{N,j}\frac{\bk{m}\bk{j}}{\left(\bk{0}\right)^2}\right)-W_{N}\fkk{m} \right)\right.+\notag\\
&- W_{N,nm}\fkk{n} \int\bound d\chi' \left(\left(\chi -\chi'\right) \left(W_{N}^{,m}-W_{N,j}\frac{\bk{m}\bk{j}}{\left(\bk{0}\right)^2}\right) -W_{N}\fkk{m} \right) +\notag\\
&+\left.\left. \frac{1}{\bk{0}}W_{N,0i}   \int_{0}^{\chi}d\chi'\left(\left( \chi - \chi' \right)\left( W_{N}^{,i}-W_{N,j}\frac{\bk{i}\bk{j}}{\left(\bk{0}\right)^2}\right)     -\bk{i}W_N\right)  \right]\right\}+\notag\\
&+\bar{n}^{i}_{a}\bar{n}^{j}_{b}\cc{4}\left\{-\ichi  \left(4W_N\int^{\chi}_0   d\chi'\chi' W_{N,ij}\right)\right.+\notag\\
&+\intchi\left[ -2  \chi W_{N,ij}^2 -4W_{N,l}\fkk{l} \int\bound d\chi'  \chi' W_{N,ij} \right.+\notag\\
&-4\chi\fkk{l}W_{N,lj}\int\bound  d\chi'W_{N,i}-4\chi\fkk{l}W_{N,il}\int\bound d\chi' W_{N,j}+\notag\\
& -4W_{N,ij}\int\bound d\chi' \left(W_N + \left(\chi -\chi'\right) W_{N,m}\fkk{m} \right) +\notag\\
&+4\bar{n}^{s}_{c}\bar{n}^{rc} W_{N,is}\int\bound d\chi'\left(\chi - \chi'\right) \chi' W_{N,rj}  +\notag\\
&\left.\left.+\chi 4W_{N,ijm} \int\bound d\chi' \left(  \left(\chi -\chi'\right)\left(W_{N}^{,m}-W_{N,l}\frac{\bk{m}\bk{l}}{\left(\bk{0}\right)^2}\right)  -W_N\fkk{m}\right) \right]\right\}.\label{eq:D4WW}
\end{align}
The couplings between $U_N$ and $W_N$ are denoted by $\DUW_{ab}$. They are all proportional to $\delta_{ab}$: 
\begin{align}
    \DUW_{ab}=&\cc{4}\delta_{ab}\left\{4\int^{\chi_S}_{0} d\chi\left( U_{N,i}\fkk{i}  \int^{\chi}_{0}d\chi'W_{N}\right) -4 \int^{\chi_S}_{0} d\chi U_{N,i}\fkk{i} W_{N}+\right.\notag\\
    &-4 \int^{\chi_S}_{0} d\chi\left[ U_{N,i} \int_{0}^{\chi}d\chi'\left(\chi -\chi'\right)\left( W_{N}^{,i}-W_{N,j}\frac{\bk{i}\bk{j}}{\left(\bk{0}\right)^2}\right)\right]  +   \notag\\
    &\left.+4\intchi\left[  U_{N,i}  \int_{0}^{\chi}d\chi'\left( W_{N}^{,i}-W_{N,j}\frac{\bk{i}\bk{j}}{\left(\bk{0}\right)^2}\right)  \right]\right\}.
\end{align}
The term $\DVW_{ab}$ contains all couplings between $V_N$ and $W_N$. Since it contains also a part proportional to $n_a n_b$ it contributes also to the shear and rotation:
\begin{align}
\DVW_{ab}=&\delta_{ab}\cc{4}\left\{-V_{NS}\int^{\chi_S}_0 d\chi \left(\chi_S -\chi\right) W_{N,m}\fkk{m}-2 V_{NS} \ichi W_N+ \right.\notag\\
&+2\chi V_{NS,m}\ichi \left(\left(\chi_S - \chi\right)\left(W_{N}^{,m}-W_{N,l}\frac{\bk{m}\bk{l}}{\left(\bk{0}\right)^2}\right)   -W_N\fkk{m}\right)+\notag\\
&+\ichi\left[2\chi V_{N,m}\frac{\bk{m}}{\bk{0}} W_{N} -4    \chi V_{N,m}\int\bound d\chi' \left(W_{N}^{,m}-W_{N,l}\frac{\bk{m}\bk{l}}{\left(\bk{0}\right)^2}\right)\right] \notag\\
&    +\intchi \left[2 \frac{\bk{m}}{\bk{0}} V_{N,m}W_{N}+ 2\frac{\bk{m}}{\bk{0}} \chi W_{N} \dchi V_{N,m}+\right.\notag\\
&\left.\left.+2\chi V_{N,m} \left(W_{N}^{,m}-W_{N,l}\frac{\bk{m}\bk{l}}{\left(\bk{0}\right)^2}\right)\right]\right\}+\notag\\
&+\bar{n}^{i}_{a}\bar{n}^{j}_{b}\cc{4}  2V_{NS} \intchi \chi W_{N,ij} \, .
\end{align}
Finally the term $\Dh_{ab}$ contains the tensor contributions
\begin{align}
\Dh_{ab}=&\delta_{ab}\cc{4}\left[\frac{1}{2\left(\bk{0}\right)^2} \ichi h_{ij}\bk{i}\bk{j}-\frac{1}{2(\bk{0})^3} \intchi h_{ij,l}\bk{i}\bk{j}\bk{l}\right]+\notag\\
&\bar{n}^{i}_{a}\bar{n}^{j}_{b}\cc{4}\left\{ \frac{1}{2}\chi_S h_{ij}+\right.  \ichi\left[-h_{ij} -\frac{\chi}{2}\left(h_{jp, i}+h_{i p,j}\right)\fkk{p}\right]+\notag\\
&+\intchi\left[ \frac{1}{2}\left(h_{jp, i}+h_{i p,j}\right)\fkk{p}+\frac{\chi}{2} h_{mp,i j}\frac{k^{p}k^{m}}{\left(\bk{0}\right)^2}\right\}\label{eq:DD4h}.
\end{align}
\subsection{Redshift Perturbations}
\label{sec:32}

In the previous section~\ref{sec:31} we have calculated the Jacobi mapping $\DD_{ab}$ in an non-expanding universe, as a function of the coordinate $\chi_S$. We now use eqs.~\eqref{eq:tildeD} and~\eqref{eq:Dz} to calculate the Jacobi mapping $\tilde\DD_{ab}$ in an expanding universe, as a function of the observed redshift $z_S$. Let us start by calculating the redshift perturbations up to order $\Oo{4}$. 

The redshift of a photon emitted at $S$ and measured at $O$ is given by 
\begin{align}
1+z_S=&\frac{\big(\tilde g_{\mu\nu}\tilde k^{\mu}\tilde u^{\nu}\big)_{S}}{\big(\tilde g_{\mu\nu}\tilde k^{\mu}\tilde u^{\nu}\big)_{O}}
=\frac{1}{a_S}\frac{\big(g_{\mu\nu}k^{\mu}u^{\nu}\big)_{S}}{\big( g_{\mu\nu} k^{\mu} u^{\nu}\big)_{O}}=\frac{1}{a_S}(1+\delta f)\, ,
\end{align}
where we have defined
\begin{equation}
\delta f\equiv\frac{\big(g_{\mu\nu}k^{\mu}u^{\nu}\big)_{S}}{\big( g_{\mu\nu} k^{\mu} u^{\nu}\big)_{O}} -1\, , \label{eq:deff} 
\end{equation}
and we have used that under conformal transformation the photon wave vector transforms as $\tilde k^\mu=k^\mu/a^2$ and the four-velocity as $\tilde u^\mu=u^\mu/a$ (see e.g.\,\cite{Bonvin:2005ps} for more detail). Note that we normalise the scale factor to $a_O=1$. Using that
$1+\bar{z}_S=1/a_S$ we can write
\begin{equation}
1+z_S=(1+\bar{z}_S)(1+\delta f)=(1+z_S-\delta  z_S)(1+\delta f)\, ,  \end{equation}
leading to
\begin{equation}
\delta  z_S=(1+ z_S)\frac{\delta f}{1+\delta f}\, . \label{eq:zgen}
\end{equation}
The perturbation $\delta f$ depends on the metric potentials and the peculiar velocity at both the source and observer positions. However, as argued at the beginning of section~\ref{sec:31} the metric perturbations at the observer do not contribute to the observed shear, convergence and rotation. The observer velocity would generate a global dipole variation in the convergence, which is degenerated with the vector contribution at the observer. This dipole can be subtracted from the observables, and we therefore do not consider it here.

Using eq.~\eqref{eq:deff}, the perturbation $\delta f$ is calculated as a function of $\chi_S$. We need to express it in terms of the observed redshift $z_S$. We obtain
\begin{align}
\delta z_S=&(1+z_S)\delta F(\chi_S)=(1+z_S)\delta F(z_S-\delta z_S)\label{eq:Fz}\\
=&(1+z_S)\left[\delta F(z_S) -\frac{d}{dz_{S}}\delta F(z_S)\delta z_S
+\frac{1}{2}\frac{d^2}{dz_{S}^2}\delta F(z_S)\delta z_{S}^2
-\frac{1}{3!}\frac{d^3}{dz_{S}^3}\delta F(z_S)\delta z_{S}^3\right]\!+\Oo{5}\,, \notag
\end{align}
where we have defined
\begin{equation}
\delta F=\frac{\delta f}{1+\delta f}
\simeq\delta f-\delta f^2+\delta f^3-\delta f^4+\dots\, .\label{eq:deltaF}
\end{equation}
The perturbation $\delta f$ depends on the four velocity $u^{\mu}$. In the PF formalism it is given by \cite{Milillo:2015cva} 
\begin{align}
u^{i}=&\frac{v^{i}}{c}u^{0}\\
u^{0}=&1+\frac{1}{c^2}\left(U_N+\frac{1}{2}v^2\right)+\frac{1}{c^4}\left(\frac{1}{2}U^2_N+2U_P +v^2 V_N +\frac{3}{2}v^2 U_N+\frac{3}{8}v^{4}-B_{Ni}v^{i}\right)\, ,
\end{align}
where $v^2=\sum_{ij}\delta_{ij}v_iv_j$.
Inserting this into~\eqref{eq:deff} and neglecting terms at the observer we obtain 
\begin{align}
\delta f(\chi_S)
=&\left(g_{00}k^0 u^0+g_{0i}k^{0}u^{i}+g_{0i}k^{i}u^{0}+g_{ij}k^i u^j\right)(\chi_S)-1\notag\\
=&e^{-2U_N\cc{2}-4U_P\cc{4}}\frac{k^{0}}{\bk{0}}u^{0}+ B_{Ni}\cc{3}\frac{k^{0}}{\bk{0}}\frac{1}{c}v^{i}u^{0} +B_{Ni}\cc{3}\frac{k^{i}}{\bk{0}}u^{0}  \notag\\
& -    e^{2V_N \cc{2}+4V_P\cc{4}}\delta_{ij}\frac{k^{i}}{\bk{0}}\frac{1}{c}v^{i}u^{0}-1\, , \label{eq:fpert}
\end{align}
where all terms are evaluated at the source position $\chi_S$.

The detailed derivation of $\delta z_S$ is presented in appendix~\ref{app:rp}. Here we show only the result up to order $\Oo{3}$. The result at order $\Oo{4}$ is very long and can be found in eq. \eqref{eq:deltaz4}. We obtain
\begin{align}
\delta z_S^{(1)}=&- (1+z_S)   \frac{1}{c}\vp \, ,\label{eq:deltaz1}\\
    \delta z_S^{(2)}=&(1+z_S)\cc{2}\left( U_{NS}-2\int^{\chi_S}_0 d\chi W_{N,0}+\frac{1}{2}v^2_S -\vp^2 +\frac{c}{\HH_S }\vp' \vp \right)\, \text{, and}\label{eq:deltaz2}\\
    \delta z_S^{(3)} =&(1+z_S)\cc{3}\left\{   \int^{\chi_S}_0 d\chi B_{Ni,0}\bk{i}  -     \vp \left(V_{NS}+\frac{1}{2}v^2_S\right)-v^{i}_S\delta_{ij}\int^{\chi_S}_{0}d\chi W^{,i}_{N}\right.\notag\\
        &+2\left(U_{NS}-2\int^{\chi_S}_{0} d\chi W_{N,0}+\frac{1}{2}v^2_S\right)\vp-\vp^3 +\notag\\
        &-\frac{c}{\HH_S}\vp'\left( U_{NS}-2\int^{\chi_S}_{0} d\chi W_{N,0}+\frac{1}{2}v^2_S -\vp^2 +\frac{1}{\HH_S }\vp' \vp \right)+\notag\\
        &+\frac{c}{\HH_S}\left(\frac{d U_{NS}}{d\chi} -2W_{NS,0} +v_S v'_S-2\vp \vp' \right)   \vp+\notag\\
        &\left.+\frac{c}{2\HH_S}\left[ \vp'\left( 1+\frac{\HH'_Sc}{\HH^2_S} \right)-\frac{c}{\HH_S}\vp''  \right]\vp^2\right\}\, ,\label{eq:deltaz3}
\end{align}
where $\vp \equiv \delta_{ij}v^{i}_S\fkk{j}$ is the radial component of the source peculiar velocity.

We can now express the Jacobi mapping as a function of $z_S$. Substituting the expressions for $\delta z_S$ in \eqref{eq:deltaz1}-\eqref{eq:deltaz3} and \eqref{eq:deltaz4} as well as the expressions of the derivatives \eqref{eq:Dprime}-\eqref{eq:Dz43} into eq.~\eqref{eq:Dz}, we obtain $\tDD_{ab}(z_S)$ up to order $\Oo{4}$.

We see that the redshift perturbations generate a new order $\mathcal{O}\left(\frac{1}{c}\right)$ in the expansion, proportional to the galaxy peculiar velocity $v_{S\parallel}$:
\begin{align}
\tDD_{ab}^{(1)}
=&\left(\frac{c}{\HH_S\chi_S}-1\right)\frac{\chi_S}{1+z_S}        \frac{v_{S\parallel}}{c}\delta_{ab}\label{eq:DZ1}\, ,
\end{align}
with $\HH\equiv\frac{1}{a} \frac{da}{d\eta}$. Note that because we define $\HH$ using the time derivative and not the derivative w.r.t.\ $\chi$, we obtain an additional factor $c$: $da/d\eta=-c\, da/d\chi$. This factor does not influence the order of the expression.
The contribution in eq.~\eqref{eq:DZ1} has been called Doppler magnification, and it is the dominant contribution to the convergence at low redshift~\cite{Bonvin:2008ni,Bacon:2014uja,Bonvin:2016dze}.
Note that in standard perturbation theory, since the peculiar velocity is a perturbative quantity it contributes to the Jacobi mapping at the same order as the gravitational potentials. In the PF framework however, the peculiar velocity is non-perturbative, but it is always weighted by a factor $1/c$. As such it is of lower order than the Newtonian gravitational potentials in the expansion $1/c$. This illustrates nicely the difference between the PF formalism and standard perturbation theory. The Doppler term is usually neglected in lensing analyses, first because at lowest order it does not contribute to the shear, and second because at high redshift, its contribution to the convergence is subdominant with respect to the Newtonian contribution of order $1/c^2$. From the PF formalism we see however that the velocity contribution will dominate in the regime where $v_{S\parallel}/c$ is larger than the lensing potential integrated along the photon trajectory, see eq.~\eqref{eq:DD2}.

At order $\Oo{2}$ we obtain 
\begin{align}
\tDD_{ab}^{(2)}
&=\tDD_{ab}^{(2)}\left(z_{S}\right)+\cc{2}\frac{\chi_S}{1+z_S}\left\{\left(1-\frac{c}{\HH_S \chi_S}\right)\left( U_{NS}-2\int^{\chi_S}_0 d\chi W_{N,0}+\frac{1}{2}v^2_S+\right.\right.\notag\\
&\hspace{1cm}\left. +\vp^2-\frac{c}{\HH_S }\vp' \vp \right)\left.+\left(1-\frac{\HH'_S c^2}{2\chi_S\HH^3_S} -\frac{3c}{2\HH_S \chi_S}\right)\vp^2\right\}\delta_{ab}\label{eq:DZ2}\, ,
\end{align}
where a prime denotes a derivative with respect to $\chi$. In the first line, $\tDD_{ab}^{(2)}\left(z_{S}\right)$ is obtained from eq.~\eqref{eq:DD2} where the background coordinate $\chi_S$ can be replaced by its value at the observed redshift $z_S$. We see that at this order, the Jacobi mapping is not only affected by the radial component of the peculiar velocity $v_{S\parallel}$ but also by its transverse part through $v_S^2=v_{S\parallel}^2+v_{S\perp}^2$. Note that since the redshift corrections at this order are proportional to $\delta_{ab}$ they will only affect the convergence, and leave the shear and rotation unchanged.

At the order $\Oo{3}$ we obtain 
\begin{align}
\tDD_{ab}^{(3)}=& \tDD_{ab}^{(3)}(z_S)+ \cc{3}\frac{\chi_S}{1+z_S}\delta_{ab}\left\{\vp\left(2-\frac{\HH'_S c^2}{\chi_S\HH^3_S} -\frac{3c}{\HH_S \chi_S}\right)\left[ U_{NS}-2\int^{\chi_S}_0 d\chi W_{N,0}+\right.\right.\notag\\
&\left.+\frac{1}{2}v^2_S -\vp^2 +\frac{c}{\HH_S }\vp' \vp \right]+ \vp^3 \left( 1  -\frac{11c}{6\chi_S  \HH_S}+\frac{\HH''_S c^3}{6\chi_S  \HH^4_S}-\frac{ \HH'^2 c^3}{2\chi_S  \HH^5_S}-\frac{ \HH'_S c^2}{\chi_S  \HH_S^3} \right)+\notag\\
&-\left(\frac{c}{\HH_S \chi_S}-1\right)\left[   \int^{\chi_S}_0 d\chi B_{Ni,0}\bk{i}-v^{i}_S\delta_{ij}\int^{\chi_S}_{0}d\chi W_N^{,i} -\vp^3 +\right.\notag\\
        &+\left(2U_{NS}-2V_{NS}-4\int^{\chi_S}_{0} d\chi W_{N,0}+\frac{1}{2}v^2_S\right)\vp +\notag\\
        &-\frac{c}{\HH_S}\vp'\left( U_{NS}-2\int^{\chi_S}_{0} d\chi W_{N,0}+\frac{1}{2}v^2_S -\vp^2 +\frac{1}{\HH_S }\vp' \vp \right)+\notag\\
        &+\frac{c}{\HH_S}\left(\frac{d U_{NS}}{d\chi} -2W_{NS,0} +v_S v'_S-2\vp \vp' \right)   \vp+\notag\\
        &+\frac{c}{2\HH_S}\left( \vp'\left( 1+\frac{\HH'_S c}{\HH^2_S} \right)-\frac{c}{\HH_S}\vp''  \right)\vp^2+\notag\\
         &\left.+\vp\int^{\chi_{S}}_{0}d\chi \left(W_{N}+(\chi_S-\chi) W_{N,i}\fkk{i}\right)\right]+\notag\\
            &+\left.\frac{c \vp}{\HH_S} \left(V_{NS}+\chi_{S}\frac{dV_{NS}}{d\chi_S}- 2W_{NS}-2\int^{\chi_{S}}_{0}d\chi  W_{N,i}\fkk{i}\right)\right\}\notag\\
            %
    &+\bar{n}^{i}_{a}\bar{n}^{j}_{b}\cc{3}\frac{\chi_S}{1+z_S}\vp\left[-\frac{2}{\chi_S}\int_{0}^{\chi_{S}}d\chi\left(\chi_{S}-\chi\right)\chi W_{N,ij} +\right.\notag\\
    &\left.+\frac{c}{\chi_S\HH_S}2\int_{0}^{\chi_{S}}d\chi\chi W_{N,ij}\,\right]. \label{eq:DZ3}
\end{align}
Without redshift perturbations, only the vector potential $B_N^i$ contributes to the Jacobi mapping at the order $\Oo{3}$, see eq.~\eqref{eq:DD3}. However, since the peculiar velocity comes at order $\mathcal{O}\left(\frac{1}{c}\right)$, we obtain couplings between the velocity and the Newtonian potentials that also contribute at this order, as well as terms cubic in the velocity. Note that at this order the peculiar velocity modifies not only the convergence, but also the shear.

Finally, at the order $\Oo{4}$ we obtain
\begin{align}
\tDD_{ab}^{(4)}= &\tDD_{ab}^{(4)}\left(z_{S}\right)-\frac{d}{d\bar{z}_{S}}\bar{\tDD}_{ab}\left(z_{S}\right)\delta z_{S}^{(4)}-\frac{d}{d\bar{z}_{S}}\tDD_{ab}^{(2)}\left(z_{S}\right)\delta z_{S}^{(2)}+\notag\\
&-\frac{d}{d\bar{z}_{S}}\tDD_{ab}^{(3)}\left(z_{S}\right)\delta z_{S}^{(1)}+\frac{1}{2}\frac{d^2}{d\bar{z}_{S}^2}\tilde{\bar{\DD}}_{ab}\left(z_{S}\right)\left(\delta z_{S}^{(2)}\right)^2+\notag\\
&+\frac{1}{2}\frac{d^2}{d\bar{z}_{S}^2}\tilde{\DD}^{(2)}_{ab}\left(z_{S}\right)\left(\delta z_{S}^{(1)}\right)^2+\frac{d^2}{d\bar{z}_{S}^2}\bar{\tDD}_{ab}\left(z_{S}\right)\delta z_{S}^{(1)}\delta z^{(3)}\notag\\
&-\frac{1}{2}\frac{d^3}{d\bar{z}_{S}^3}\bar{\tDD}_{ab}\left(z_{S}\right)\delta z_{S}^{(2)}\left(\delta z_{S}^{(1)}\right)^2+\frac{1}{4!}\frac{d^4}{d\bar{z}_{S}^4}\bar{\tDD}_{ab}\left(z_{S}\right)\left(\delta z_{S}^{(1)}\right)^4,  \label{eq:Dz44}
\end{align}
where we list the individual terms of \eqref{eq:Dz44} in the appendix in equation \eqref{eq:DDz4a} - \eqref{eq:DDz4b}.


\section{Extraction of the Convergence, Shear and Rotation}
\label{sec:sec4}

As shown in section \ref{sec:sec3}, the convergence, shear and rotation can be expressed in terms of the spin-0 and spin-2 components of the Jacobi mapping $\tilde D_{ab}$, see eqs.~\eqref{eq:kappa} and~\eqref{eq:gamma}. Following~\cite{2010PhRvD..81h3002B, 2012PhRvD..86b3001B, 5246828490804498bbad0dac6a987a72, 1966JMP.....7..863N}, we first introduce spin operators on the sphere. In weak lensing, the use of these operators has the advantage that we do not rely on the small-angle or flat-sky approximation.

\subsection{Spin Operators on a Sphere}
\label{sec:sec41}
We want to describe the shear, convergence and rotation on the sphere of the sky. To each image we can associate a unit vector at the observer, $e^i_r$, pointing in the direction of the image. This vector is equal to the wave vector of the photon at the observer $e^i_r=k_O^i=\bar k^i$ (recall that we use $\chi=c(\eta_O-\eta)$ as time coordinate, so that $k_O^i$ points from the observer to the image). We then define at the observer the angular unit vectors in spherical coordinates $e^{i}_{\theta}$ and $e^{i}_{\phi}$ that are orthogonal to $e^i_r$. We can identify $e^{i}_{\theta}=\bar n_1^i$ and $e^{i}_{\phi}=\bar n_2^i$. Note that $e^{i}_{\theta}$ and $e^{i}_{\phi}$ are orthogonal to the photon wave vector at the observer, and are then parallel transported along the background geodesics. Therefore they do not live in the screen defined by $n_a^\mu$ along the geodesics. 

A spin-$s$ field on the sphere of the sky ${}_{s}X$ is defined as a field which transforms as ${}_{s}X\rightarrow e^{i\alpha s}{}_{s}X$ when $e^{i}_{\theta}$ and $e^{i}_{\phi}$ are rotated by an angle $\alpha$ around $e_r^i$.
We introduce the unit vectors $\eplus{i}$ and $\eminus{i}$ defined as 
\begin{align}
e^{i}_{\pm}=e_{\theta}^{i}\pm ie^{i}_{\phi}\, .
\end{align}
In the 2D basis $({\bf e}_\theta,{\bf e}_\phi)$, the vector ${\bf e}_\pm$ have components $e^a_\pm=(1,\pm i)$. The spin-0 and spin-2 component of $\DD_{ab}$ are then given by
\begin{align}
{}_{2}\DD&=\eplus{a}\eplus{b}\DD_{ab}\, , \label{eq:2Dee} \\
{}_{0}\DD&=\eminus{a}\eplus{b}\DD_{ab}\,.\label{eq:0Dee}
\end{align}
We see immediately that any term proportional to $\delta_{ab}$ will contribute only to the real part of ${}_{0}\DD$, i.e.\,to the convergence since
\begin{align}
\eplus{a}\eplus{b}\delta_{ab}=0\,,\quad \mbox{and}\quad\eminus{a}\eplus{b}\delta_{ab}=2\, .
\end{align}
The terms proportional to $\bar n_a^i \bar n_b^j$ have however a more complicated structure. We have indeed
\begin{equation}
\eplus{a}\eplus{b}\bar n_a^i \bar n_b^j=\eplus{i}\eplus{j}\, ,
\end{equation}
which contributes to ${}_{2}\DD$ and
\begin{equation}
\eminus{a}\eplus{b}\bar n_a^i \bar n_b^j=\eminus{i}\eplus{j}\, ,
\end{equation}
which contributes to both the real and imaginary part of ${}_{0}\DD$, i.e.\;to the convergence and the rotation.

The vectors $e^i_\pm$ acting on partial derivative $\partial_i$ give rise to
derivative operators on the sky $\slpa$ and $\bar{\slpa}$, which increase and decrease the spin $s$ by 1, respectively:
\begin{align}
&\slpa {}_{s}X\equiv -\sin^{s}\theta\left(\partial_{\theta}+\frac{i}{\sin\theta}\partial_{\phi}\right)\sin^{-s}\theta\, {}_{s}X=-\left(\partial_{\theta}+\frac{i}{\sin\theta}\partial_{\phi}\right){}_{s}X+s\cot\theta \,{}_{s}X\, ,\\
&\bar{\slpa} {}_{s}X\equiv -\sin^{-s}\theta\left(\partial_{\theta}-\frac{i}{\sin\theta}\partial_{\phi}\right)\sin^{s}\theta\, {}_{s}X=-\left(\partial_{\theta}-\frac{i}{\sin\theta}\partial_{\phi}\right){}_{s}X-s\cot\theta\, {}_{s}X\,.
\end{align}
The derivatives $\slpa$ and $\bar{\slpa}$ are effectively angular covariant derivatives on a sphere. In particular we have
\begin{align}
&e_+^i\partial_i X=-\frac{1}{\chi}\slpa X\,, 
&&e_-^i\partial_i X=-\frac{1}{\chi}\bar\slpa X\,,\\
&e_+^ie_+^j\partial_i\partial_j X=\frac{1}{\chi^2}\slpa^2 X\,,\quad
&&e_-^ie_-^j\partial_i\partial_j X=\frac{1}{\chi^2}\bar\slpa^2 X\,.
\end{align}
If we apply both $\bar{\slpa}$ and $\slpa$ consecutively, the spin $s$ remains unchanged and we obtain an expression corresponding to the angular Laplace operator in spherical coordinates $\bar\slpa\slpa X=\slpa\bar\slpa X$. We can show that~\footnote{Note that there is a typo in the equivalent equation (40) in~\cite{2010PhRvD..81h3002B}. The second sign should be a $+$.}
\begin{equation}
e_+^ie_-^j\partial_i\partial_j X=
e_-^ie_+^j\partial_i\partial_j X=
\frac{1}{\chi^2}\bar\slpa\slpa X +\frac{2}{\chi}e_r^i\partial_i X\, .   
\end{equation}

Finally, let us note that the PF metric contains vector and tensor potentials, which can be decomposed into spin fields:
\begin{align}
B^{i}=&B_re_r^{i}+\frac{1}{2}{}_{-1}B \eplus{i}+\frac{1}{2}{}_{1}B\eminus{i}\, ,\\
 h^{ij}=&h_{rr}\left(e_r^{i}e_r^{j}-\frac{1}{2}\eplus{\left(i\right.}\eminus{\left.j\right)}\right)+{}_{-1}h_r \eplus{\left(i\right.}e^{\left.j\right)}_r+{}_{1}h_r \eminus{\left(i\right.}e^{\left.j\right)}_r+\frac{1}{4}{}_{-2}h\eplus{i}\eplus{j}+\frac{1}{4}{}_{2}h\eminus{i}\eminus{j}\,,
\end{align}
where $B_r$ and $h_{rr}$ are spin-0 functions and ${}_{s}B$, ${}_{s}h$ and ${}_{s}h_r$ are spin-$s$ functions. Note that these components are not independent, since $B^i$ and $h^{ij}$ are divergenceless.

In appendix~\ref{app:Dspin} we present the derivation of ${}_{0}\DD$ and ${}_{2}\DD$ up to order $\Oo{4}$, using eqs.~\eqref{eq:2Dee} and~\eqref{eq:0Dee}. Here we only display the final results for the reduced shear $g$, the convergence $\kappa$ and the rotation $\omega$, which are obtained from ${}_{0}\DD$ and ${}_{2}\DD$ using eqs.~\eqref{eq:kappa},~\eqref{eq:gamma} and~\eqref{eq:rs}.

\subsection{The reduced shear $g$}
\label{sec:43}
The reduced shear $g$ is measured from the ellipticity of galaxies. It is given by 
\begin{align}
g=\frac{\gamma}{1-\kappa}=-\frac{{}_{2}\tDD}{\textrm{Re}\left[{}_{0}\tDD\right]}\, .\label{eq:gg}
\end{align}
At order $\Oo{2}$, the shear and reduced shear are equal, since $\bar \kappa=0$. We have 
\begin{align}
g^{(2)}=-\frac{{}_{2}\tDD^{(2)}}{{}_{0}\bar{\tDD}}=-\cc{2}\int_{0}^{\chi_{S}}d\chi\frac{\chi_{S}-\chi}{\chi_S \chi} \slpa^2 W_{N}\, .\label{eq:g2}
\end{align}
This is the standard Newtonian expression for the shear, written in terms of derivatives on the sphere. 

At order $\Oo{3}$ both ${}_{2}\tDD$ and ${}_{0}\tDD$ contribute to the reduced shear. Since the imaginary part of ${}_{0}\tDD$ is of order $\Oo{4}$ (see Appendix~\ref{app:Dspin}) we can write
\begin{align}
g^{(3)}=&-\frac{{}_{2}\tDD^{(2)}+{}_{2}\tDD^{(3)}}{{}_{0}\bar{\tDD}+{}_{0}\tDD^{(1)}}=-\frac{1}{{}_{0}\bar{\tDD}}\left(-\frac{{}_{0}\tDD^{(1)}}{{}_{0}\bar{\tDD}}{}_{2}\tDD^{(2)}+{}_{2}\tDD^{(3)}\right) \notag    \\
=&\cc{3}\left\{\int_{0}^{\chi_{S}} d\chi'\frac{\chi_{S}-\chi}{2\chi_S\chi} \left[ \frac{d}{d\chi} \left(\chi \slpa _{1}B\right)+\slpa^2 B_{r}\right]+ \right.\notag\\
&\left.-\frac{c}{\HH_S\chi_S^2}\vp \int_{0}^{\chi_{S}}d\chi  \slpa^2 W_{N}    \right\}.\label{eq:g3}
\end{align}
We see that the dominant correction to the Newtonian expression~\eqref{eq:g2} is due to two different effects: the vector potential $B_N^i$ and the peculiar velocity of the galaxies $v_{S\parallel}$. The vector potential in the PF approximation has been computed from N-body simulations on non-linear scales~\cite{Thomas:2015kua,Bruni:2013mua,Thomas:2014aga}. It was found that the power spectrum of the vector field $B^{i}_N/c^3$ is of the order of $10^{-5}$ the power spectrum of the scalar potential $W_N/c^2$ over a range of scales and redshifts. Comparing eq.~\eqref{eq:g3} with eq.~\eqref{eq:g2} we see that part of the vector contribution enters in the reduced shear with exactly the same kernel as the Newtonian scalar part. As such we expect that the impact of the vector potential on the reduced shear will be of the order of $\sim \sqrt{10^{-5}}g^{(2)}\sim 3\times 10^{-3}g^{(2)}\sim 3\times 10^{-5}$, since $g^{(2)}$ is of order $10^{-2}$~\cite{Dodelson:2005ir}. The requirement to measure cosmological parameters from a survey like Euclid with 1\% precision is that additive systematics to the shear remain below $3\times 10^{-4}$~\cite{Amara:2007as}. This is only one order of magnitude above the expected contribution from the vector modes. A careful calculation of the reduced shear power spectrum from the vector modes is therefore necessary to assess precisely its amplitude. In particular, since the ratio between the vector potential and the scalar potential grows in the strongly non-linear regime (see Fig.~5 and 6 of~\cite{Thomas:2015kua}), the transverse derivatives in~\eqref{eq:g2} and~\eqref{eq:g3} will not act in the same way and may therefore enhance the vector contribution. We will compute this in a future work.

The second contribution at order $\Oo{3}$ is due to the peculiar velocity of galaxies $v_{S\parallel}$, coupled to the standard Newtonian shear. This contribution is due to two effects. First, the reduced shear $g$ is measured as a function of redshift, which is affected by the source peculiar velocity. To understand this effect, let us assume that we measure $g$ for two different galaxies that are at the same redshift. One of the galaxies has no peculiar velocity, whereas the other has a velocity directed towards the observer. As a consequence, the second galaxy is physically situated at a larger distance than the first one. The impact of a given lens on the two galaxies will then be different, since the distance between the lens and the source is different. The second velocity contribution to $g$ simply comes from the fact that the shear at second order $\gamma^{(2)}$ is divided by the convergence at first order $\kappa^{(1)}$ which is affected by peculiar velocity. This effect reflects the fact that peculiar velocities change the apparent size of galaxies, which has then an impact on the reduced shear. Note that in standard perturbation theory, this term appears at the next order, i.e.\;at the same order as the lens-lens coupling and Born correction, see eq.~\eqref{eq:g4}. However here since velocities are of order $\mathcal{O}\left({\frac{1}{c}}\right)$, this coupling is already present at order $\Oo{3}$.

From the second line of~\eqref{eq:g3} we see that the velocity contribution is of the order of 
\begin{equation}
\left(\frac{c}{\HH_S \chi_S}-1\right)\frac{v_{S\parallel}}{c} g^{(2)}\sim \left(\frac{c}{\HH_S \chi_S}-1\right)10^{-3}g^{(2)}\sim\left(\frac{c}{\HH_S \chi_S}-1\right) 10^{-5}\, ,\label{eq:g3vel}
\end{equation}
where we have used that in average the peculiar velocity is of order $v_{S\parallel}/c\sim 10^{-3}$~\cite{Johnson:2014kaa,Watkins:2008hf}. We see that the importance of this contribution depends strongly on the redshift: at small redshift, the prefactor in~\eqref{eq:g3vel} becomes large and the velocity contribution dominates over the vector contribution. For example at $z_S= 0.1$ the prefactor is $\sim 10$ so that the velocity contribution reaches $10^{-4}$. Around redshift 1.6, the prefactor vanishes and the velocity does not contribute at all. At larger redshift, the prefactor slowly decreases towards -1. Finally let us mention that the vector contribution is integrated along the line of sight and therefore it does not depend much on the size of the redshift bins in which the reduced shear is measured. On the other hand, the velocity contribution enters at the source. As a consequence, this contribution becomes strongly suppressed if the reduced shear is averaged over thick redshift bins. Hence for photometric lensing surveys, where thick redshift bins are used, we expect the vector contribution to be the dominant contribution at order $\Oo{3}$.

At order $\Oo{4}$, the reduced shear contains contributions from the shear up to order $\Oo{4}$ and from the convergence up to order $\Oo{2}$. We obtain
\begin{align}
g^{(4)}=&-\frac{{}_{2}\tDD}{{}_{0}\tDD}=
-\frac{{}_{2}\tDD^{(2)}+{}_{2}\tDD^{(3)}+{}_{2}\tDD^{(4)}}{{}_{0}\tDD^{(0)}+{}_{0}\tDD^{(1)}+{}_{0}\tDD^{(2)}}\notag\\
=&\cc{4}\Bigg\{ -\int_0^{\chi_S}d\chi\frac{\chi_S-\chi}{\chi_S\chi}\slpa^2 \left(2W_{P}+ W_{N}^2\right)\label{eq:g4}\\
&-\int_0^{\chi_S}d\chi\frac{\chi_S-\chi}{\chi_S\chi^2}\slpa\left[\slpa^2 W_{N}\int\bound d\chi'  \frac{\chi-\chi'}{\chi'}\bar{\slpa} W_{N}
+\slpa\bar{\slpa}W_{N}\int\bound d\chi'  \frac{\chi-\chi'}{\chi'}\slpa W_{N}\right]\notag\\
&+\int_0^{\chi_S}d\chi\frac{\chi_S-\chi}{\chi_S\chi}\slpa^2 W_{N}\cdot
\left[\int^{\chi_{S}}_{0}d\chi' \frac{\chi_S-\chi'}{\chi'\chi_S}  \bar{\slpa}\slpa W_{N}-\frac{2}{\chi_S}\int^{\chi_{S}}_{0}d\chi'W_{N}\right]\notag\\
&+2\int_0^{\chi_S}d\chi\frac{\chi_S-\chi}{\chi_S\chi}\left[\frac{1}{\chi} W_{N}\int\bound d\chi' \slpa^2 W_{N}  - \frac{1}{\chi}\slpa W_N\int\bound  d\chi' \frac{\chi-\chi'}{\chi'}\slpa W_N \right.\notag\\
&\left.-\slpa^2\left(W_{N,0}\int\bound  W_{N}d\chi'\right)\right]
+\frac{2}{\chi_S}\int_0^{\chi_S}\!\!d\chi\!\left[W_N\int\bound \frac{d\chi'}{\chi'}\slpa^2 W_{N}+\frac{1}{\chi}\slpa^2\left(W_{N}\int\bound\!\!d\chi'  W_{N}\right)\right]\notag\\
&-\frac{1}{4} {}_{2}h_S-\frac{1}{2}\int_0^{\chi_S}d\chi\left(\frac{\chi_S-\chi}{\chi_S\chi}\frac{1}{2}\slpa^2h_{rr}+\frac{1}{\chi}\slpa{}_{1}h_r\right)+\notag\\
&+\left[2\int_{0}^{\chi_{S}}d\chi\frac{\chi_{S}-\chi}{\chi_S \chi} \slpa^2 W_{N} +\frac{c}{\HH_S \chi_S}\int_{0}^{\chi_{S}}d\chi\frac{1}{\chi_S} \slpa^2 W_{N}\,\right]\left( U_{NS}-2\int^{\chi_S}_0 d\chi W_{N,0}\right)\notag\\
&+\left(v^2_S+\frac{c}{\HH_S}\vp\vp'\right)\int_{0}^{\chi_{S}}d\chi\frac{\chi_{S}-\chi}{\chi_S \chi} \slpa^2 W_{N}  \, \notag\\
&+\frac{1}{2\HH_S \chi_S^2}\ichi \slpa^2 W_N \left(\frac{\chi_S}{\chi}\frac{2c}{\HH_S}\vp \vp'+\vs^2   \right)\notag\\
&+\vp^2\left[-\frac{ c^2}{\HH^2_S}\frac{1}{2\chi_S^2}\slpa^2 W_{NS} -3\left(1-\frac{c}{2\HH_S\chi_S}+\frac{c^2\HH'_S}{2\HH_S^3\chi_S}    \right)\int_{0}^{\chi_{S}}d\chi\frac{\chi_{S}-\chi}{\chi_S \chi} \slpa^2 W_{N} \right.\notag\\
&+\left.\frac{1}{2\chi_S}\frac{c}{\HH_S}\left(3+\frac{\HH'_S c}{2\HH^2_S} -\frac{2c}{\HH_S\chi_S}\right) \int_{0}^{\chi_{S}}d\chi\frac{1}{\chi}\slpa^2  W_{N}\right]\notag\\
&+\vp\left[ \int_{0}^{\chi_{S}}d\chi \frac{\left(\chi_{S}-\chi\right)}{\chi_S\chi}\left(\frac{d}{d\chi}\left(\chi\slpa{}_{1}B^{N}\right)+\slpa^2 B_{r}^N\right)\right.\notag\\
&\left.+\frac{c}{2\HH_S \chi_S}\int_{0}^{\chi_{S}}d\chi \frac{1}{\chi_S}\left(\frac{d}{d\chi}\left(\chi\slpa{}_{1}B^{N}\right)+\slpa^2 B_{r}^N\right)\right]\Bigg\}\, .\notag
\end{align}
The first line is the standard shear contribution, where the Newtonian potential $W_N$ has been replaced by the relativistic potential $W_P$ and the square of $W_N$. This term encodes the fact that large-scale structures along the photon trajectory are not completely described by the Newtonian potential $W_N$, and that $W_P$ and $W_N^2$ both give corrections to the potential felt by the photons.  The second line contains the lens-lens coupling and correction to Born approximation~\cite{Cooray:2002mj}. These terms have four transverse derivatives of the potential, and they are therefore expected to dominate at small scales. The third line contains the product between the shear and the convergence at order $\Oo{2}$. The first term in this line also has four transverse derivatives and is therefore of the same order of magnitude as the lens-lens coupling and post-Born correction. Note that the boundary term in the convergence, proportional to $V_{NS}$ (see eq.~\eqref{eq:kappaO2}) cancels with a similar term in ${}_2\tDD$ and does not contribute to the reduced shear. Lines 4 and 5 contain various couplings along the photon trajectory. These terms have been computed for the first time using standard perturbation theory up to second order in~\cite{2010PhRvD..81h3002B} and the expressions agree. Line 6 contains the contribution from the tensor modes, which also appear at second order in SPT. Finally, the last 6 lines contain the contributions due to redshift perturbations. In line 7, we have the redshift perturbations due to gravitational redshift and the integrated Sachs-Wolfe effect. In the following 3 lines, we have the Doppler contributions coupled with the scalar potential. Since the velocity is of order $1/c$, the reduced shear at order $\Oo{4}$ contains contribution from the second order Doppler, i.e.\ from both $\vp$ and $v_{S\perp}$ through the transverse Doppler effect. Finally, in the last two lines we have couplings between the first order Doppler contribution and the vector potential.

\subsection{The convergence $\kappa$}
\label{sec:42}
We now calculate the convergence, i.e.\ the part of the Jacobi map which modifies only the size of the galaxy. As discussed in section~\ref{sec:sec3} it is given by the real part of the spin-0 contribution
\begin{equation}
\kappa=1-\frac{1+z_S}{2\chi_{S}}\textrm{Re}\big[{}_{0}\tilde\DD\big]\, .
\end{equation}
At order $\Oo{1}$ the convergence becomes
\begin{align}
\kappa^{(1)}=&\left(1-\frac{c}{\HH_S \chi_S}\right)     \frac{v_{S\parallel}}{c}\label{eq:kappaO1}\, .
\end{align}
This contribution, called Doppler magnification, has been derived in~\cite{Bonvin:2008ni} for the first time and studied in detail in~\cite{Bacon:2014uja,Bonvin:2016dze}. Since it is directly sensitive to the galaxy peculiar velocity, it provides an alternative way of measuring velocities, independently from redshift-space distortions, and to test theories of modified gravity~\cite{Andrianomena:2018aad}. In the PF formalism, this term is the dominant contribution to the convergence. As shown in~\cite{Bacon:2014uja,Bonvin:2016dze} this is effectively the case at low redshift $z\leq 0.5$. At high redshift however, the order $\Oo{2}$ derived below dominates over the Doppler magnification, because the deviations generated by $\bar{\slpa}\slpa W_{N}$ accumulate along the photon trajectory, whereas the peculiar velocity decreases with redshift. Nevertheless, the Doppler term is still measurable in this regime due to its dipole around overdensities \cite{Bonvin:2016dze}.

At order $\Oo{2}$ the convergence is given by
\begin{align}
\kappa^{(2)}=&\cc{2}\left[-\int^{\chi_{S}}_{0}d\chi   \frac{\chi_S-\chi}{\chi\chi_S}\, \bar{\slpa}\slpa W_{N}
-V_{NS} +  \frac{2}{\chi_S} \int^{\chi_{S}}_{0}d\chi W_{N}\right.\notag\\
&+\left(\frac{c}{\HH_S \chi_S}-1\right)\left(U_{NS}-2\int_0^{\chi_S} d\chi W_{N,0}+\frac{1}{2}v^2_S-\frac{c}{\HH_S}\vp'\vp\right)\notag\\
&\left.-\left(2-\frac{5c}{2\HH_S\chi_S}-\frac{c\HH'_S}{2\HH^3_S\chi_S}    \right)  v_{S\parallel}^2\right]\label{eq:kappaO2}
\end{align}
The first term in~\eqref{eq:kappaO2} is the standard Newtonian contribution to the convergence. Since it contains two transverse derivatives, it dominates over the other terms when one correlates galaxies at small separations. This term is the only one which changes the apparent size of galaxies through a real focusing of the light beam. The other two terms in the first line modify the length of the geodesic between the source and the observer, and consequently they change the apparent size of galaxies. The terms in the second and third line are due to the fact that we observe the size of galaxies as a function of redshift, which is a perturbed quantity. In particular, the first term in the second line is the contribution from gravitational redshift and the second one is the integrated Sachs-Wolfe contribution. The terms proportional to peculiar velocities in the second and third lines are second-order Doppler contributions.  These contributions are sensitive not only to the radial part of the peculiar velocity $v_{S\parallel}$ but also to its transverse part since $v_S^2=v_{S\parallel}^2+v_{S\perp}^2$. Note that one contribution depends also on the time derivative of the peculiar velocity $\vp'$, which contributes to the redshift perturbation at second-order, see eq.~\eqref{eq:deltaz2}.

The convergence at order $\Oo{3}$ is given by
\begin{align}
\kappa^{(3)}=&\cc{3}\left\{\int^{\chi_S}_0 d\chi\left[\frac{\chi_{S}-\chi}{2\chi_S\chi} \bar{\slpa}\slpa B_{Nr}+ \frac{1}{4\chi}\left(\bar{\slpa}{}_{1}B_N+\slpa{}_{-1}B_N\right)-\frac{1}{\chi_S}B_{Nr}+\right.\right.\notag\\
%
 &\left. +\left(\frac{c}{\HH_S \chi_S}-1\right)   B_{Nr,0}\right]+\vp\Bigg[\left(-4   +\frac{\HH'_S c^2}{\chi_S\HH^3_S}  +\frac{5c}{\HH_S \chi_S}\right) U_{NS}  \notag\\
%
&     +\int^{\chi_S}_0 d\chi \left(6-\frac{2\HH'_S c^2}{\chi_S\HH^3_S} -\frac{10c}{\HH_S \chi_S}+\frac{\chi}{\chi_S}+\frac{c \chi}{\HH_S \chi_S^2}\right) W_{N,0} +\notag\\
 	 &+\int_{0}^{\chi_{S}}d\chi\left(\frac{\chi_S-\chi}{\chi\chi_S} -\frac{c}{\chi \chi_S\HH_S}\right)  \bar{\slpa}\slpa W_N   +2V_{NS}   + W_{NS}    +\notag\\
&     +\frac{c}{\HH_S}\left(\frac{d U_{NS}}{d\chi} -2W_{NS,0}  \right) +\frac{c}{\HH_S \chi_S}     \Big[ -3V_{NS}    + W_{NS}+\notag\\
&  -\chi_{S}\frac{dV_{NS}}{d\chi_S} +\int^{\chi_{S}}_{0}d\chi \frac{2}{\chi_S}W_{N}       +\frac{c}{\HH}\left(\frac{d U_{NS}}{d\chi} -2W_{NS,0}  \right) \Big]\Bigg]+\notag\\
        &+\vp'\left(\frac{c}{\HH_S \chi_S}-1\right)  \frac{c}{\HH_S}\left( -U_{NS}   +2\int^{\chi_S}_{0} d\chi W_{N,0} \right)+\notag\\
        &    +\left(\frac{c}{\HH_S \chi_S}-1\right)  \left(\frac{1}{2}\vpl \ichi \frac{1}{\chi}\bar{\slpa} W_N    +\frac{1}{2}\vm \ichi \frac{1}{\chi}\slpa W_N\right) +\notag\\
%
	&+\left(   \frac{\HH'_S c^2}{\chi_S\HH^3_S}      +\frac{4c}{\HH_S \chi_S}     -3\right)\vp\frac{1}{2}\vs^2  +\frac{1}{2} \frac{c}{\HH_S}\vp^2 \vp'\left(   3\frac{\HH'_S c^2}{\chi_S\HH^3_S}    + \right.\notag\\
&\left.    +\frac{5c}{\HH_S \chi_S}   -\frac{\HH'_S c}{\HH^2_S\chi_S}  -3 \right)  -\left( \frac{c}{\HH_S\chi_S}-1\right) \frac{c}{\HH} \vp'\left(\frac{1}{2}\vs^2  +\frac{1}{\HH_S }\vp' \vp \right)+\notag\\
        &+\left(\frac{c}{\HH_S \chi_S}-1\right)  \frac{c}{\HH} \vp\vs \vs'  -\left(\frac{c}{\HH_S \chi_S}-1\right)  \frac{c^2}{2\HH_S^2}\vp'' \vp^2 +\notag\\
        &+ \vp^3 \left(2   -\frac{8c}{\HH_S \chi_S}     -\frac{\HH''_S c^3}{6\chi_S  \HH^4_S}+\frac{ \HH'^2 c^3}{2\chi_S  \HH^5_S}    \right)  \Bigg\}
.\label{eq:kappaO3}
\end{align}
As for the reduced shear, at this order the convergence contains two types of contributions. First, contributions from the vector potential $B^i_N$. The dominant contribution at small scales is given by the first term, which contains two transverse derivatives, and is equivalent to the shear contribution in eq.~\eqref{eq:g3}. In addition, since the convergence is a spin-0 field it contains contributions from the spin-1 and -1 part of $B^i_N$, on which the transverse operators $\slpa$ and $\bar\slpa$ act once. The second type of contributions to the convergence are due to the coupling between the first order Doppler contribution and the convergence at second order. The spin-1 and -1 contributions $\vpl$ and $\vm$, respectively, stem from the decomposition of the peculiar velocity field $\vs^{i}$ into $\vs^{i}=\vp^{i}+\frac{1}{2}\vm \eplus{i}+\frac{1}{2}\vpl \eminus{i}$ and occur when the vector field $\vs^{i}$ is coupled with the derivative of the scalar potential $W_N$. Finally, the convergence contains also a pure Doppler contribution, proportional to the velocity cubed, in the last line of~\eqref{eq:kappaO3}. 

The expression for the convergence at order $\Oo{4}$ is grouped into various terms according to the potentials or their couplings plus various contributions from redshift perturbations: 
\begin{align}
    \kappa^{(4)}=&\kap{P}+\kap{UW}+\kap{VV}+\kap{VW}+\kap{WW}+\kap{h}+\kap{\delta z}+\kap{v}+\kap{v^2}+\kap{v^4}\label{eq:kappasplit}.
\end{align}
The superscripts $(UW)$, $(VW)$, $(VV)$, and $(WW)$ refer to the couplings of the Newtonian potentials $U_N$, $V_N$, and $W_N$, whereas the superscript $(P)$ denotes the contributions of the post-Friedmann potentials $U_P$, $V_P$, and $W_P$. The last four terms in eq.~\eqref{eq:kappasplit} with the superscripts $(\delta z)$, $(v)$, $(v^2)$, and $(v^4)$ refer to the terms that are introduced via the redshift perturbations in eq.~\eqref{eq:Dz44}. In particular $(\delta z)$ regroups all redshift perturbations not due to peculiar velocity, whereas the other terms regroup the velocity terms at each relevant order. 

The first contribution of eq.~\eqref{eq:kappasplit} reads
\begin{align}
\kap{P}=& \cgen{2}{4}\left(\ichi   \frac{2}{\chi_S}W_P    -V_{P}    -\ichi \frac{\chi_S-\chi}{\chi_S\chi}\bar{\slpa}\slpa W_{P}\right)\, ,
\label{eq:kappaP}
\end{align}
and is of purely relativistic origin. Note that $\kap{P}$ takes on the same form as $\kappa^{(2)}$ in eq.~\eqref{eq:kappaO2} with the relativistic potentials $2V_P$ and $2W_P$ replacing the Newtonian potentials $V_N$ and $W_N$. The terms derived from the redshift perturbations in $\kappa^{(2)}$ have their relativistic analogue in $\kap{\delta z}$.

The next contributions $\kap{UW}$, $\kap{VV}$, $\kap{VW}$, and $\kap{WW}$ collect the coupling terms with the Newtonian potentials $U_N$, $V_N$, and $W_N$: 
\begin{align}
    \kap{UW}=&\frac{1}{\chi_S}\cc{4}\left\{
\int^{\chi_S}_{0} d\chi \left[-4 W_{N}\left(U_{NS}  +U_N +\left(\chi_S -\chi\right)\dchi U_N  \right) +\right.\right.\notag\\
&\left.-4U_{N,0} \left(   W_{N} - \left(\chi_S -\chi\right)\int^{\chi}_{0}d\chi'W_{N}  \right)\right]+\notag\\
    &+2 \int^{\chi_S}_{0} d\chi \left(\slpa U_{N} \int_{0}^{\chi}d\chi'\frac{\chi-\chi'}{\chi \chi'}\bar{\slpa}W_{N}    +\bar{\slpa} U_{N} \int_{0}^{\chi}d\chi'\frac{\chi-\chi'}{\chi \chi'} \slpa W_{N}\right) +   \notag\\
    &\left.-2  \ichi \frac{\chi_S-\chi}{\chi} \left(\slpa U_{N} \int_{0}^{\chi}d\chi' \frac{1}{\chi'}\bar{\slpa}W_{N}  +\bar{\slpa} U_{N} \int_{0}^{\chi}d\chi' \frac{1}{\chi'} \slpa W_{N}\right)\right\},\label{eq:kappaUW}\\[10pt]
    \kap{VV}=&\cc{4}\left[2\ichi \frac{\left(  \chi_S - \chi\right)\chi}{\chi_S}\left(\dchi V_N\right)^2-\frac{1}{4}V^2_{NS}\right],\label{eq:kappaVV}\\[10pt]
\kap{VW}=& \cc{4}\left\{\ichi  \frac{1}{\chi_S}\left[ W_N\left(V_{NS}  +2\chi \frac{d}{d\chi_S} V_{NS}   -2\chi V_{NS,0} - 2\chi_S  \dchi V_{N}\right) +\right.\right.\notag\\
&\left.  +2 W_{N}\chi_S V_{N,0} +2\slpa V_{N}\int^{\chi}_0 d\chi' \frac{1}{\chi'}\bar{\slpa} W_{N}    +2\bar{\slpa} V_{N}\int^{\chi}_0 d\chi' \frac{1}{\chi'}\slpa W_{N}   \right] +\notag\\
&+\ichi \frac{\chi_S - \chi}{\chi_S} \left[       - 2 \chi W_{N}  \left( \ddchi V_{N}   -\dchi V_{N,0}\right)-\slpa V_{N} \frac{1}{\chi}\bar{\slpa} W_{N} \right.\notag\\
&-\bar{\slpa} V_{N} \frac{1}{\chi}\slpa W_{N}    -\slpa V_{NS} \frac{1}{\chi}\bar{\slpa} W_{N}     -\bar{\slpa} V_{NS} \frac{1}{\chi}\slpa W_{N}-  V_{NS} \frac{1}{\chi}\bar{\slpa}\slpa    W_{N} +\notag\\
&\left.\left.    +V_{NS}W_{N,0}   \right]\right\}\, ,\label{eq:kappaVW}
\end{align}
and
\begin{align}
\kap{WW}=& \cc{4}\left\{\ichi\frac{1}{\chi_S}\left[ 4 W_N^2  - 4W_{NS}W_N  +8 W_{N,0}  \int^{\chi}_{0}d\chi'W_{N}\right.\right.+\notag\\
&- 10  \slpa W_N   \int_{0}^{\chi}d\chi' \frac{\chi-\chi'}{\chi \chi'}\bar{\slpa} W_N -   10  \bar{\slpa} W_N     \int_{0}^{\chi}d\chi' \frac{\chi-\chi'}{\chi \chi'}\slpa W_N   +\notag\\
& +  4 W_N  \int\bound d\chi' \frac{\chi - \chi'}{\chi \chi'}\bar{\slpa}\slpa W_{N}     +  2 W_N  \int\bound   d\chi'\frac{1}{ \chi'}\bar{\slpa}\slpa  W_{N}    +\notag  \\
&\left.  +  2 \frac{1}{\chi} \slpa \bar{\slpa}\left(W_N \int\bound \dc' W_{N} \right)  \right]  +  \ichi \frac{\chi_S - \chi}{\chi_S} \Big[ -4 W_{N,0}^2      +\notag\\
& -   \bar{\slpa}\slpa W_N^2  \frac{1}{\chi} -8 W_{N} W_{N,0}  +2\left(\dchi W_N\right)^2    -4 W_{N,0}\dchi W_N +\notag\\
&-4W_{N,00} \int\bound d\chi' W_{N}   -\frac{2}{\chi^2} W_N  \int\bound d\chi' \bar{\slpa}\slpa W_{N}  +\notag\\
&+6 \slpa W_N   \int_{0}^{\chi}d\chi' \frac{1}{\chi \chi'}\bar{\slpa} W_N +6 \bar{\slpa} W_N     \int_{0}^{\chi}d\chi' \frac{1}{\chi \chi'}\slpa W_N+\notag\\
& +  8 \slpa   W_{N,0} \int_{0}^{\chi}d\chi' \frac{\chi-\chi'}{\chi \chi'}\bar{\slpa} W_N + 8 \bar{\slpa}   W_{N,0} \int_{0}^{\chi}d\chi' \frac{\chi-\chi'}{\chi \chi'}\slpa W_N +\notag\\
&+2  \slpa W_N   \int_{0}^{\chi}d\chi' \frac{\chi - \chi'}{\chi^2\chi'}  \bar{\slpa} W_N        +2  \bar{\slpa} W_N     \int_{0}^{\chi}d\chi'   \frac{\chi - \chi'}{\chi^2\chi'}\slpa W_N+\notag\\
&-2 \bar{\slpa}\slpa \left( W_{N,0}   \int\bound  d\chi' \frac{1}{ \chi'}  W_{N}\right) +2 \bar{\slpa}\slpa  W_{N,0}   \int\bound  d\chi'  \frac{\chi - \chi'}{ \chi\chi'}  W_{N} +\notag\\
%
& +4 \frac{1}{ \chi^2}\bar{\slpa}\slpa  W_{N}\int\bound d\chi' W_N  -4 \frac{1}{ \chi^2}\bar{\slpa}\slpa  W_{N}\int\bound d\chi' \left(\chi-\chi'\right) W_{N,0} \notag\\
%
&-\frac{1}{2}\slpa^2 W_{N}\int\bound d\chi'\frac{\chi - \chi'}{\chi^2 \chi'} \bar{\slpa}^2 W_{N}       -\frac{1}{2}\bar{\slpa}^2 W_{N}\int\bound d\chi'\frac{\chi - \chi'}{\chi^2 \chi'}  \slpa^2 W_{N} +\notag\\
&  - 2 W_{N,0}\int\bound d\chi' \frac{\chi - \chi'}{\chi \chi'}\bar{\slpa}\slpa W_{N}     - \slpa\left( \bar{\slpa}\slpa W_N       \int\bound d\chi'\frac{\chi-\chi'}{\chi^2 \chi'}\bar{\slpa} W_{N}\right)   + \label{eq:kappaWW} \\
&   \left.\left.-  \slpa \bar{\slpa}^2 W_N      \int\bound d\chi'\frac{\chi-\chi'}{\chi^2\chi'}\slpa W_{N}\notag\right]\right\}.
\end{align}
The tensor potential $h_{ij}$ contributes to the convergence in the following way: 
\begin{align}
\kap{h}=&\cc{4}\frac{1}{4}\left[ h_{rr}\left(\chi_S\right)-\ichi \frac{1}{\chi}\left(\slpa {}_{-1}h_{r}+\bar{\slpa} {}_{1}h_{r} -h_{rr}\right)\right.+\notag\\
&\left.+\ichi \frac{\chi_S - \chi}{\chi_S\chi}\left(\slpa \bar{\slpa} h_{rr}-\chi h_{rr,0}\right)\right].\label{eq:kappah}
\end{align}
Finally, the redshift perturbations are split into four different groups: the first group is denoted by $\kap{\delta z}$ and refers to the redshift perturbations independent of the peculiar velocity, while the other groups $\kap{v}$, $\kap{v^2}$, and $\kap{v^4}$ refer to the terms dependent on the peculiar velocity. 
\begin{align}
    \kap{\delta z}=&
\cc{4}\Bigg\{
 \left[   \frac{1}{\chi_S}\int^{\chi_{S}}_{0}d\chi \left( 2W_{N}-\left(\chi_{S}-\chi-\frac{c}{\HH_s}\right)  \frac{1}{\chi}\bar{\slpa}\slpa W_N \right) \right.+\notag\\
%
            &\left.+\frac{c }{\chi_S\HH_S} \left(\chi_{S}\frac{dV_{NS}}{d\chi_S}- 2 W_{NS}\right)\right]\left( U_{NS}-2\int^{\chi_S}_0 d\chi W_{N,0} \right)+\notag\\
         -&   \left(1-\frac{\HH'_S c^2}{2\chi_S\HH^3_S}-\frac{3c}{2\HH_S \chi_S}\right)\left( U_{NS}-2\int^{\chi_S}_0 d\chi W_{N,0}\right)^2+\notag\\
&+\left(\frac{c}{\HH_S \chi_S}-1\right)\left\{2U_{PS}-4\int^{\chi_S}_{0}d\chi W_{P,0}-\frac{5}{2}U_{NS}^2   -\frac{1}{2}\int^{\chi_S}_{0}d\chi h_{rr,0}+\right.\Bigg.\notag\\
       & -2 \left(\int^{\chi_S}_{0}d\chi W_{N,0}\right)^2+ 4\left(U_{NS}\int^{\chi_S}_{0} d\chi W_{NS}\right)_{,0}+ 4W_{NS,0}\ichi W_N+\notag\\
       &+2 \frac{1}{\chi_S}\bar{\slpa} U_{NS} \int_{0}^{\chi_S}d\chi\left(\chi_S-\chi\right)\frac{1}{\chi}\slpa W_N    +2 \frac{1}{\chi_S}\slpa U_{NS} \int_{0}^{\chi_S}d\chi\left(\chi_S-\chi\right)\frac{1}{\chi}\bar{\slpa} W_N   +   \notag\\
		&-4\frac{d}{\chi_S} U_{NS}  \int^{\chi_S}_{0}d\chi W_{N} +V_{NS}\left(U_{NS}-2\ichi W_{N,0}\right)+\notag\\
&+\frac{c}{\HH_S}\left(  \frac{d U_{NS}}{d\chi_S}U_{NS}-2\frac{d U_{NS}}{d\chi_S}\int^{\chi_S}_{0}d\chi W_{N,0} -2W_{NS,0} U_{NS}    +2W_{NS,0}2\int^{\chi_S}_{0}d\chi W_{N,0} \right)\notag\\  
	   &+\int^{\chi_S}_{0} d\chi\left(  4 W_N \dchi U_{N}   - 4U_{N,0} W_{N} -4   W_{N,0} W_{N} - 4  W_{N,00}\int^{\chi}_{0}d\chi'W_{N} +\right.\notag\\
        %
        &-2 \slpa U_{N}  \int_{0}^{\chi}d\chi' \frac{1}{\chi \chi'}\bar{\slpa}W_N    -2   \bar{\slpa} U_{N}  \int_{0}^{\chi}d\chi'\frac{1}{\chi \chi'}\slpa W_N  +   \notag\\
	&- 2\slpa W_{N,0}\int_{0}^{\chi}d\chi\frac{\chi - \chi'}{\chi chi'}\bar{\slpa} W_N   -2\slpa W_N   \int_{0}^{\chi}d\chi\frac{\chi - \chi'}{\chi^2 \chi'}\bar{\slpa} W_N  +\notag\\
	&\left. \left.   -2\bar{\slpa} W_{N,0} \int_{0}^{\chi}d\chi'\frac{\chi - \chi'}{\chi \chi'}\slpa W_N         -2\bar{\slpa} W_N    \int_{0}^{\chi}d\chi'\frac{\chi - \chi'}{\chi^2 \chi'}\slpa W_N     \right)\right\}\Bigg\},
\end{align}
\begin{align}
    \kap{v}=&\cc{4}\Bigg\{
\vp\left[ \int_{0}^{\chi_{S}}d\chi \left(\frac{\chi_{S}-\chi}{\chi_S \chi}-\frac{c}{\chi_S \chi \HH_S}\right)\frac{1}{2}\left( 2\chi B_{Nr,0}  -\frac{d}{d\chi}\left(\chi \slpa {}_{1}B_N\right)   -\slpa^2B_{Nr}\right)\right.+\notag\\
&\left.-\frac{c}{\chi_S \HH_S}B_{NSr}  +\int^{\chi_S}_0 d\chi  \left( B_{Nr} +\left(2-\frac{\HH'_S c^2}{\chi_S \HH^3_S} -\frac{3c}{\chi_S\HH_S }\right)B_{Nr,0}\right) \right]+\notag\\
&-\left(\frac{c}{\HH_S \chi_S}  -1\right)
\left[	 \vpl\left(      \frac{1}{2}{}_{-1}B_{NS}   +\frac{1}{2}\int^{\chi_S}_{0}d\chi \frac{1}{\chi}\left(\bar{\slpa} B_{Nr} +{}_{-1}B_{N} \right)       \right)\right.+\notag\\
	&\left.+\vm\left(    \frac{1}{2}{}_{1}B_{NS}   +\frac{1}{2}\int^{\chi_S}_{0}d\chi  \frac{1}{\chi}  \left(\slpa B_{Nr}   +{}_{1}B_{N} \right)\right)\right] \Bigg\},
\end{align}
\begin{align}
    \kap{v^2}=&\cc{4}\Bigg\{\left(\frac{c}{\HH_S \chi_S}  -1\right)\left\{
	\vp^{\prime 2}\left( 2\frac{c^2}{\HH_S^2 }\int^{\chi_S}_{0}d\chi W_{N,0}  -\frac{c^2}{\HH_S^2 } U_{NS} \right)    +    \right.\notag\\
&+ \left(\vs \vs' +\frac{\vp c}{\HH_S}\vp'' \right)\frac{c}{\HH_S}\left( U_{NS}-2\int^{\chi_S}_0 d\chi W_{N,0}\right)  +\notag\\
&    -\frac{c}{\HH_S }\vp'\vm\frac{1}{2}\int^{\chi_S}_{0}d\chi' \frac{1}{\chi'}\slpa W_{N}    -\frac{c}{\HH_S }\vp'\vpl\frac{1}{2}\int^{\chi_S}_{0}d\chi' \frac{1}{\chi'}\bar{\slpa} W_{N}  +  \notag\\
	&  +\vp\frac{c}{\HH_S}\vm'\frac{1}{2}\int^{\chi_S}_{0}d\chi' \frac{1}{\chi'}\slpa W_{N}   +\vp\frac{c}{\HH_S}\vpl'\frac{1}{2}\int^{\chi_S}_{0}d\chi' \frac{1}{\chi'}\bar{\slpa} W_{N}  +  \notag\\
	& +\vp\vm\frac{1}{2} \frac{1}{\chi_S}\slpa W_{NS}    +\vp\vpl \frac{c}{2\HH_S\chi_S}\bar{\slpa} W_{NS}   \left. \right\}+\notag\\
+&\left[  \left(\frac{c}{\HH_S \chi_S}  -1\right)\left(  \frac{c}{\HH_S}\frac{d U_{NS}}{d\chi}   -2\frac{c}{\HH_S}W_{NS,0} \right) + \frac{2}{\chi_S}\int^{\chi_{S}}_{0}d\chi  W_{N}    -3 V_{NS}+\right.\notag\\
&+\frac{c }{\chi_S\HH_S} \left(\chi_{S}\frac{dV_{NS}}{d\chi_S}   +4 W_{NS}\right)  - \int^{\chi_{S}}_{0}d\chi  \left(\frac{\chi_{S}-\chi}{\chi_S \chi}-\frac{c}{\chi_S \chi \HH_s}\right) \bar{\slpa}\slpa W_N +\notag\\
            &\left.+\left(4-\frac{\HH'_S c^2}{\chi_S\HH^3_S}-\frac{5c}{\HH_S \chi_S}\right) 2\int^{\chi_S}_0 d\chi W_{N,0}   -\left(1-\frac{\HH'_S c^2}{\chi_S\HH^3_S}+\frac{c}{\HH_S \chi_S}\right)U_{NS} \right]\frac{1}{2}\vs^2 +\notag\\
+&\left[   \frac{2}{\chi_S}\int^{\chi_{S}}_{0}d\chi W_{N}    -  \int^{\chi_{S}}_{0}d\chi \left(\frac{\chi_{S}-\chi}{\chi_S \chi}-\frac{c}{\chi_S \chi \HH_s}\right)  \bar{\slpa}\slpa W_N \right.+\notag\\
&+\frac{c }{\HH_S}  \frac{dV_{NS}}{d\chi_S}        - V_{NS} +\left(   -1 +\frac{\HH'_S c^2}{\chi_S\HH^3_S} +\frac{2c}{\chi_S \HH_S}  +\frac{\HH'_S c}{ \HH_S^2 }\right)U_{NS}     +  \notag\\
            &\left.  + \left(   3  -\frac{7c}{\HH_S \chi_S} -\frac{2\HH'_S c^2}{\chi_S\HH^3_S} -2\frac{\HH'_S c}{ \HH_S^2 }\right)\int^{\chi_S}_0 d\chi W_{N,0}  \right]\left( \frac{c}{\HH_S }\vp' \vp \right)+\notag\\
&+\left[    V_{NS}\frac{c}{2\HH_S \chi_S}\left(   1+\frac{\HH'c}{\HH^2} \right)  -3V_{NS}  +2  W_{NS}\frac{c}{\HH_S \chi_S}\left(   4+\frac{\HH'c}{\HH^2} \right) +\right. +\notag\\
&+\frac{1}{2}\left(\chi_S\frac{d}{d\chi_S}V_{NS}\frac{c}{\HH_S \chi_S}      +\int^{\chi_{S}}_{0}d\chi  \frac{1}{\chi}\bar{\slpa}\slpa W_N  \frac{c}{\HH_S \chi_S}\right)\left(   1+\frac{\HH'c}{\HH^2}   \right)  +\notag\\
	&+\left(\frac{c}{\HH_S \chi_S}  -1\right)   \left(5\frac{c}{\HH_S}+\frac{\HH'_Sc^2}{\HH^3_S} - \frac{1}{2}\frac{c^2}{\HH_S^2} \frac{d}{d\chi_S}   \right)\frac{1}{2}\frac{d U_{NS}}{d\chi_S} +3\frac{c}{\HH_S}\frac{d W_{NS}}{d\chi_S}   +\notag\\
&   +\left(8\frac{c}{\HH_S}+\frac{\HH'_Sc^2}{\HH^3_S} - \frac{1}{2}\frac{c^2}{\HH_S^2} \frac{d}{d\chi_S}  -5\frac{c^2}{\HH_S^2 \chi_S}  -\frac{\HH'_Sc^3}{\HH^4_S\chi_S}   + \frac{1}{2}\frac{c^3}{\HH_S^3\chi_S} \frac{d}{d\chi_S}   \right)W_{NS,0}      + \notag\\
&+\left(-1-\frac{7c}{2\chi_S  \HH_S}    -\frac{\HH''_S c^3}{2\chi_S  \HH^4_S}   +\frac{3 \HH'^2 c^3}{2\chi_S  \HH^5_S}       -\frac{\HH'_S c^2}{\chi_S\HH^3_S}  \right) U_{NS}+\notag\\
&+\left(  \frac{\HH''_S c^3}{\chi_S  \HH^4_S}   +2   -\frac{3 \HH'^2 c^3}{\chi_S  \HH^5_S}    -\frac{ \HH'_S c^2}{\chi_S  \HH_S^3}      -\frac{2c}{\HH_S \chi_S}  \right)  \int^{\chi_S}_{0} d\chi W_{N,0}  +\notag\\
    &\left.+\frac{c^2}{\HH^2_S \chi_S}\left(   -2  \frac{d}{d\chi_S}W_{NS}   +\frac{1}{2}\bar{\slpa}\slpa W_{NS}  -\frac{d}{d\chi_S}V_{NS}   -\frac{1}{2}\chi_S\frac{d^2}{d\chi_S^2}V_{NS}\right)\right]\vp^2+\notag\\
&  +\left(2-\frac{\HH'_S c^2}{\chi_S\HH^3_S} -\frac{3c}{\HH_S \chi_S}\right)\vp  \left( \vm\frac{1}{2}\int^{\chi_S}_{0}d\chi' \frac{1}{\chi'}\slpa W_{N}   + \vpl\frac{1}{2}\int^{\chi_S}_{0}d\chi' \frac{1}{\chi'}\bar{\slpa} W_{N}\right)  \Bigg\},
\end{align}
and 
\begin{align}
    \kap{v^4}=&\cc{4}\Bigg\{\left(\frac{c}{\HH_S \chi_S}-1\right)\Bigg[
    \frac{1}{2}\left( 1+ \frac{\HH'_Sc}{\HH^2_S} \right)\frac{c}{\HH_S} \left(\vp^2\vs \vs'  -\vp^3 \vp''\right)      +\notag\\
 	&     +   \frac{c}{2\HH_S}\vs \vs' \vs^2    -\frac{c^2}{2\HH_S^2}\vp^2\vs \vs''  +\frac{c^3}{6\HH_S^3}\vp^3 \vp'''   -\frac{c^2}{1\HH_S^2 }\vp^{\prime 2}\vs^2  +    \notag\\
	 &   -\frac{c^2}{\HH_S^3 }\vp^{\prime 3}\vp   -\frac{1}{2}\frac{c^2}{\HH_S^2}\vp^2\vs^{\prime 2}    +\frac{c^2}{2\HH_S^2}\vp\vp'' \vs^2   +\frac{3c^2}{2\HH_S^2}\frac{c}{\HH_S}\vp^2\vp'\vp''   +\notag\\
	&  +\vp^2 \vp^{\prime 2}\frac{c}{\HH_S} \int^{\chi_S}_{0}d\chi W_{N,0}  \Bigg]
%
   +\left(     \frac{1}{2}   +\frac{\HH'_S c^2}{\chi_S\HH^3_S}     -\frac{\HH'_S c^2}{2\HH^2_S}   \right)  \frac{c}{\HH_S }\vp' \vp \vs^2       +\notag\\
  & -\left(     3-\frac{\HH'_S c^2}{\chi_S\HH^3_S}-\frac{4c}{\HH_S \chi_S}\right)  \frac{1}{8}\vs^4  
  -\left(     \frac{3}{2}   +\frac{\HH'_S c^2}{\chi_S\HH^3_S} -\frac{3\HH'_S c}{2\HH^2_S}  -\frac{c}{\HH_S \chi_S}\right)      \frac{c^2}{\HH_S^2 }\vp^{\prime 2} \vp^2  +\notag\\    
& +\left(  \frac{\HH''_S c^3}{2\chi_S  \HH^4_S}  -\frac{5c}{2\chi_S  \HH_S}-\frac{3 \HH'^2 c^3}{2\chi_S  \HH^5_S}-\frac{3 \HH'_S c^2}{\chi_S  \HH_S^3}         \right)\frac{1}{2}\vs^2 \vp^2+\notag\\
        &+\left(2-\frac{\HH'_S c^2}{\chi_S\HH^3_S} -\frac{3c}{\HH_S \chi_S}\right)\frac{c}{\HH_S}\left[     -\vp' \vp\frac{1}{2}\vs^2   +\vp^{\prime 2} \vp^2\frac{1}{\HH_S }+ \vp^2\frac{d U_{NS}}{d\chi}   \right.+\notag\\
        &  - \vp^22W_{NS,0} + \vp^2 \vs \vs'    \left.  -\vp^3\frac{c}{\HH_S}\vp''  \right]
+\left(\frac{17c}{6\chi_S  \HH_S}   -\frac{4}{3}   -\frac{2\HH''_S c^3}{3\chi_S  \HH^5_S} +\right.\notag\\
&\left.+\frac{\HH''_S c^2}{6  \HH^4_S}       +\frac{ \HH^{\prime 2} c^3}{3\chi_S  \HH^5_S}    -\frac{ \HH'_S c^2}{2\chi_S  \HH_S^3}     +\frac{3\HH'_Sc}{2\HH^2_S}      +\frac{\HH^{\prime 2}_S c^2}{6\HH^4_S}        \right)\frac{c}{\HH_S }\vp' \vp^3 \notag\\
& -\left(\frac{\HH'_S c^2}{24\chi_S\HH^3_S}    -\frac{c}{12\chi_S  \HH_S}     -\frac{c^3 \HH''_S}{12\chi_S  \HH^4}       +\frac{1 \HH'^2 c^3}{4\chi_S  \HH^5_S}       -\frac{\HH_S^{\prime \prime \prime}c^4}{24\chi_S  \HH_S^5}         +\right.\notag\\
&\left.-\frac{15c^4 \HH^{\prime 3}_S}{24\chi_S  \HH_S^7}       +\frac{5 c^4\HH'_S \HH''_S}{12\chi_S  \HH^6}      \right)\vp^4\Bigg\}.\label{eq:kappav4}
\end{align}

 The convergence at second-order in standard perturbation theory has been computed in~\cite{Umeh:2014ana}. We expect some of the terms in our formalism to be equivalent to the SPT result, while others will be different, due to the different counting of perturbations.

%
\subsection{The rotation $\omega$}
\label{sec:rotation}
The rotation $\omega$ is related to the imaginary part of the spin-0 component ${}_{0}\tDD$ via eq.~\eqref{eq:kappa}, which is proportional to the anti-symmetric part of $\tDD_{ab}$, see eq.~\eqref{eq:0D} . From eqs.~\eqref{eq:DD2}, \eqref{eq:DD3}, \eqref{eq:DZ1},~\eqref{eq:DZ2} and~\eqref{eq:DZ3} we see that up to order $\Oo{3}$, $\tDD_{ab}$ is symmetric and that there is therefore no rotation at those orders. At order $\Oo{4}$ on the other hand, there is a anti-symmetric contribution generated by the coupling $\RR^{(2)}_{ac}\DD^{(2)}_{cb}$ in eq.~\eqref{eq:S}. We obtain (see appendix~\ref{app:Dspin} for more detail)
\begin{align}
\omega^{(4)}=&\frac{1}{2\chi_Sc^4}\intchi \frac{1}{\chi^2}\left(\slpa^2W_N\int^{\chi}_0 d\chi' \frac{\chi - \chi'}{\chi'} \bar{\slpa}^2 W_N  \right.\notag\\
&\left.  -\bar{\slpa}^2W_N\int^{\chi}_0 d\chi' \frac{\chi - \chi'}{\chi'} \slpa^2 W_N\right).
\end{align}
 We see that the only terms that contribute to the rotation at order $\Oo{4}$ are the lens-lens coupling and the post-Born correction, i.e.\ the terms with four transverse derivatives, which dominate at small scales. The rotation contributes in principle to the ellipticity orientation, as discussed in~\cite{2010PhRvD..81h3002B}. However, since the shear is at least of order $\Oo{2}$ and the rotation is of order $\Oo{4}$, the contribution to the ellipticity is of order $\Oo{6}$. This represents a very small contribution to the ellipticity B-mode.

\subsection{Kaiser-Squires relation}
The shear and the convergence are usually assumed to obey the Kaiser-Squires relation \cite{1993ApJ...404..441K}, given by
\begin{align}
\langle |\gamma |^2\rangle=\langle \kappa^2 \rangle.\label{eq:KSR}
\end{align}
This relation can be related to the fact that at small scales the standard contribution to the shear~\eqref{eq:gg} and to the convergence (first term in~\eqref{eq:kappaO2}) obey
\begin{equation}
\bar\slpa^2\gamma=\bar\slpa\slpa\kappa\, .   
\end{equation}
This relation is violated by a number of different terms. First, at order $\mathcal{O}\left(\frac{1}{c}\right)$, the Doppler magnification~\eqref{eq:kappaO1} breaks this relation. As pointed out in~\cite{Bonvin:2008ni}, this could provide an alternative way of measuring the peculiar velocity, by combining the measured shear with the measured convergence in order to isolate the Doppler contribution. At order $\Oo{2}$, the relativistic effects in the convergence, as well as the redshift contributions to the convergence also violate this relation. At order $\Oo{3}$, the vector potential has one term which satisfies the Kaiser-Squires relation, and other terms which violate it. Finally, at order $\Oo{4}$ a large number of terms break this relation as well.

The dominant source of violation among all these terms will depend on the range of redshift, the scales considered and the size of the redshift bins. For example, at small redshifts and for thin redshift bins the Doppler magnification will probably be the dominant source of violation. On the other hand, if the shear and convergence are averaged over large redshift bins, the Doppler contribution (as well as all contributions at the source) will quickly become negligible and other effects will come into play. In particular, in this regime the Shapiro time delay, which contributes to the convergence (last term in the first line of eq.~\eqref{eq:kappaO2}) but not to the shear at order $\Oo{2}$, will become relevant. In addition, the vector potential contributes to the convergence through an integral along the line-of-sight (first two terms in eq.~\eqref{eq:kappaO3}) in a different way than to the shear and therefore this violation will survive in the case of thick redshift bins. In a future work, we will calculate in detail the amount of violation induced by the Shapiro time delay and the vector potential to determine which one dominates and if the violation is large enough to be detected. We expect the Shapiro time delay to dominate at large scales, and the vector potential at small scales.

\section{Conclusion}
\label{sec:con}
Weak gravitational lensing is becoming a powerful tool to map the Universe. It provides a thorough insight into the distribution of matter, including dark matter, and will help us constrain dark energy and modified gravity. Most of the weak lensing analyses use standard perturbation theory to describe correlations on large scales, where the perturbations are expected to be small, and non-linear prescriptions on small scales. However, weak lensing is a gravitational effect which mixes large and small scales by integrating inhomogeneities along the line of sight. Furthermore, future surveys will cover wide parts of the sky and deliver high-precision data both at small non-linear scales, and at large cosmological scales, where relativistic effects become relevant. Hence a careful analysis of weak lensing requires a formalism able to model at the same time relativistic effects and effects at small scales arising from the fully non-linear matter distribution.  

In this paper we computed the convergence $\kappa$, the reduced shear $g$, and the rotation $\omega$ up to order $\Oo{4}$ in the Post-Friedmann formalism. Our results provide a systematic and consistent description of weak lensing observables, including scalar, vector and tensor modes as well as galaxies' peculiar velocities. 

At lowest order in the PF expansion, $\mathcal{O}\left(\frac{1}{c}\right)$, the only non-vanishing observable is the convergence. It is affected by peculiar velocities, which modify the apparent distance between the source and the observer, and consequently the apparent galaxy size. This effect, called Doppler magnification~\cite{Bonvin:2008ni,Bacon:2014uja,Bonvin:2016dze}, dominates at redshift $z\leq 0.5$ and can be used along redshift-space distortions to test modifications of gravity~\cite{Andrianomena:2018aad}. 

At order $\Oo{2}$, we recovered the standard Newtonian expression for the reduced shear at linear order in perturbation theory, expressed in terms of two transverse derivative operators on the sphere~\cite{Castro:2005bg}. The convergence at order $\Oo{2}$ on the other hand contains various contributions. First, there is the standard Newtonian expression, with two transverse derivatives, similar to the one for the shear. Then we found a Sachs-Wolfe contribution and a Shapiro time delay contribution, which modify the length of the geodesics, and consequently the apparent size of galaxies. These terms agree with the expression at linear order in perturbation theory~\cite{Bonvin:2008ni,2010PhRvD..81h3002B}. Finally, the convergence contains also redshift perturbations, which modify the apparent distance between the source and the observer. In the PF formalism, the square of the velocity contributes at order $\Oo{2}$. Hence, contrary to standard PT, the convergence at this order is sensitive to the transverse part of the peculiar velocity $v_\perp$, through the transverse Doppler effect. 

At order $\Oo{3}$, we obtained a contribution from the vector potential $B^i_N$. This effect contributes both to the reduced shear and to the convergence. A quick comparison between the vector contribution and the scalar contribution based on results from numerical simulations~\cite{Bruni:2013mua,Thomas:2015kua} seems to indicate that the contribution from the vector potential is just below the precision on shear measurements required to measure cosmological parameters with 1\% precision. A full computation of the impact of this effect on the shear correlation function is therefore necessary to assess its importance for future surveys. 

Finally at order $\Oo{4}$, we found a host of couplings contributing to gravitational lensing observables. First, the lens-lens coupling and the post-Born corrections affect the reduced shear, the convergence, and also the rotation. These terms contain four transverse derivatives of the gravitational potentials and dominate therefore at small scales. Then, both the reduced shear and the convergence are affected by various relativistic couplings. Our expression for the relativistic couplings in the reduced shear agrees with the result obtained at second-order in PT~\cite{2010PhRvD..81h3002B}. The convergence at second-order in PT has been computed in~\cite{Umeh:2014ana}. A full comparison of the results is beyond the scope of this paper, due to the complexity of the expressions. 
In addition, due to the difference between perturbation theory and the PF framework, the reduced shear contains Doppler contributions at order $(v/c)^2$, and the convergence at order $(v/c)^4$ that are not present at second-order in PT. 

We then used our expressions for the reduced shear and the convergence to identify violations to the Kaiser-Squires relation. In the case of thin redshift bins, the dominant contribution which violates this relation is the Doppler magnification at order $\mathcal{O}\left(\frac{1}{c}\right)$~\cite{Bonvin:2008ni}. With thick redshift bins however, this contribution (along with all contributions at the source) becomes negligible and we are left with a contribution from the Shapiro time delay, which will dominate at large scales, and a contribution from the gravito-magnetic frame-dragging vector field, which we expect to dominate at small scales. In a future work we will investigate how this could be used to detect the contribution from the vector potential in weak gravitational lensing analyses. 

To conclude, let us mention that our framework does not rely on General Relativity and that it is valid for any metric theory of gravity where light propagates on null geodesics. As such the relativistic and non-linear effects calculated in this work can in principle be used to test the theory of gravity.

\section*{Acknowledgments}
This paper is based upon work from COST action CA15117 (CANTATA), supported by COST (European Cooperation in Science and Technology). HG thanks the Faculty of Technology of the University of Portsmouth for support during her PhD studies. CB thanks Ruth Durrer for useful discussions and acknowledges funding by the Swiss National
Science Foundation. MB and DB are supported by the UK STFC Grant No. ST/N000668/1.
\appendix
\section{Derivation of $\DD_{ab}^{(4)}$}\label{App:B}
The Jacobi mapping $\DD_{ab}^{(4)}$ is the solution of the following differential equation
\begin{align}
\frac{d^2}{d\chi^2}\DD_{ab}^{(4)}=&-\left(\frac{1}{k^{0}}\frac{dk^0}{d\chi}\right)^{(4)}\delta_{ab}-\left(\frac{1}{k^{0}}\frac{dk^0}{d\chi}\right)^{(2)}\frac{d}{d\chi}\DD_{ab}^{(2)}+\notag\\
&+\left(\frac{1}{\left(k^0\right)^2}\RR_{a}^{\, c}\right)^{(4)}\chi\delta_{cb}+\left(\frac{1}{\left(k^0\right)^2}\RR_{a}^{\, c}\right)^{(2)}\DD_{cb}^{(2)}\label{eq:DDD4}
\end{align}
In this section, we will derive the expressions for each term and subsequently combine them. The first term in eq.~\eqref{eq:DDD4} reads
\begin{align}
-\left(\frac{1}{k^0}\frac{d}{d\chi}k^{0}\right)^{(4)}
=&- 4\dd{}{\chi}\left(U_P-W_P\right)\cc{4}-4W_{P,i}\fkk{i}\cc{4}+\cc{4}\frac{1}{2\bk{0}} \dd{h_{ij}}{\chi}\bk{i}\bk{j}+\notag\\
&-\cc{4}\frac{1}{2(\bk{0})^2} h_{ij,l}\bk{i}\bk{j}\bk{l}       +\cgen{4}{4}\frac{d}{d\chi} \left( U_{N,i}\fkk{i}  \int^{\chi}_{0}d\chi''W_{N}\right) +\notag\\
 &-\cgen{4}{4}  U_{N,i}\fkk{i} W_{N}-\cgen{4}{4}\left(W_{N,0i} \fkk{i} \int^{\chi}_{0}d\chi''W_{N}\right)+\notag\\
&-\cgen{4}{4}\frac{d}{d\chi}\left[  U_{N,i} \int_{0}^{\chi}d\chi'\int^{\chi'}_{0}d\chi''\left( W_{N}^{,i}-W_{N,j}\frac{\bk{i}\bk{j}}{\left(\bk{0}\right)^2}\right)\right]  +   \notag\\
&+\cgen{4}{4}  U_{N,i}  \int_{0}^{\chi}d\chi'\left( W_{N}^{,i}-W_{N,j}\frac{\bk{i}\bk{j}}{\left(\bk{0}\right)^2}\right)  +   \notag\\
&+\cgen{4}{4}\frac{1}{\bk{0}}W_{N,0i}   \int_{0}^{\chi}d\chi'\int^{\chi'}_{0}d\chi''\left( W_{N}^{,i}-W_{N,j}\frac{\bk{i}\bk{j}}{\left(\bk{0}\right)^2}\right) 
.\label{eq:teil1}
\end{align}
%
The second term in \eqref{eq:DDD4} becomes
\begin{align}
-\left(\frac{1}{k^{0}}\frac{dk^0}{d\chi}\right)^{(2)}\frac{d}{d\chi}\DD_{ab}^{(2)}=&-\frac{1}{\bk{0}}\frac{d}{d\chi}k^{0(2)}\left( -k^{0(2)}\delta_{ab}+\int d\chi' \RR_{ab}\chi'  \right)\notag\\
=&\frac{1}{2\bk{0}}\frac{d}{d\chi}\left(k^{0(2)}\right)^2\delta_{ab}-\frac{1}{\bk{0}}\frac{d}{d\chi}k^{0(2)}\int d\chi' \RR_{ab}\chi'\notag\\
=&\frac{1}{2}\frac{d}{d\chi}\left(2U_N-2W_N+2\int\bound W_{N,l}\fkk{l}\right)^{2}\delta_{ab}-\notag\\
& - 4\dchi\left[\left(U_N-W_N\right)\int  \bar{n}^{i}_{a}\bar{n}^{j}_{b}\chi W_{N,ij}d\chi' \right] +4\left(U_N-W_N\right) \bar{n}^{i}_{a}\bar{n}^{j}_{b}\chi W_{N,ij} \notag\\
&- 2\dchi\left(U_N-W_N\right)\left(-V_N+\chi \dchi V_N\right)\delta_{ab}\notag\\
&-2W_{N,l}\fkk{l} \left[\int 2 \bar{n}^{i}_{a}\bar{n}^{j}_{b}\chi W_{N,ij}d\chi'+\left(-V_N+\chi \dchi V_N\right)\delta_{ab}\right].\label{eq:teil2}
\end{align}

The third term in \eqref{eq:DDD4} is given by
\begin{align}
\left(\frac{1}{k^0 k^0}\RR_{ac}\right)^{(4)}\chi\delta^c_b=&\chi\left\{\delta_{ab}\cc{4}\left(2\frac{d^2}{d\chi^2}V_{P}-\frac{d}{d\chi}V_{N}\frac{d}{d\chi}V_{N}+2\dchi\left(U_N-W_N\right)\frac{d V_N}{d\chi}+2W_{N,l}\fkk{l}\frac{d V_N}{d\chi} \right)+\right.\notag\\
&+\bar{n}^{i}_{a}\bar{n}^{j}_{b}\cc{4}\left[  4W_{P,ij}-4 W_{N,ij}^2 +4W_{N,i} W_{N,j}-4W_{N,ij}V_N\right.-\notag\\
&\left.-4W_{N,lj}\fkk{l}\int\bound W_{N,i}d\chi-4W_{N,il}\fkk{l}\int\bound W_{N,j}d\chi\right]+\notag\\
&\left.+\frac{1}{2}\left[\ddchi h_{ij}-\dchi \left(h_{jp, i}+h_{i p,j}\right)\bk{p}+ h_{mp,i j}\frac{\bk{p}\bk{m}}{\left(\bk{0}\right)^2}\right] \right\}\label{eq:teil3}
\end{align}

The last term in \eqref{eq:DDD4} yields
\begin{align}
\left(\frac{1}{k^0 k^0}\RR_{ac}\right)^{(2)}\DD^{(2)c}_{\enspace b}=&\left(\frac{d^2}{d\chi^2}V_N \delta_{ac}+2\bar{n}^{q}_{a}\bar{n}^{p}_{c}W_{N,qp}\right)\left[\chi V_N\delta_{b}^{c}-2\int\bound d\chi'W_N \delta_{b}^{c}  +\right.\notag\\
& -2\int\bound \int^{\chi'}_0 d\chi'\dc'' W_{N,m}\fkk{m}\delta_{b}^{c}+\left.2\int^{\chi}_0 \int^{\chi'}_0  d\chi'd\chi'' \chi''\bar{n}^{q}_{c}\bar{n}^p_b W_{N,qp}\right]\notag\\
=&\delta_{ab}\left[\frac{\chi}{2}\frac{d^2}{d\chi^2} V_N^2-\chi\left(\frac{dV_N}{d\chi}\right)^2  -2\frac{d^2}{d\chi^2}\left( V_N \int\bound d\chi'W_N\right)+\right.\notag\\
&+2\dchi \left(V_N W_N\right)+2W_N\frac{d}{d\chi}V_N -2\frac{d^2}{d\chi^2}\left(V_N\int\bound \int^{\chi'}_0  d\chi'd\chi'' W_{N,m}\fkk{m} \right)  \notag\\
&\left.  +4\frac{d}{d\chi}\left(V_N\int\bound d\chi' W_{N,m}\fkk{m} \right)   -2V_N W_{N,m}\bk{m}\right]+\notag\\
&+\bar{n}^{i}_{a}\bar{n}^j_b \left[2\frac{d^2}{d\chi^2}\left(V_N \int\bound\int^{\chi'}_0 d\chi'd\chi'' \chi'' W_{N,ij}\right)\right.+\notag\\
&-4\frac{d}{d\chi}\left(V_N \int\bound d\chi' \chi' W_{N,ij}\right)   +4V_N  \chi W_{N,ij}  -4W_{N,ij}\int\bound d\chi' W_N\notag\\
& -4W_{N,ij}\int\bound\int^{\chi'}_0 \dc'\dc'' W_{N,m}\fkk{m}  +\notag\\
&\left.+4\bar{n}^{s}_{c}\bar{n}^{rc} W_{N,is}\int\bound\int^{\chi'}_0 d\chi'd\chi'' \chi'' W_{N,rj}\right]\label{eq:teil4}
\end{align}
Next, we expand $\Ss_{ab}\left(x^{i}_{\text{pert}}\right)=\Ss_{ab}(x^{i})+\delta x^{j}\cdot \delta\left(\Ss_{ab}\right)_{j}\big|_{x}$. Considering the definition of the function $\Ss_{ab}$ in \eqref{eq:S}, we see that we obtain two additional terms
\begin{align}
\frac{1}{\left(\bk{0}\right)^2}\left(\RR_{ac}^{(2)}\bar{\DD}^{c}_{\hspace{1mm}b}\right){}_{,j}\delta x^{j}\quad \text{and} \quad -\frac{1}{\bk{0}}\delta_{ab}\left(\frac{d k^{0(2)}}{d\chi}\right){}_{,j}\delta x^{j}\label{eq:twoterms}
\end{align}
which read
\begin{align}
&\frac{1}{\left(\bk{0}\right)^2}\left(\RR_{ac}^{(2)}\bar{\DD}^{c}_{\hspace{1mm}b}\right)_{,j}\delta x^{j}=\notag\\
=&\bar{n}^{i}_{a}\bar{n}^{j}_{b}\cc{4}\left[-\chi 4W_{N,ijm}\int\bound W_{N}d\chi'\fkk{m}  +\chi 4W_{N,ijm} \int\bound \int_{0}^{\chi'}d\chi''d\chi' \left(W_{N}^{,m}-W_{N,l}\frac{\bk{m}\bk{l}}{\left(\bk{0}\right)^2}\right) \right]+\notag\\
&+    \delta_{ab}\cc{4}\left\{-2\frac{d^{2}}{d\chi^{2}}\left(\chi V_{N,m}\int\bound W_{N}d\chi'\fkk{m}\right)+2\dchi\left(2 V_{N,m}\int\bound W_{N}d\chi'\fkk{m} +\chi V_{N,m}\fkk{m} W_{N} \right)+\right.\notag\\
&-2 \fkk{m} V_{N,m}W_{N}+ 2\fkk{m} \chi W_{N} \dchi V_{N,m}+2\frac{d^{2}}{d\chi^{2}}\left[\chi V_{N,m}\int\bound \int_{0}^{\chi'}\left(W_{N}^{,m}-W_{N,l}\frac{\bk{m}\bk{l}}{\left(\bk{0}   \right)^2  }\right)d\chi''d\chi'\right]+\notag\\
&-4\dchi \left[ V_{N,m}\int\bound \int_{0}^{\chi'}\left(W_{N}^{,i}-W_{N,j}\frac{\bk{m}\bk{j}}{\left(\bk{0}   \right)^2 }\right)d\chi''d\chi'+    \chi V_{N,m}\int\bound \left(W_{N}^{,m}-W_{N,l}\frac{\bk{m}\bk{l}}{\left(\bk{0}   \right)^2  }\right)d\chi' \right]+\notag\\
&\left.+4V_{N,m}\int\bound \dc \left(W_{N}^{,m}  -W_{N,l}\frac{\bk{m}\bk{l}}{\left(\bk{0}   \right)^2  }\right)+2\chi V_{N,m} \left(W_{N}^{,m}   -W_{N,l}\frac{\bk{m}\bk{l}}{\left(\bk{0}   \right)^2  }\right)
\right\}\label{eq:dS1}
\end{align}
and
\begin{align}
-\frac{1}{\bk{0}}\left(\frac{d}{d\chi}\bar{\DD}_{ab}\frac{d k^{0(2)}}{d\chi}\right){}_{,m}\delta x^{m}=&\delta_{ab}\cc{4}\left\{ 4\dchi\left( U_{N,m}  \int\bound d\chi'W_{N}\fkk{m}\right)-4U_{N,m}\fkk{m}W_{N}+    \right.\notag\\
& -4\dchi\left(W_{N,m}\int\bound d\chi'W_{N}\fkk{m}\right)+4W_{N,m}\fkk{m}W_{N}+\notag\\
&+4W_{N,nm}\frac{\bk{n}\bk{m}}{\left(\bk{0}\right)^2}\int\bound d\chi'W_{N}  +\notag\\
&-4 \dchi\left[ U_{N,m} \int\bound\int_{0}^{\chi'}d\chi''d\chi'  \left(W_{N}^{,m}-W_{N,j}\frac{\bk{m}\bk{j}}{\left(\bk{0}\right)^2}\right)   \right]+\notag\\
&+4 U_{N,m} \int\bound d\chi' \left(W_{N}^{,m}-W_{N,j}\frac{\bk{m}\bk{j}}{\left(\bk{0}\right)^2}\right)     +\notag\\
&   +4\dchi\left[W_{N,m} \int\bound\int_{0}^{\chi'} d\chi''d\chi'  \left(W_{N}^{,m}-W_{N,j}\frac{\bk{m}\bk{j}}{\left(\bk{0}\right)^2}\right)   \right]+\notag\\
&-4 W_{N,m} \int\bound  d\chi' \left(W_{N}^{,m}-W_{N,j}\frac{\bk{m}\bk{j}}{\left(\bk{0}\right)^2}\right) +\notag\\
&\left.-4W_{N,nm}\fkk{n}   \int\bound\int_{0}^{\chi'}d\chi''d\chi'\left(W_{N}^{,m}-W_{N,j}\frac{\bk{m}\bk{j}}{\left(\bk{0}\right)^2}\right)      \right\}.
\label{eq:dS2}
\end{align}
We substitute \eqref{eq:teil1} - \eqref{eq:teil4} into \eqref{eq:DDD4} and add the terms \eqref{eq:dS1} and \eqref{eq:dS2}. Furthermore, we perform the integrals to obtain an expression for $\DD^{(4)}_{ab}$. For clarity, we split $\DD^{(4)}_{ab}$ into two parts, the first referring to the terms of $\DD^{(4)}_{ab}$ involving the delta function $\delta_{ab}$ and the second part referring to the terms with $\bar{n}^{i}_{a}\bar{n}^{j}_{b}$:
\begin{align}
\DD^{(4)}_{\mathbf{\delta_{ab}}ab}(\chi_S) =&\delta_{ab}\cc{4}\left\{-V_{NS}\int^{\chi_S}_0  \int\bound d\chi d\chi' W_{N,m}\fkk{m}-2 V_N \ichi W_N+ \right.\notag\\
& -2\chi_S V_{NS,m}\ichi  W_{N}\fkk{m}+\chi_S\left( 2V_{PS}+\frac{1}{2} V_{NS}^2\right)+\notag\\
&+2\chi_S V_{NS,m}\intchi\left(W_{N}^{,m}-W_{N,l}\frac{\bk{m}\bk{l}}{\left(\bk{0}\right)^2}\right)+\notag\\
%
&+\ichi\left[4\left( U_{N,i}\fkk{i}\int^{\chi}_{0}d\chi'W_{N}\right) -4W_P    -4\int\bound \dc' W_{P,i}\fkk{i} +\right.\notag\\
& +2W_N^2+2\left(\int\bound d\chi W_{N,l}\fkk{l}\right)^2+4W_N\int\bound d\chi W_{N,l}\fkk{l}+\frac{1}{2\bk{0}} h_{ij}\bk{i}\bk{j}+\notag\\
 & -4  U_{N,i} \int_{0}^{\chi}d\chi'\left(\chi - \chi'\right)\left( W_{N}^{,i}-W_{N,j}\frac{\bk{i}\bk{j}}{\left(\bk{0}\right)^2}\right) + \notag\\
&-4  W_{N,m}\int\bound \dc \left(\chi - \chi'\right)\left(W_{N}^{,m}-W_{N,j}\frac{\bk{m}\bk{j}}{\left(\bk{0}\right)^2}\right)+2\chi V_{N,m}\fkk{m} W_{N}  \notag\\
&\left.  -4    \chi V_{N,m}\int\bound \dc' \left(W_{N}^{,m}-W_{N,l}\frac{\bk{m}\bk{j}}{\left(\bk{0}\right)^2}\right) +4W_{N,m}\int\bound \dc' W_{N}\fkk{m}\right]\notag\\
&+\intchi\left[-2\chi \left(\frac{d V_N}{d\chi}\right)^2 -4 U_{N,i}\fkk{i} W_{N}+\right.\notag\\
%
 &-4W_{N,0i} \fkk{i} \int^{\chi}_{0}d\chi'W_{N}+4  U_{N,i}  \int_{0}^{\chi}d\chi'\left( W_{N}^{,i}-W_{N,j}\frac{\bk{i}\bk{j}}{\left(\bk{0}\right)^2}\right)+   \notag\\
&+4 W_{N,0i}   \int_{0}^{\chi'}\dc'\left(\chi - \chi'\right)\left( W_{N}^{,i}-W_{N,j}\frac{\bk{i}\bk{j}}{\left(\bk{0}\right)^2}\right)+ \notag\\
& -4W_{N,m}\fkk{m}W_{N}+4W_{N,nm}\frac{\bk{n}\bk{m}}{\left(\bk{0}\right)^2}\int\bound W_{N}d\chi'  +\notag\\
%
&+4 W_{N,m} \int\bound\left(W_{N}^{,m}-W_{N,j}\frac{\bk{m}\bk{j}}{\left(\bk{0}\right)^2}\right)d\chi' -\frac{1}{2(\bk{0})^2} h_{ij,l}\bk{i}\bk{j}\bk{l}+\notag\\
&-4W_{N,nm}\fkk{n}   \int\bound d\chi' \left(\chi - \chi'\right) \left(W_{N}^{,m}-W_{N,j}\frac{\bk{m}\bk{j}}{\left(\bk{0}\right)^2}\right)   \notag\\
%
&\left.+2 \fkk{m} V_{N,m}W_{N}+ 2\fkk{m} \chi W_{N} \dchi V_{N,m}+2\chi V_{N,m} \left(W_{N}^{,m}-W_{N,l}\frac{\bk{m}\bk{j}}{\left(\bk{0}\right)^2}\right)\right\}\label{eq:S4d}
\end{align}
and
\begin{align}
\DD^{(4)}_{\mathbf{\bar{n}^{i}_{a}\bar{n}^{j}_{b}}ab}=&\bar{n}^{i}_{a}\bar{n}^{j}_{b}\left[ 2V_N \int\bound\int^{\chi'}_0 d\chi'd\chi'' \chi'' W_{N,ij}+\frac{1}{2}\chi_S h_{ij}+\right.\notag\\
&+\ichi\left[ - 4W_N\int  \chi W_{N,ij}d\chi' -h_{ij}-\frac{\chi}{2}\left(h_{jp, i}+h_{i p,j}\right)\fkk{p}\right]+\notag\\
&+\intchi\left[ -4W_{N,l}\fkk{l} \int  \chi W_{N,ij}d\chi' \right. + 2\chi\left(2W_{P,ij}- W_{N,ij}^2\right)-\notag\\
&-4\chi\fkk{l}W_{N,lj}\int\bound W_{N,i}d\chi-4\chi\fkk{l}W_{N,il}\int\bound W_{N,j}d\chi+\notag\\
& -4W_{N,ij}\int\bound d\chi' W_N -4W_{N,ij}\int\bound\int^{\chi'}_0 W_{N,m}\fkk{m}  +\notag\\
&+4\bar{n}^{s}_{c}\bar{n}^{rc} W_{N,is}\int\bound\int^{\chi'}_0 d\chi'd\chi'' \chi'' W_{N,rj}-\chi 4W_{N,ijm}\int\bound W_{N}d\chi'\fkk{m} +\notag\\
& +\chi 4W_{N,ijm} \int\bound \int_{0}^{\chi'}d\chi''d\chi'\left(W_{N}^{,m}-W_{N,l}\frac{\bk{m}\bk{l}}{\left(\bk{0}\right)^2  }\right) \notag\\
&\left.+\frac{\chi}{2}\left(h_{jp, i}+h_{i p,j}\right)\fkk{p}+\frac{\chi}{2} h_{mp,i j}\frac{\bk{p}\bk{m}}{\left(\bk{0}\right)^2}\right]
\label{eq:S4nn}
\end{align}
with
\begin{align}
2&\left(U_N-W_N+\int\bound d\chi W_{N,l}\bk{l}\right)^2-V_N^2+4V_N W_N+4V_N \int\bound d\chi W_{N,l}\bk{l}=\notag\\
&2W_N^2+V_N^2+2\left(\int\bound d\chi W_{N,l}\bk{l}\right)^2+4W_N \int\bound d\chi W_{N,l}\bk{l}
\end{align}
and
\begin{align}
&\dchi V_N^2-2\dchi \left(U_N -W_N\right)\left(\chi\dchi V_N-V_N\right)-2W_{N,l}\bk{l}\left(\chi\dchi V_N-V_N\right)+\notag\\
&-2V_N \dchi W_N-2V_N W_{N,m}\bk{m}+2\chi\dchi\left(U_N-W_N\right)\frac{d V_N}{d\chi}+2\chi W_{N,l}\bk{l}\frac{d V_N}{d\chi} =0
\end{align}
\section{Redshift perturbations}
\label{app:rp}
In this section we will compute the different orders of $\delta z_S$, $\delta F$, as well as $\delta f$ using 
\begin{align}
\delta z_S=&(1+z_S)\left[\delta F(z_S) -\frac{d}{dz_{S}}\delta F(z_S)\delta z_S
+\frac{1}{2}\frac{d^2}{dz_{S}^2}\delta F(z_S)\delta z_{S}^2
-\frac{1}{3!}\frac{d^3}{dz_{S}^3}\delta F(z_S)\delta z_{S}^3\right]\!+\label{eq:fzs} \\
&+\Oo{5}
\end{align}
and
\begin{equation}
\delta F=\frac{\delta f}{1+\delta f}= \delta f-\delta f^2+\delta f^3-\delta f^4+\dots\, ,
\end{equation}
and 
\begin{align}
    \delta f\equiv \frac{g_{\mu\nu}k^{\mu}u^{\nu}|_{S}}{g_{\mu\nu}k^{\mu}u^{\nu}|_{O}}-1.
\end{align}
$\delta f$ up to order $\Oo{4}$ reads
\begin{align}
\delta f^{(1)}=& -    \frac{1}{c}\vp\label{eq:B5}\\ 
\delta f^{(2)}=&\cc{2}\left(U_{NS}-2\ichi W_{N,0}+\frac{1}{2}v^2_S\right) \\
\delta f^{(3)}=&\cc{3}\left[\int^{\chi_S}_0 d\chi B_{Ni,0}\bk{i}  -     \vp \left(V_{NS}+\frac{1}{2}v^2_S\right)-v^{i}_S\delta_{ij}\int^{\chi_S}_{0}d\chi W^{,i}_{N}\right] \\
\delta f^{(4)}=&\cc{4}\left\{2U_{PS}-4\int^{\chi_S}_{0}d\chi W_{P,0}-\frac{3}{2}U_{NS}^2+2 \left(\int^{\chi_S}_{0}d\chi W_{N,0}\right)-\frac{1}{2}\int^{\chi_S}_{0}d\chi h_{ij,0}\frac{\bk{i}\bk{j}}{\left(\bk{0}\right)^2}+\right.\notag\\
&+v^2_S\left(V_{NS}+\frac{1}{2}U_{NS}+\frac{3}{8}\right)-B_{NSi}v^{i}+v^{i}_S\int^{\chi}_{0}d\chi B_{Nm}^{\hspace{3mm},i}\fkk{m}+\notag\\
&  -\cgen{4}{4}\left( U_{NS,i}\fkk{i}  \int^{\chi}_{0}d\chi W_{N}\right) +\cgen{4}{4}\ichi  U_{N,i}\fkk{i} W_{N}+\notag\\
 &+\cgen{4}{4}\ichi\left(W_{N,0i} \fkk{i} \int^{\chi}_{0}d\chi'W_{N}\right)+\notag\\
&+\cgen{4}{4}  U_{NS,i} \int_{0}^{\chi_S}\int^{\chi}_{0}d\chi d\chi'\left( W_{N}^{,i}-W_{N,j}\frac{\bk{i}\bk{j}}{\left(\bk{0}\right)^2}\right)  +   \notag\\
&-\cgen{4}{4}\ichi \left[  U_{N,i}  \int_{0}^{\chi}d\chi'\left( W_{N}^{,i}-W_{N,j}\frac{\bk{i}\bk{j}}{\left(\bk{0}\right)^2}\right)  \right]+   \notag\\
&\left.-\cgen{4}{4}\ichi\left[W_{N,0i}   \int_{0}^{\chi}\int^{\chi''}_{0}\dc d\chi'\left( W_{N}^{,i}-W_{N,j}\frac{\bk{i}\bk{j}}{\left(\bk{0}\right)^2}\right)  \right]\right\}
\end{align}
For $\delta F$ up to order $\Oo{4}$, we obtain
\begin{align}
    \delta F^{(1)}=&\delta f^{(1)}= -    \frac{1}{c}\vp\\ 
    \delta F^{(2)}=&\delta f^{(2)}-\delta f^{(1)2}\notag\\
        =&\cc{2}\left(U_{NS}-2\ichi W_{N,0}+\frac{1}{2}v^2_S -\vp^2\right)\\ 
    \delta F^{(3)}=&\delta f^{(3)} - 2 \delta f^{(2)}\delta f^{(1)}+\delta f^{(1)3}\\
        =&\cc{3}\left[\int^{\chi_S}_0 d\chi B_{Ni,0}\bk{i}  -     \vp \left(V_{NS}+\frac{1}{2}v^2_S\right)-v^{i}_S\delta_{ij}\int^{\chi_S}_{0}d\chi W^{,i}_{N}\right.\notag\\
        &\left.+2\left(U_{NS}-2\ichi W_{N,0}+\frac{1}{2}v^2_S\right)\vp-\vp^3\right]\\
    \delta F^{(4)}=&\delta f^{(4)}-\delta f^{(2)2}+3\delta f^{(2)}\delta f^{(1)2}-\delta^{(1)4}\\
        =&\cc{4}\left\{2U_{PS}-4\int^{\chi_S}_{0}d\chi W_{P,0}-\frac{3}{2}U_{NS}^2+2 \left(\int^{\chi_S}_{0}d\chi W_{N,0}\right)+\right.\notag\\
        &-\frac{1}{2}\int^{\chi_S}_{0}d\chi h_{ij,0}\frac{\bk{i}\bk{j}}{\left(\bk{0}\right)^2}   +v^2_S\left(V_{NS}+\frac{1}{2}U_{NS}+\frac{3}{8}\right)-B_{NSi}v^{i}+\notag\\
        &  +v^{i}_S\int^{\chi}_{0}d\chi B_{Nm}^{\hspace{3mm},i}\fkk{m}   -\cgen{4}{4}\left( U_{NS,i}\fkk{i}  \ichi W_{N}\right) +\notag\\
         &+\cgen{4}{4}\ichi  U_{N,i}\fkk{i} W_{N}+\cgen{4}{4}\ichi \left(W_{N,0i} \fkk{i} \int^{\chi}_{0}d\chi'W_{N}\right)+\notag\\
        &+\cgen{4}{4}  U_{NS,i} \int^{\chi_S}_0 \int^{\chi'}_{0}\dc d\chi'\left( W_{N}^{,i}-W_{N,j}\frac{\bk{i}\bk{j}}{\left(\bk{0}\right)^2}\right)  +   \notag\\
        &-\cgen{4}{4}\ichi\left[  U_{N,i}  \int_{0}^{\chi}d\chi'\left( W_{N}^{,i}-W_{N,j}\frac{\bk{i}\bk{j}}{\left(\bk{0}\right)^2}\right)  \right]+   \notag\\
        &-\cgen{4}{4}\ichi\left[W_{N,0i}   \int_{0}^{\chi}\int^{\chi'}_{0}\dc' d\chi''\left( W_{N}^{,i}-W_{N,j}\frac{\bk{i}\bk{j}}{\left(\bk{0}\right)^2}\right)  \right]+\notag\\
        &-\left(U_{NS}-2\ichi W_{N,0}+\frac{1}{2}v^2_S \right)^2+\notag\\
        &\left.+3\left(U_{NS}-2\ichi W_{N,0}+\frac{1}{2}v^2_S \right)\vp^2-\vp^4\right\}.
\end{align}
The derivatives w.r.t.\ $z_S$ in \eqref{eq:fzs} read 
\begin{align}
    -\frac{d}{dz_{S}}\delta F(z_S)^{(1)}=& -\frac{d\chi_S}{dz_{S}} \frac{d}{d\chi_S} \delta F(z_S)^{(1)} =\frac{a^2_S}{a'_S} \vp'\frac{1}{c}=-a_S\frac{c}{\HH_S c}\vp'\\
     -\frac{d}{dz_{S}}\delta F(z_S)^{(2)}=&- \frac{d\chi_S}{dz_S}\frac{d}{\chi_S}\delta F(z_S)^{(2)}=\frac{a^2_S}{a'_S}\frac{d}{\chi_S}\delta F(z_S)^{(2)}\notag\\
        =&-\frac{a_Sc}{\HH_S}\cc{2}\left(\frac{d U_{NS}}{d\chi_S} -2W_{NS,0} +\vs v'_S-2\vp \vp'  \right)\\
     -\frac{d}{dz_{S}}\delta F(z_S)^{(3)}=&-\frac{a_Sc}{\HH_S}\cc{3}\left[      B_{NSi,0}\bk{i}  -     \vp' \left(V_{NS}+\frac{1}{2}v^2_S\right)\right.+\notag\\
        &- \vp \left(\frac{d}{d\chi_S}V_{NS}+v_Sv_S'\vp'\right) -v^{i\prime}_S\delta_{ij}\int^{\chi_S}_{0}d\chi W^{,i}_{N}\notag\\
         &-v^{i}_S\delta_{ij} W^{,i}_{NS}+2\left(\frac{d}{d\chi_S}U_{NS}-2 W_{NS,0}+v_S v'_S\right)\vp+\notag\\
         &\left.+2\left(U_{NS}-2\ichi W_{N,0}+\frac{1}{2}v^2_S\right)\vp'-3\vp^2\vp'\right]\\
    \frac{d^2}{dz_S^2}\delta F^{(1)}=&\frac{d\chi_S}{dz_S}\frac{d}{d\chi_S}\left(a_S \frac{c}{c\HH_S}\vp'  \right)=\frac{a^2_Sc}{c\HH_S}\left[ \vp'\left( 1+\frac{\HH'_Sc}{\HH^2_S} \right)-\frac{c}{\HH_S}\vp''  \right]\\
    \frac{d^2}{dz_S^2}\delta F^{(2)}=&-\cc{2}\frac{a^2_Sc}{\HH_S}\left\{\left(1+\frac{\HH'_Sc}{\HH^2_S}  \right)\left[\frac{d U_{NS}}{d\chi_S}-2 W_{NS,0}+v_Sv_S'-2\vp \vp'  \right]+\right.\notag\\
        &\left.-\frac{c}{\HH_S}\left[  \frac{d^2 U_{NS}}{d\chi_S^2}-2 \frac{dW_{NS,0}}{d\chi_S}+v_S^{\prime 2}+v_Sv_S''-2\vp^{\prime2}-2\vp \vp''  \right]  \right\}\\
    \frac{d^3}{dz_S^3}\delta F^{(1)}=&\frac{a^3_Sc}{c\HH_S}\left\{\left(2+\frac{\HH'_Sc}{\HH^2_S}  \right)\left[\vp'\left(1+\frac{\HH'_Sc}{\HH^2_S}   \right)-\frac{c}{\HH_S}\vp'' \right]+\right.\\
    &\left.-\frac{c}{\HH_S}\left[\vp''\left(1+\frac{\HH'_Sc}{\HH^2_S}  \right)+\vp'\left(\frac{\HH''_Sc}{\HH^2_S}+\frac{2\HH_S^{\prime 2}}{\HH^2_S}   \right) -\frac{\vp'''c}{\HH_S}+\frac{\HH'_S \vp''c}{\HH^2_S}  \right]   \right\}\label{eq:B21}
\end{align}
Note that the prime denotes the derivative w.r.t.\ the parameter $\chi$. However, we define $\HH\equiv \dot{a}(\eta)/a(\eta)$ using the derivative w.r.t.\ the conformal time $\eta$. Since $\frac{d}{d\chi}=-\frac{1}{c}\frac{d}{d\eta}$, every $\HH$ comes with a factor $\left(-\frac{1}{c}\right)$. This factor $c$ does not change the order of the expression because it only appears due to the convention we choose for $\HH$.

Using \eqref{eq:B5} - \eqref{eq:B21}, we obtain for $\delta z_S$ up to order $\Oo{4}$ the following:
\begin{align}
    \delta z_S^{(1)}=&(1+z_S)\delta F^{(1)}=- (1+z_S)   \frac{1}{c}\vp \label{eq:deltaz11}\\
    \delta z_S^{(2)}=&(1+z_S)\left(\delta F^{(2)}-\frac{d}{dz_S}\delta F^{(1)}\delta z^{(1)}    \right)\notag\\
        =&(1+z_S)\cc{2}\left[ U_{NS}-2\int^{\chi_S}_0 d\chi W_{N,0}+\frac{1}{2}v^2_S -\vp^2 +\frac{c}{\HH_S }\vp' \vp \right]\label{eq:deltaz22}\\
    \delta z_S^{(3)}=&(1+z_S)\left(\delta F^{(3)} -\frac{d}{dz_S} \delta F^{(1)}\delta z^{(2)} -\frac{d}{dz_S} \delta F^{(2)}\delta z^{(1)} +\frac{1}{2}\frac{d^2}{dz_S^2} \delta F^{(1)}\delta z^{(1)2}   \right)\\
        =&(1+z_S)\cc{3}\left\{   \int^{\chi_S}_0 d\chi B_{Ni,0}\bk{i}  -     \vp \left(V_{NS}+\frac{1}{2}v^2_S\right)-v^{i}_S\delta_{ij}\int^{\chi_S}_{0}d\chi W^{,i}_{N}\right.\notag\\
        &+2\left(U_{NS}-2\int^{\chi_S}_{0} d\chi W_{N,0}+\frac{1}{2}v^2_S\right)\vp-\vp^3 +\notag\\
        &-\frac{c}{\HH_S}\vp'\left[ U_{NS}-2\int^{\chi_S}_{0} d\chi W_{N,0}+\frac{1}{2}v^2_S -\vp^2 +\frac{1}{\HH_S }\vp' \vp \right]+\notag\\
        &+\frac{c}{\HH_S}\left[\frac{d U_{NS}}{d\chi} -2W_{NS,0} +v_S v'_S-2\vp \vp' \right]   \vp+\notag\\
        &\left.+\frac{c}{2\HH_S}\left[ \vp'\left( 1+\frac{\HH'_Sc}{\HH^2_S} \right)-\frac{c}{\HH_S}\vp''  \right]\vp^2\right\}\\
    \delta z_S^{(4)}=&(1+z_S)\left[\delta F^{(4)}- \frac{d}{dz_S} \delta F^{(1)}\delta z^{(3)}-\frac{d}{dz_S} \delta F^{(2)}\delta z^{(2)}-\frac{d}{dz_S} \delta F^{(3)}\delta z^{(1)} +\right.\\
    &\left.+ \frac{1}{2}\frac{d^2}{dz_S^2} \delta F^{(2)}\delta z^{(1)2}+ \frac{d^2}{dz_S^2} \delta F^{(1)}\delta z^{(1)}\delta z^{(2)}-\frac{1}{6}\frac{d^3}{dz_S^3} \delta F^{(1)}\delta z^{(1)3} \right]\notag\\
        =&(1+z_S)\cc{4}\left\{2U_{PS}-4\int^{\chi_S}_{0}d\chi W_{P,0}-\frac{5}{2}U_{NS}^2-2 \left(\int^{\chi_S}_{0}d\chi W_{N,0}\right)^2+\right.\notag\\
       & + 4U_{NS}\int^{\chi_S}_{0} d\chi W_{NS,0}+4 U_{NS,i} \int_{0}^{\chi_S}d\chi\left(\chi_S-\chi\right)\left( W_{N}^{,i}-W_{N,j}\frac{\bk{i}\bk{j}}{\left(\bk{0}\right)^2}\right)  +   \notag\\
       &+3\left(U_{NS}-2\int^{\chi_S}_{0} d\chi W_{N,0}+\frac{1}{2}v^2_S \right)\vp^2-\vp^4+\notag\\
        &-\frac{1}{2}\int^{\chi_S}_{0}d\chi h_{ij,0}\frac{\bk{i}\bk{j}}{\left(\bk{0}\right)^2}+v^2_S\left(V_{NS}-\frac{1}{2}U_{NS}+\frac{1}{8}v^2_S-2\int^{\chi_S}_{0} d\chi W_{N,0}\right)+\notag\\
        & -B_{NSi}v^{i}_S+v^{i}_S\int^{\chi_S}_{0}d\chi B_{Nm}^{\hspace{3mm},i}\fkk{m} -4U_{NS,i}\fkk{i}  \int^{\chi_S}_{0}d\chi''W_{N} +\notag\\
         &+\int^{\chi_S}_{0} d\chi\left[  4U_{N,i}\fkk{i} W_{N}+4\left(W_{N,0i} \fkk{i} \int^{\chi}_{0}d\chi'W_{N}\right)+\right.\notag\\
        &-4  U_{N,i}  \int_{0}^{\chi}d\chi'\left( W_{N}^{,i}-W_{N,j}\frac{\bk{i}\bk{j}}{\left(\bk{0}\right)^2}\right)+   \notag\\
        &\left.-4W_{N,0i}   \int_{0}^{\chi}d\chi'\left(\chi - \chi'\right)\left( W_{N}^{,i}-W_{N,j}\frac{\bk{i}\bk{j}}{\left(\bk{0}\right)^2}\right)\right]+\notag\\
        &-\frac{c}{\HH_S }\vp'\left\{   \int^{\chi_S}_0 d\chi B_{Ni,0}\bk{i}  -     \vp \left(V_{NS}+\frac{1}{2}v^2_S\right)-v^{i}_S\delta_{ij}\int^{\chi_S}_{0}d\chi W^{,i}_{N}\right.\notag\\
        &+2\left(U_{NS}-2\int^{\chi_S}_{0}d\chi W_{N,0}+\frac{1}{2}v^2_S\right)\vp-\vp^3 +\notag\\
        &+\frac{c}{\HH_S }\vp'\left[ U_{NS}-2\int^{\chi_S}_{0}d\chi W_{N,0}+\frac{1}{2}v^2_S -\vp^2 +\frac{1}{\HH_S }\vp' \vp \right]+\notag\\
        &-\frac{c}{\HH_S}\left[\frac{d U_{NS}}{d\chi} -2W_{NS,0} +v_s v'_S-2\vp \vp'  \right]   \vp+\notag\\
        &\left.+\frac{c}{2\HH_S}\left[ \vp'\left( 1+\frac{\HH'_Sc}{\HH^2_S} \right)-\frac{c}{\HH_S}\vp''  \right]\vp^2\right\}+\frac{c}{\HH_S}\left[\frac{d U_{NS}}{d\chi} -2W_{NS,0} +v_s v'_S\right.\notag\\
        &\left.-2\vp \vp' \right]\left[ U_{NS}-2\int^{\chi_S}_{0}d\chi W_{N,0}+\frac{1}{2}v^2_S -\vp^2 -\frac{c}{\HH_S }\vp' \vp \right]+\notag\\
        &+\frac{c}{\HH_S}\left[      B_{NSi,0}\bk{i}  -     \vp' \left(V_{NS}+\frac{1}{2}v^2_S\right)- \vp \left(\frac{d}{d\chi}V_{NS}+v_Sv_S'\vp'\right)\right.+\notag\\
        & -v^{i\prime}_S\delta_{ij}\int^{\chi_S}_{0}d\chi W^{,i}_{N} -v^{i}_S\delta_{ij} W^{,i}_{NS}+2\left(\frac{d}{d\chi}U_{NS}-2 W_{NS,0}+v_S v'_S\right)\vp+\notag\\
         &\left.+2\left(U_{NS}-2\int^{\chi_S}_{0}d\chi W_{N,0}+\frac{1}{2}v^2_S\right)\vp'-3\vp^2\vp'\right]\vp+\notag\\
         &+\frac{1}{2}\frac{\vp^2c}{\HH_S}\left\{\left(1+\frac{\HH'_Sc}{\HH^2_S}  \right)\left[\frac{d U_{NS}}{d\chi_S}-2 W_{NS,0}+\left(v_Sv_S'-2\vp \vp'\right)  \right]+\right.\notag\\
        &\left.-\frac{c}{\HH_S}\left[  \frac{d^2 U_{NS}}{d\chi_S^2}-2 \frac{dW_{NS,0}}{d\chi_S}+\left(v_S^{\prime 2}+v_Sv_S''-2\vp^{\prime2}-2\vp \vp''\right)  \right]  \right\}+\notag\\
        &-\frac{\vp c}{\HH_S}\left[ \vp'\left( 1+\frac{\HH'_S c}{\HH^2_S} \right)-\frac{c}{\HH_S}\vp''  \right]\left[ U_{NS}-2\int^{\chi_S}_0 d\chi W_{N,0}+\frac{1}{2}v^2_S +\right.\notag\\
        &\left.-\vp^2 +\frac{c}{\HH_S }\vp' \vp \right]+\frac{\vp^3}{6}\frac{c}{\HH_S}\left\{\left(2+\frac{\HH'_S c}{\HH^2_S}  \right)\left[\vp'\left(1+\frac{\HH'_S c}{\HH^2_S}   \right)-\frac{c}{\HH_S}\vp'' \right]+\right.\notag\\
    &\left.-\frac{c}{\HH_S}\left[\vp''\left(1+\frac{\HH'_S c}{\HH^2_S}  \right)+\vp'\left(\frac{\HH''_S c}{\HH^2_S}+\frac{2\HH_S^{\prime 2}}{\HH^2_S}   \right) -\frac{\vp''' c}{\HH_S}+\frac{\HH'_S \vp'' c}{\HH^2_S}  \right]   \right\}.\label{eq:deltaz4}
\end{align}

From eq.~\eqref{eq:Dz} we see that to calculate $\tilde D_{ab}$ up to order $\Oo{4}$ we need the first derivative of $\tilde D_{ab}$ up to order $\Oo{3}$, the second derivative up to order $\Oo{2}$ and the third and fourth derivatives for the background $\tilde D_{ab}$ only. 

\begin{align}
\frac{d}{dz_{S}}\tDD_{ab}\left(z_{S}\right)&=\frac{\chi_S}{(1+z_S)^2}\left(\frac{c}{\HH_S \chi_S}-1\right)\delta_{ab}\label{eq:Dprime}\\
&+\frac{1}{(1+z_S)^2}\left[-\left(1+z_S\right) \delta \tDD_{ab}+\frac{c}{\HH_S}\frac{d}{d\chi_{S}}\big((1+z_S)\delta\tDD_{ab}\big)\right]\notag\\
\frac{d^2}{dz_{S}^2}\tilde{\DD}_{ab}(z_S)=& \frac{\chi_S}{(z_S+1)^3}\left(2-\frac{\HH'_S c^2}{\chi_S\HH^3_S} -\frac{3c}{\HH_S \chi_S}\right)\delta_{ab}+\\
&+\frac{1}{\left(z_S+1\right)^3}\left[2(1+z_S)\delta \tDD_{ab}-\frac{d}{d\chi_S}\left((1+z_S)\delta\tDD_{ab}\right)\frac{c}{\HH_S}\left(   3+\frac{\HH'c}{\HH^2}     \right)+\right.\notag\\
&\left.+\frac{c^2}{\HH^2_S}\frac{d^2}{d\chi_S^2}\left((1+z_S)\delta\tDD_{ab}\right)\right]\\
%
\frac{d^3}{dz_{S}^3}\tilde{\bar{\DD}}_{ab}(z_S)=&\delta_{ab}\frac{\chi_S}{\left(z_S+1\right)^4} \left( \frac{11c}{\chi_S  \HH_S}-6-\frac{\HH''_Sc^3}{\chi_S  \HH^4_S}+\frac{3 c^3\HH'^2}{\chi_S  \HH^5_S}+\frac{6 c^2\HH'_S}{\chi_S  \HH_S^3}   \right)\\
\frac{d^4}{dz_{S}^4}\tilde{\bar{\DD}}_{ab}(z_S)=&\delta_{ab}\frac{\chi_S}{\left(z_S+1\right)^5}\left(24 -\frac{\HH_S^{\prime \prime \prime}c^4}{\chi_S  \HH_S^5}+\frac{10c^3 \HH''_S}{\chi_S  \HH^4}-\frac{15c^4 \HH^{\prime 3}_S}{\chi_S  \HH_S^7}-\frac{30 c^3\HH_S^{\prime 2}}{\chi_S  \HH_S^5}+\right.\notag\\
&\left.-\frac{35 c^2 \HH'_S}{\chi_S  \HH_S^3}+\frac{10 c^4\HH'_S \HH''_S}{\chi_S  \HH^6}-\frac{50c}{\chi_S  \HH} \right).\label{eq:Dz43}
\end{align}
The expression for the Jacobi mapping $\tDD_{ab}(z_S)$ in terms of the redshift $z_S$ in \eqref{eq:Dz} reads for the orders $\mathcal{O}\left(\frac{1}{c}\right)$ - $\Oo{4}$
\begin{align}
\tDD_{ab}^{(1)}\left(\chi_{S}\right)=& \tDD_{ab}^{(1)}\left(\bar{z}_{S}\right)=-\frac{d}{d\bar{z}_{S}}\bar{\tDD}_{ab}\left(z_{S}\right)\delta z_{S}^{(1)}\label{eq:tDDz1}\\
 \tDD_{ab}^{(2)}\left(\chi_{S}\right) =&\tDD_{ab}^{(2)}\left(z_{S}\right)-\frac{d}{d\bar{z}_{S}}\bar{\DD}_{ab}\left(z_{S}\right)\delta z_{S}^{(2)}+\frac{1}{2}\frac{d^2}{d\bar{z}_{S}^2}\bar{\tDD}_{ab}\left(z_{S}\right)\delta z_{S}^{(1)2},\\
 \tDD_{ab}^{(3)}\left(\bar{z}_{S}\right)=&\tDD^{(3)}_{ab}\left(z_{S}\right)-\frac{d}{d\bar{z}_{S}}\bar{\tDD}_{ab}\left(z_{S}\right)\delta z_{S}^{(3)}   -\frac{d}{d\bar{z}_{S}}\tDD^{(2)}_{ab}\left(z_{S}\right)\delta z_{S}^{(1)}+\notag\\
&    +\frac{d^2}{d\bar{z}_{S}^2}\bar{\tDD}_{ab}\left(z_{S}\right)\delta z_{S}^{(1)}\delta z_{S}^{(2)}-\frac{1}{6}\frac{d^3}{d\bar{z}_{S}^3}\bar{\tDD}_{ab}\left(z_{S}\right)\delta z_{S}^{(1)3},\enspace \text{and}\\
  \tDD_{ab}^{(4)}\left(\bar{z}_{S}\right)=&\tDD_{ab}^{(4)}\left(z_{S}\right)-\frac{d}{d\bar{z}_{S}}\bar{\tDD}_{ab}\left(z_{S}\right)\delta z_{S}^{(4)}-\frac{d}{d\bar{z}_{S}}\tDD_{ab}^{(2)}\left(z_{S}\right)\delta z_{S}^{(2)}+\notag\\
&-\frac{d}{d\bar{z}_{S}}\tDD_{ab}^{(3)}\left(z_{S}\right)\delta z{S}^{(1)}+\frac{1}{2}\frac{d^2}{d\bar{z}_{S}^2}\tilde{\bar{\DD}}_{ab}\left(z_{S}\right)\left(\delta z_{S}^{(2)}\right)^2+\notag\\
&+\frac{1}{2}\frac{d^2}{d\bar{z}_{S}^2}\tilde{\DD}^{(2)}_{ab}\left(z_{S}\right)\left(\delta z_{S}^{(1)}\right)^2+ \frac{d^2}{d\bar{z}_{S}^2}\bar{\tDD}_{ab}\left(z_{S}\right)\delta z_{S}^{(1)}\delta z^{(3)}+\notag\\
&-\frac{1}{2}\frac{d^3}{d\bar{z}_{S}^3}\bar{\tDD}_{ab}\left(z_{S}\right)\delta z_{S}^{(2)}\left(\delta z_{S}^{(1)}\right)^2 +\frac{1}{4!}\frac{d^4}{d\bar{z}_{S}^4}\bar{\tDD}_{ab}\left(z_{S}\right)\left(\delta z_{S}^{(1)}\right)^4 . \label{eq:Dz4}
\end{align}
Due to the length of the full expression of $\tDD_{ab}(z_S)$ in \eqref{eq:Dz4} we list the different terms in this section of the appendix:
\begin{align}
-\frac{d}{d\bar{z}_{S}}&\bar{\tDD}_{ab}\left(z_{S}\right)\delta z_{S}^{(4)}=
\frac{\chi_S}{(1+z_S)}\left(1-\frac{c}{\HH_S \chi_S}\right)\delta_{ab}\cc{4}\left\{2U_{PS}-4\int^{\chi_S}_{0}d\chi W_{P,0}-\frac{5}{2}U_{NS}^2+\right.\notag\\
       & -2 \left(\int^{\chi_S}_{0}d\chi W_{N,0}\right)^2+ 4U_{NS}\int^{\chi_S}_{0} d\chi W_{NS,0}+\notag\\
       &+4 U_{NS,i} \int_{0}^{\chi_S}d\chi\left(\chi_S-\chi\right)\left( W_{N}^{,i}-W_{N,j}\frac{\bk{i}\bk{j}}{\left(\bk{0}\right)^2}\right)  +   \notag\\
       &+3\left(U_{NS}-2\int^{\chi_S}_{0} d\chi W_{N,0}+\frac{1}{2}v^2_S \right)\vp^2-\vp^4+\notag\\
        &-\frac{1}{2}\int^{\chi_S}_{0}d\chi h_{ij,0}\frac{\bk{i}\bk{j}}{\left(\bk{0}\right)^2}+v^2_S\left(V_{NS}-\frac{1}{2}U_{NS}+\frac{1}{8}v^2_S-2\int^{\chi_S}_{0} d\chi W_{N,0}\right)+\notag\\
        & -B_{NSi}v^{i}_S+v^{i}_S\int^{\chi_S}_{0}d\chi B_{Nm}^{\hspace{3mm},i}\fkk{m} -4U_{NS,i}\fkk{i}  \int^{\chi_S}_{0}d\chi''W_{N} +\notag\\
         &+\int^{\chi_S}_{0} d\chi\left[  4U_{N,i}\fkk{i} W_{N}+4\left(W_{N,0i} \fkk{i} \int^{\chi}_{0}d\chi'W_{N}\right)+\right.\notag\\
        &-4  U_{N,i}  \int_{0}^{\chi}d\chi'\left( W_{N}^{,i}-W_{N,j}\frac{\bk{i}\bk{j}}{\left(\bk{0}\right)^2}\right)+   \notag\\
        &\left.-4W_{N,0i}   \int_{0}^{\chi}d\chi'\int^{\chi'}_{0}d\chi''\left( W_{N}^{,i}-W_{N,j}\frac{\bk{i}\bk{j}}{\left(\bk{0}\right)^2}\right)\right]+\notag\\
        &-\frac{c}{\HH_S }\vp'\left\{   \int^{\chi_S}_0 d\chi B_{Ni,0}\bk{i}  -     \vp \left(V_{NS}+\frac{1}{2}v^2_S\right)-v^{i}_S\delta_{ij}\int^{\chi_S}_{0}d\chi W^{,i}_{N}\right.\notag\\
        &+2\left(U_{NS}-2\int^{\chi_S}_{0}d\chi W_{N,0}+\frac{1}{2}v^2_S\right)\vp-\vp^3 +\notag\\
        &+\frac{c}{\HH_S }\vp'\left[ U_{NS}-2\int^{\chi_S}_{0}d\chi W_{N,0}+\frac{1}{2}v^2_S -\vp^2 +\frac{1}{\HH_S }\vp' \vp \right]+\notag\\
        &-\frac{c}{\HH_S}\left[\frac{d U_{NS}}{d\chi} -2W_{NS,0} +v_S v'_S-2\vp \vp'  \right]   \vp+\notag\\
        &\left.+\frac{c}{2\HH_S}\left[ \vp'\left( 1+\frac{\HH'_Sc}{\HH^2_S} \right)-\frac{c}{\HH_S}\vp''  \right]\vp^2\right\}+\frac{c}{\HH_S}\left[\frac{d U_{NS}}{d\chi} -2W_{NS,0} +v_S v'_S\right.\notag\\
        &\left.-2\vp \vp' \right]\left[ U_{NS}-2\int^{\chi_S}_{0}d\chi W_{N,0}+\frac{1}{2}v^2_S -\vp^2 -\frac{c}{\HH_S }\vp' \vp \right]+\notag\\
        &+\frac{c}{\HH_S}\left[      B_{NSi,0}\bk{i}  -     \vp' \left(V_{NS}+\frac{1}{2}v^2_S\right)\right.+\notag\\
        &- \vp \left(\frac{d}{d\chi}V_{NS}+v_Sv_S'\right) -v^{i\prime}_S\delta_{ij}\int^{\chi_S}_{0}d\chi W^{,i}_{N}\notag\\
         &-v^{i}_S\delta_{ij} W^{,i}_{NS}+2\left(\frac{d}{d\chi}U_{NS}-2 W_{NS,0}+v_S v'_S\right)\vp+\notag\\
         &\left.+2\left(U_{NS}-2\int^{\chi_S}_{0}d\chi W_{N,0}+\frac{1}{2}v^2_S\right)\vp'-3\vp^2\vp'\right]\vp+\notag\\
         &+\frac{1}{2}\frac{\vp^2c}{\HH_S}\left\{\left(1+\frac{\HH'_Sc}{\HH^2_S}  \right)\left[\frac{d U_{NS}}{d\chi_S}-2 W_{NS,0}+\left(v_Sv_S'-2\vp \vp'\right)  \right]+\right.\notag\\
        &\left.-\frac{c}{\HH_S}\left[  \frac{d^2 U_{NS}}{d\chi_S^2}-2 \frac{dW_{NS,0}}{d\chi_S}+\left(v_S^{\prime 2}+v_Sv_S''-2\vp^{\prime2}-2\vp \vp''\right)  \right]  \right\}+\notag\\
        &-\frac{\vp c}{\HH_S}\left[ \vp'\left( 1+\frac{\HH'_S c}{\HH^2_S} \right)-\frac{c}{\HH_S}\vp''  \right]\left[ U_{NS}-2\int^{\chi_S}_0 d\chi W_{N,0}+\frac{1}{2}v^2_S -\vp^2 +\frac{c}{\HH_S }\vp' \vp \right]+\notag\\
        &+\frac{\vp^3}{6}\frac{c}{\HH_S}\left\{\left(2+\frac{\HH'_S c}{\HH^2_S}  \right)\left[\vp'\left(1+\frac{\HH'_S c}{\HH^2_S}   \right)-\frac{c}{\HH_S}\vp'' \right]+\right.\notag\\
    &\left.-\frac{c}{\HH_S}\left[\vp''\left(1+\frac{\HH'_S c}{\HH^2_S}  \right)+\vp'\left(\frac{\HH''_S c}{\HH^2_S}+\frac{2\HH_S^{\prime 2}}{\HH^2_S}   \right) -\frac{\vp''' c}{\HH_S}+\frac{\HH'_S \vp'' c}{\HH^2_S}  \right]   \right\},\label{eq:DDz4a}\\
    -\frac{d}{d\bar{z}_{S}}&\tDD_{ab}^{(2)}\left(z_{S}\right)\delta \tilde{z}_{S}^{(2)}=\delta_{ab}\frac{\chi_S}{1+z_S}\left[ \left(V_{NS} -  2\frac{1}{\chi_S}\int^{\chi_{S}}_{0}d\chi \Big[W_{N}+(\chi_S-\chi) W_{N,i}\fkk{i}\Big]\right)\right.+\notag\\
%
            &-\left.\frac{c }{\chi_S\HH_S} \left(V_{NS}+\chi_{S}\frac{dV_{NS}}{d\chi_S}- 2 \Big[W_{NS}+\int^{\chi_{S}}_{0}d\chi  W_{N,i}\fkk{i}\Big]\right)\right]\left[ U_{NS}-2\int^{\chi_S}_0 d\chi W_{N,0}+\right.\notag\\
            &\left.+\frac{1}{2}v^2_S -\vp^2 +\frac{c}{\HH_S }\vp' \vp \right]+\notag\\
            %
    &-\bar{n}^{i}_{a}\bar{n}^{j}_{b}\frac{\chi_S}{1+z_S}\left[\frac{1}{\chi_S}\left(2\int_{0}^{\chi_{S}}d\chi\left(\chi_{S}-\chi\right)\chi W_{N,ij} \right)\right.\notag\\
    &\left.-\frac{c}{\chi_S\HH_S}2\int_{0}^{\chi_{S}}d\chi\chi W_{N,ij}\,\right]\left[ U_{NS}-2\int^{\chi_S}_0 d\chi W_{N,0}+\frac{1}{2}v^2_S -\vp^2 +\frac{c}{\HH_S }\vp' \vp \right],\\
 -\frac{d}{d\bar{z}_{S}}&\tDD_{ab}^{(3)}\left(z_{S}\right)\delta \tilde{z}_{S}^{(1)}=\delta_{ab}\frac{\chi_S}{1+z_S}\left[-\frac{\vp}{\chi_S} \int_{0}^{\chi_{S}}d\chi\left(\chi_{S}-\chi\right) B_{Ni,j}\frac{\bk{j} \bk{i}}{\left(\bk{0}\right)^2}\right.\notag\\
    &\left.+\frac{c\vp}{\chi_S\HH_S}\int_{0}^{\chi_{S}}d\chi B_{Ni,j}\frac{\bk{j} \bk{i}}{\left(\bk{0}\right)^2}\right]+\notag\\
    &+\bar{n}^{i}_{a}\bar{n}^{j}_{b}\frac{\chi_S}{1+z_S}\left[-\frac{\vp}{\chi_S} \int_{0}^{\chi_{S}}d\chi \left(\chi_{S}-\chi\right)\chi\left(\frac{dB^{N}_{(i,j)}}{d\chi}-\frac{\bk{m}}{\bk{0}}B_{m,ij}^N\right)\right.\notag\\
    &\left.+\frac{c\vp}{\chi_S\HH_S}\int_{0}^{\chi_{S}}d\chi \chi\left(\frac{dB^{N}_{(i,j)}}{d\chi}-\frac{\bk{m}}{\bk{0}}B_{m,ij}^N\right)\right],\\
 \frac{1}{2}\frac{d^2}{d\bar{z}_{S}^2}&\tilde{\bar{\DD}}_{ab}\left(z_{S}\right)\left(\delta \tilde{z}_{S}^{(2)}\right)^2=\delta_{ab}\frac{\chi_S}{z_S+1}\left(1-\frac{\HH'_S c^2}{2\chi_S\HH^3_S}-\frac{3c}{2\HH_S \chi_S}\right)\left[ U_{NS}+\right.\notag\\
    &\left.-2\int^{\chi_S}_0 d\chi W_{N,0}+\frac{1}{2}v^2_S -\vp^2 +\frac{c}{\HH_S }\vp' \vp \right]^2,\\
\frac{1}{2}\frac{d^2}{d\bar{z}_{S}^2}&\tilde{\DD}^{(2)}_{ab}\left(z_{S}\right)\left(\delta \tilde{z}_{S}^{(1)}\right)^2=\delta_{ab}\frac{\chi_S}{z_S+1}\left[\V_{NS} - 2 \frac{1}{\chi_S} \int^{\chi_{S}}_{0}d\chi \Big[W_{NS}+(\chi_S-\chi) W_{N,i}\fkk{i}\Big]+\right.\notag\\
    &-\left(V_{NS}+\chi_S\frac{d}{d\chi_S}V_{NS}  - 2  \Big[W_{NS}+\int^{\chi_{S}}_{0}d\chi W_{NS,i}\fkk{i}\Big]   \right)\frac{c}{2\HH_S \chi_S}\left(   3+\frac{\HH'c}{\HH^2}     \right)+\notag\\
    &\left.+\frac{c^2}{2\HH^2_S \chi_S}\left(2\frac{d}{d\chi_S}V_{NS}+\chi_S\frac{d^2}{d\chi_S^2}V_{NS}  - 2  \Big[\frac{d}{d\chi_S}W_{NS}+ W_{NS,i}\fkk{i}\Big]   \right)\right]\vp^2\notag\\
    &+\bar{n}^{i}_a \bar{n}^{j}_b\frac{\chi_S}{z_S+1}\left[2\frac{1}{\chi_S}\int_{0}^{\chi_{S}}d\chi\left(\chi_{S}-\chi\right)\chi W_{N,ij}-\int_{0}^{\chi_{S}}d\chi\chi W_{N,ij}\frac{c}{\HH_S\chi_S}\left(   3+\frac{\HH'c}{2\HH^2 \chi_S}     \right)+\right.\notag\\
    &\left.+\frac{c^2}{\HH^2_S}\chi_S W_{NS,ij}\right]\vp^2,\\
\frac{d^2}{d\bar{z}_{S}^2}&\bar{\tDD}_{ab}\left(z_{S}\right)\delta z_{S}^{(1)}\delta z^{(3)}=-\frac{\chi_S}{z_S+1}\delta_{ab}\vp\left(2-\frac{\HH'_S c^2}{\chi_S\HH^3_S} -\frac{3c}{\HH_S \chi_S}\right)\left\{   \int^{\chi_S}_0 d\chi B_{Ni,0}\bk{i}+\right.\notag\\
    &\left.-     \vp \left(V_{NS}+\frac{1}{2}v^2_S\right)-v^{i}_S\delta_{ij}\int^{\chi_S}_{0}d\chi W^{,i}_{N}\right.\notag\\
        &+2\left(U_{NS}-2\int^{\chi_S}_{0} d\chi W_{N,0}+\frac{1}{2}v^2_S\right)\vp-\vp^3 +\notag\\
        &-\frac{c}{\HH_S}\vp'\left[ U_{NS}-2\int^{\chi_S}_{0} d\chi W_{N,0}+\frac{1}{2}v^2_S -\vp^2 +\frac{1}{\HH_S }\vp' \vp \right]+\notag\\
        &+\frac{c}{\HH_S}\left[\frac{d U_{NS}}{d\chi} -2W_{NS,0} +v_S v'_S-2\vp \vp' \right]   \vp+\notag\\
        &\left.+\frac{c}{2\HH_S}\left[ \vp'\left( 1+\frac{\HH'_Sc}{\HH^2_S} \right)-\frac{c}{\HH_S}\vp''  \right]\vp^2\right\},\\
-\frac{1}{2}\frac{d^3}{d\bar{z}_{S}^3}&\bar{\tDD}_{ab}\left(z_{S}\right)\delta z_{S}^{(2)}\left(\delta z_{S}^{(1)}\right)^2=\delta_{ab}\frac{\chi_S}{z_S+1} \left(-\frac{11c}{2\chi_S  \HH_S}+3+\frac{\HH''_S c^3}{2\chi_S  \HH^4_S}-\frac{3 \HH'^2 c^3}{2\chi_S  \HH^5_S}+\right.\notag\\
    &\left.-\frac{3 \HH'_S c^2}{\chi_S  \HH_S^3} \right)\vp^2\left[ U_{NS}-2\int^{\chi_S}_0 d\chi W_{N,0}+\frac{1}{2}v^2_S -\vp^2 +\frac{c}{\HH_S }\vp' \vp \right],\\
\frac{1}{4!}\frac{d^4}{d\bar{z}_{S}^4}&\bar{\tDD}_{ab}\left(z_{S}\right)\left(\delta z_{S}^{(1)}\right)^4 =\frac{1}{24}\delta_{ab}\frac{\chi_S}{z_S+1}\left(24 -\frac{\HH_S^{\prime \prime \prime}c^4}{\chi_S  \HH_S^5}+\frac{10c^3 \HH''_S}{\chi_S  \HH^4}-\frac{15c^4 \HH^{\prime 3}_S}{\chi_S  \HH_S^7}+\right.\notag\\
    &\left.-\frac{30 c^3\HH_S^{\prime 2}}{\chi_S  \HH_S^5}-\frac{35 c^2 \HH'_S}{\chi_S  \HH_S^3}+\frac{10 c^4\HH'_S \HH''_S}{\chi_S  \HH^6}-\frac{50c}{\chi_S  \HH} \right)\vp^4.\label{eq:DDz4b}
\end{align}
\section{The spin-0 and spin-2 fields ${}_{0}\DD$ and ${}_{2}\DD$:}
\label{app:Dspin}
Previously we noted that the Jacobi mapping does not depend on the normalisation of $\bk{\mu}$, since it only depends on the ratio $\fkk{i}$. Hence, for the following sections we choose to normalise $\bk{\mu}$ such that $\bk{0}=\bk{i}\bk{j}\delta_{ij}=1$.

We begin with the spin-2 field ${}_{2}\DD$. Using \eqref{eq:2Dee}, ${}_{2}\DD$ reads up to order $\Oo{4}$ the following\\

\begin{align}
{}_{2}\tDD^{(2)}\left(z_{S}\right)=&\frac{1}{z_S+1} 2\cc{2}\int_{0}^{\chi_{S}}d\chi\frac{\chi_{S}-\chi}{\chi} \slpa^2 W_{N}\label{eq:2tDD2s},\\
{}_2 \tDD^{(3)}\left(z_{S}\right)=&-\frac{1}{z_S+1} \cc{3}\left\{\int_{0}^{\chi_{S}} d\chi'\frac{\chi_{S}-\chi}{\chi} \left[ \frac{d}{d\chi} \left(\chi \slpa _{1}B\right)+\slpa^2 B_{r}\right]+ \right.\notag\\
&-2\vp\int^{\chi_S}_0 d\chi \frac{\chi_S-\chi}{\chi}\slpa^2 W_N+ \notag\\
&\left.+ \frac{2c}{\HH_S }\vp \int^{\chi_S}_0 d\chi \frac{1}{\chi}\slpa^2 W_N    \right\}        \label{eq:2tDD3s},\enspace\text{and}\\
 {}_{2}\tDD^{(4)}\left(z_{S}\right)=&\frac{1}{z_S+1} \frac{1}{c^4}\left\{\frac{1}{2}\chi_S {}_{2}h\left(\chi_S\right)+\right.\ichi\left[ - 4W_N\int\bound  \frac{1}{\chi'}\slpa^2 W_{N}d\chi'+\right.\notag\\
 &\left.-4\frac{1}{\chi}\slpa^2\left(W_{N}\int\bound  W_{N}d\chi'\right)\right]+\intchi\left[ 2\frac{1}{\chi}\slpa^2 \left(2W_{P}+ W_{N}^2\right)  \right. +\notag\\
&+ 2V_{NS} \frac{1}{\chi}\slpa^2 W_{N}+2 \frac{1}{\chi^2}\slpa\left(\slpa^2 W_{N}\int\bound d\chi'  \frac{\chi-\chi'}{\chi'}\bar{\slpa} W_{N}\right)+\notag\\
&+2 \frac{1}{\chi^2}\slpa \left(\slpa\bar{\slpa}W_{N}\int\bound d\chi'  \frac{\chi-\chi'}{\chi'}\slpa W_{N}\right)+\notag\\
& + \frac{4}{\chi^2}\slpa W_N\int\bound  d\chi' \frac{\chi-\chi'}{\chi'}\slpa W_N -4\frac{1}{\chi^2} W_{N}\int\bound \slpa^2 W_{N}d\chi'+\notag\\
& +4\frac{1}{\chi}\slpa^2\left(W_{N,0}\int\bound  W_{N}d\chi'\right)+\frac{1}{2\chi}\slpa^2 h_{rr}+\frac{\chi_S}{\chi}\slpa{}_{1}h_r+\notag\\
%
    &+\left[\frac{c}{\HH_S}2\int_{0}^{\chi_{S}}d\chi\frac{1}{\chi} \slpa^2 W_{N}-2\int_{0}^{\chi_{S}}d\chi \frac{\chi_{S}-\chi}{\chi}\slpa^2 W_{N} \,\right]\Big( U_{NS}+\notag\\
    &-2\int^{\chi_S}_0 d\chi W_{N,0}+\frac{1}{2}v^2_S -\vp^2 +\frac{c}{\HH_S }\vp'\vp \Big)+\notag\\
    &-\vp \int_{0}^{\chi_{S}}d\chi \frac{\left(\chi_{S}-\chi\right)}{\chi}\left[\frac{d}{d\chi}\left(\chi\slpa{}_{1}B^{N}\right)+\slpa^2 B_{r}^N\right]+\notag\\
    &-\frac{c\vp}{\HH_S}\int_{0}^{\chi_{S}}d\chi \frac{1}{\chi}\left[\frac{d}{d\chi}\left(\chi\slpa{}_{1}B^{N}\right)+\slpa^2 B_{r}^N\right]+\notag\\
    &-\vp^2\int_{0}^{\chi_{S}}d\chi\frac{1}{\chi}\slpa^2  W_{N}\frac{c}{\HH_S}\left(   3+\frac{\HH'c}{2\HH^2}     \right)+\notag\\
    &\left.+2\vp^2\int_{0}^{\chi_{S}}d\chi\frac{\chi_{S}-\chi}{\chi} \slpa^2 W_{N}+\frac{\vp^2 c^2}{\HH^2_S}\frac{1}{\chi_S}\slpa^2 W_{NS}\right\}\label{eq:DZ44}
\end{align}

with $W_{N,r}=W_{N,i}\bk{i}$ and using \eqref{eq:C22}-\eqref{eq:C24} for
\begin{align}
\eplus{i}\eplus{j}4\bar{n}^{s}_{c}\bar{n}^{rc} & W_{N,is}\int\bound\int^{\chi'}_0 d\chi'd\chi'' \chi'' W_{N,rj}=\notag\\
=&\eplus{i}\eplus{j}4\frac{1}{2}\left(\eplus{s}\eminus{r}+\eminus{s}\eplus{r}\right) W_{N,is}\int\bound\int^{\chi'}_0 d\chi'd\chi'' \chi'' W_{N,rj}\notag\\
=&2 \frac{1}{\chi^2}\slpa^2 W_{N}\int\bound \int_{0}^{\chi'} d\chi'd\chi''  \left(\frac{1}{\chi''}\bar{\slpa}\slpa W_{N}+2W_{N,r}   \right)+\notag\\
&+2 \left(\frac{1}{\chi^2}\bar{\slpa}\slpa W_{N}+\frac{2}{\chi}W_{N,r}   \right)\int\bound \int_{0}^{\chi'} d\chi'd\chi''  \frac{1}{\chi''}\slpa^2 W_{N}\notag\\
\end{align}
and 
\begin{align}
-\eplus{i}\eplus{j}\chi 4W_{N,ijm}\int\bound W_{N}d\chi'\bk{m} =& \left(- \frac{4}{\chi}\slpa^2 W_{N,r}+\frac{8}{\chi^2}\slpa^2 W_N\right)\int\bound W_{N}d\chi' 
\end{align}
and
\begin{align}
\eplus{i}\eplus{j}&\chi 4W_{N,ijm} \int\bound \int_{0}^{\chi'}\left(W_{N}^{,m}-W_{N,l}\bk{m}\bk{l}\right)d\chi''d\chi'=\notag\\
=&\eplus{i}\eplus{j}\chi 4W_{N,ijm} \frac{1}{2}\left( \eplus{m}\eminus{n}+\eminus{m}\eplus{n}  \right)\int\bound \int_{0}^{\chi'}W_{N,n}d\chi''d\chi'\notag\\
=&+\frac{2}{\chi^2}\slpa^3 W_N \int\bound \int^{\chi'}_0 d\chi' d\chi'' \frac{1}{\chi''}\bar{\slpa}W_N+\notag\\
&+\left(\frac{2}{\chi^2}\slpa\bar{\slpa}\slpa W_N +\frac{8}{\chi}\slpa W_{N,r} - \frac{4}{\chi^2}\slpa W_N\right)\int\bound \int^{\chi'}_0 d\chi' d\chi'' \frac{1}{\chi''}\slpa W_N.
\end{align}
The last six lines in \eqref{eq:DZ44} are contributions from the redshift perturbations. 

The spin-0 field ${}_{0}\tDD$ is obtained by $\eminus{a}\eplus{b}\tDD_{ab}$. The real and imaginary part of ${}_{0}\tDD$ is related to the convergence and rotation, respectively. The combination of $\eplus{a}$ and $\eminus{b}$ neither lowers nor raises the spin $s$ and thus results in a spin-0 expression. We rearrange $\eminus{a}\eplus{b}\tDD_{ab}$ to 
\begin{align}
\eminus{a}\eplus{b}\tDD_{ab}=&\frac{1}{2}\left(\eplus{a}\eminus{b}+\eminus{a}\eplus{b}\right)\tDD_{ij}+\frac{1}{2}\left(\eminus{a}\eplus{b}-\eplus{a}\eminus{b}\right)\tDD_{ab}\\
=&\Re\left({}_{0}\tDD\right)+i\Im\left({}_{0}\tDD\right)\equiv {}_{0}\tDD_{\rm R}+{}_{0}\tDD_{\rm I}\label{eq:epluseminus}
\end{align}
and split the expression into its real and imaginary part. First, we compute the real part ${}_{0}\tDD_{\rm R}$ and obtain for the background and up to order $\Oo{3}$:
\begin{align}
{}_{0}\bar{\tDD}_{\rm R}\left(z_{S}\right)=&\frac{1}{z_S+1}2\chi_S\\
{}_{0}\tDD^{(1)}_{\rm R}\left(z_{S}\right)=&\frac{1}{z_S+1}\frac{1}{c} 2\chi_S\left(\frac{c}{\HH_S \chi_S}-1\right)     \vp\label{eq:D01}\\
{}_{0}\tDD^{(2)}_{\rm R}\left(z_{S}\right)=&\frac{1}{z_S+1}\cc{2}\left[\chi_{S}2V_{NS} - 2 \int^{\chi_{S}}_{0}d\chi \left(2W_{N}-(\chi_S-\chi) \frac{1}{\chi}\bar{\slpa}\slpa W_{N}\right)\right.+\notag\\
&+\left(1-\frac{c}{\HH_S \chi_S}\right)2\chi_S\left(U_{NS}-2\ichi W_{N,0}+\frac{1}{2}v^2_S\right)+\notag\\
&\left.+\left(1-\frac{3c}{2\HH_S\chi_S}-\frac{c^2\HH'_S}{2\HH_S^3\chi_S}    \right)2\chi_S  v_{Sk}^2\right]\label{eq:tDD2s}\enspace\text{and}\\
{}_{0} \tDD^{(3)}_{\rm R}\left(\bar{z}_{S}\right)=&\frac{1}{z_S+1} \cc{3}2\left\{-\frac{1}{4}\int^{\chi_S}_0 d\chi\left(\bar{\slpa}{}_{1}B_N+\slpa{}_{-1}B_N-4B_{Nr}\right)+\right.\notag\\
&-\int_{0}^{\chi_{S}}d\chi\frac{\chi_{S}-\chi}{2\chi} \left(\frac{1}{2}\bar{\slpa}{}_{1}B_N+\frac{1}{2}\slpa{}_{-1}B_N+\bar{\slpa}\slpa B_{Nr}  \right)+\notag\\    
 & -\left(\frac{c}{\HH_S }-\chi_S\right)   \int^{\chi_S}_0 d\chi B_{Nr,0}  \notag\\
	&+\vp\Bigg[\left(2\chi_S-\frac{\HH'_S c^2}{\HH^3_S} -\frac{3c}{\HH_S }\right)\left( U_{NS}-2\int^{\chi_S}_0 d\chi W_{N,0} \right)+\notag\\
 	 &-\int_{0}^{\chi_{S}}d\chi\left(\chi_{S}-\chi -\frac{c}{\HH_s}\right)  \frac{1}{\chi} \bar{\slpa}\slpa W_N  +\notag\\
&   -2\chi_SV_{NS}   +2\chi_SU_{NS}  - \chi_S W_{NS} +\int^{\chi_{S}}_{0}d\chi (\chi_S-\chi) W_{N,0}       +\notag\\
        &  -3\chi_S\int^{\chi_S}_{0} d\chi W_{N,0}     +\frac{c \chi_S}{\HH}\left(\frac{d U_{NS}}{d\chi} -2W_{NS,0}  \right) +\notag\\
&+\frac{c}{\HH_S }     \Bigg[ 3V_{NS}   -2U_{NS}  - W_{NS}  +\chi_{S}\frac{dV_{NS}}{d\chi_S} +\notag\\
        &  -\int^{\chi_{S}}_{0}d\chi \left(2W_{N}-(\chi_S-\chi) W_{N,0} \right)     +3\int^{\chi_S}_{0} d\chi W_{N,0} +\notag\\
&    -\frac{c}{\HH}\left(\frac{d U_{NS}}{d\chi} -2W_{NS,0}  \right) \Bigg]\Bigg]+\notag\\
        &+\vp'\left(\frac{c}{\HH_S }-\chi_S\right)  \frac{c}{\HH_S}\left( U_{NS}-2\int^{\chi_S}_{0} d\chi W_{N,0} \right)+\notag\\
        &    -\left(\frac{c}{\HH_S }-\chi_S\right)  \left(\frac{1}{2}\vpl \ichi \frac{1}{\chi}\bar{\slpa} W_N    +\frac{1}{2}\vm \ichi \frac{1}{\chi}\slpa W_N\right) +\notag\\
%
	&+\left(3\chi_S-\frac{\HH'_S c^2}{\HH^3_S} -\frac{4c}{\HH_S }\right)\vp\frac{1}{2}\vs^2   +\notag\\
&+\left(\frac{3}{2} \chi_S    -\frac{3}{2}\frac{\HH'_S c^2}{\HH^3_S} -\frac{4c}{\HH_S }  + \frac{3}{2} \frac{c}{\HH_S }      +\frac{1}{2}\frac{\HH'_S c}{\HH^2_S}  \right) \frac{c}{\HH_S}\vp^2 \vp'  +\notag\\
%
	&+ \vp^3 \left(-2\chi_S   +\frac{8c}{\HH_S }     +\frac{\HH''_S c^3}{6  \HH^4_S}-\frac{ \HH'^2 c^3}{2  \HH^5_S}    \right) +\notag\\
        %
     &+\left( \frac{c}{\HH_S}-\chi_S\right) \frac{c}{\HH} \vp'\left(\frac{1}{2}\vs^2  +\frac{1}{\HH_S }\vp' \vp \right)+\notag\\
        &-\left(\frac{c}{\HH_S }-\chi_S\right)  \frac{c}{\HH} \vp\vs \vs'  +\left(\frac{c}{\HH_S }-\chi_S\right)  \frac{c^2}{2\HH_S^2}\vp'' \vp^2   \Bigg\}
.\label{eq:tDD3s}
\end{align}
As at order $\Oo{4}$ the expression for ${}_{0}\tDD^{(4)}_{\rm R}$ is very long, we split it into nine parts: 
\begin{align}
{}_{0}\tDD^{(4)}_{\rm R}=&\tDm{P}+\tDm{VW}+\tDm{U W}+\tDm{W W}+\tDm{h}+\notag\\
&+\tDm{\delta z} +\tDm{v}+\tDm{v^2}+\tDm{v^4}   \label{eq:tDDsplit}
\end{align}
where the superscripts $(U W)$, $(V W)$, and $(W W)$ refer to the respective couplings and the superscripts $(P)$ to terms involving the quantities $U_P$, $V_P$, and $W_P$. The superscript $(h)$ denotes terms with the tensor potential $h_{ij}$. The contributions of the redshift perturbations are split into two categories, which read $\tDm{\delta z}$ and $\tDm{v}$. The superscripts $(v)$, $(v^2)$, and $(v^4)$ refer to terms with the peculiar velocity $v_S$ or its projection along the line of sight $\vp$, while the superscript $(\delta z)$ denotes all terms stemming from redshift perturbations independent of the peculiar velocity $v_S$.

We begin with $\tDm{P}$. Note that due to the form of the metric \eqref{eq:g00} - \eqref{eq:gij}, the contributions of the potentials $V_P$ and $W_P$ in $\tDm{4}$ will be of the same form as the potentials $V_N$ and $W_N$ in $\tDm{2}$ in \eqref{eq:tDD2s} without redshift perturbations. The terms with $U_P$ and $W_P$, which are derived from the redshift perturbations in \eqref{eq:Dz44}, will appear in the part $\tDm{\delta z}$.
\begin{align}
{}_{0}\tDD_{\mathbf{P}}=&\frac{1}{z_S+1} \cc{4}\left[\chi_S4V_{P}-\ichi\left(8W_P-4\frac{\chi_S-\chi}{\chi}\bar{\slpa}\slpa W_{P}\right)\right]
\label{eq:tDDP0}.
\end{align}
\begin{align}
    \tDm{UW}=&\frac{1}{z_S+1}\cc{4}\left\{
\int^{\chi_S}_{0} d\chi 8\left[U_{NS} W_{N}  -U_N W_N  -U_{N,0}   \int^{\chi}_{0}d\chi'W_{N}  +\right.\right.\notag\\
&\left.   +\left(\chi_S -\chi\right)\left(U_{N,0} W_{N}-W_{N}\dchi U_N\right)\right]+\\
    &-4 \int^{\chi_S}_{0} d\chi\frac{1}{\chi} \slpa U_{N} \int_{0}^{\chi}d\chi'\left(\chi-\chi'\right)\frac{1}{\chi}\bar{\slpa}W_{N}  +\notag\\
&-4 \int^{\chi_S}_{0} d\chi  \frac{1}{\chi}\bar{\slpa} U_{N} \int_{0}^{\chi}d\chi'\left(\chi-\chi'\right) \frac{1}{\chi} \slpa W_{N} +   \notag\\
    &+4  \intchi \frac{1}{\chi} \slpa U_{N} \int_{0}^{\chi}d\chi' \frac{1}{\chi}\bar{\slpa}W_{N}+\notag\\
&\left.+4\intchi \frac{1}{\chi}\bar{\slpa} U_{N} \int_{0}^{\chi}d\chi' \frac{1}{\chi} \slpa W_{N}\right\}.
\end{align}
For $\tDm{VV}$ and $\tDm{VW}$, we obtain
\begin{align}
    \tDm{VV}=\cc{4}\left[\frac{\chi_S}{2}V^2_{NS}-4\intchi \chi'\left(\dchi V_N\right)^2\right]
\end{align}
and
\begin{align}
\tDm{VW}=&\frac{1}{z_S+1} 2\cc{4}\left\{\ichi\left[-2 V_{NS} W_N -2\chi W_{N}\dchi V_{NS}   +2\chi V_{NS,0}W_{N} + 2\chi  W_{N}\dchi V_{N} +\right.\right.\notag\\
&\left.  -2 W_{N}\chi V_{N,0} -2\slpa V_{N}\int^{\chi}_0 d\chi' \frac{1}{\chi'}\bar{\slpa} W_{N}-2\bar{\slpa} V_{N}\int^{\chi}_0 d\chi' \frac{1}{\chi'}\slpa W_{N} +  V_{NS} W_{N}\right] \notag\\
&    +\intchi \left[  2W_{N} \dchi V_{N}   -2W_{N}V_{N,0}+ 2 \chi W_{N}  \left( \ddchi V_{N}   -\dchi V_{N,0}\right)\right.\notag\\
&+\slpa V_{N} \frac{1}{\chi}\bar{\slpa} W_{N} +\bar{\slpa} V_{N} \frac{1}{\chi}\slpa W_{N}+\slpa V_{NS} \frac{1}{\chi}\bar{\slpa} W_{N}+\bar{\slpa} V_{NS} \frac{1}{\chi}\slpa W_{N}+\notag\\
&\left.\left.+  V_{NS} \frac{1}{\chi}\bar{\slpa}\slpa    W_{N}     -V_{NS}W_{N,0}   \right]\right\}\, .
\end{align}
respectively. The next two terms contain the couplings of the lensing potentials $W_N-W_N$ as well as $h_{ij}$:
\begin{align}
\tDm{WW}=&\frac{1}{z_S+1} \cc{4}\left\{\ichi\left[ -8 W_N^2  + 8W_{NS}W_N  -16 W_{N,0}  \int^{\chi}_{0}d\chi'W_{N}\right.\right.+\notag\\
&+  \frac{24}{\chi}  \slpa W_N   \int_{0}^{\chi}d\chi' \frac{\chi-\chi'}{\chi'}\bar{\slpa} W_N +   \frac{24}{\chi}  \bar{\slpa} W_N     \int_{0}^{\chi}d\chi' \frac{\chi-\chi'}{\chi'}\slpa W_N   +\\
& -   \frac{4}{\chi} W_N  \int\bound d\chi' \frac{\chi - \chi'}{\chi'}\bar{\slpa}\slpa W_{N}         +4\slpa \bar{\slpa} W_N  \int\bound d\chi' W_N\frac{\chi - \chi'}{\chi \chi'} +\notag  \\
&\left.    -  4 W_N  \int\bound   \frac{1}{ \chi'}\bar{\slpa}\slpa  W_{N}d\chi' -  4  \slpa \bar{\slpa}\left(W_N \int\bound \frac{1}{\chi'}W_{N}d\chi \right)  \right]\\
&  +  \intchi \left[  16W_{N} W_{N,0}      +8W_{N,00} \int\bound W_{N}d\chi' +\right. \notag\\
& -\frac{24}{\chi}  \slpa W_N   \int_{0}^{\chi}d\chi' \frac{1}{\chi'}\bar{\slpa} W_N  -\frac{24}{\chi}  \bar{\slpa} W_N     \int_{0}^{\chi}d\chi' \frac{1}{\chi'}\slpa W_N+\notag\\
& -  \frac{16}{\chi} \slpa   W_{N,0} \int_{0}^{\chi}d\chi' \frac{\chi-\chi'}{\chi'}\bar{\slpa} W_N -  \frac{16}{\chi} \bar{\slpa}   W_{N,0} \int_{0}^{\chi}d\chi' \frac{\chi-\chi'}{\chi'}\slpa W_N +\notag\\
&+\frac{4}{\chi^2}  \slpa W_N   \int_{0}^{\chi}d\chi' \frac{\chi-\chi'}{\chi'}\bar{\slpa} W_N        +\frac{4}{\chi^2}  \bar{\slpa} W_N     \int_{0}^{\chi}d\chi' \frac{\chi-\chi'}{\chi'}\slpa W_N+\notag\\
&+4 \bar{\slpa}\slpa \left( W_{N,0}   \int\bound   \frac{1}{ \chi'}  W_{N}d\chi'\right) -4 \bar{\slpa}\slpa  W_{N,0}   \int\bound   \frac{1}{ \chi'}  W_{N}d\chi'  +\notag\\
&   +\frac{1}{\chi}\slpa W_{N} \int\bound \frac{1}{\chi'}\bar{\slpa}W_{N}d\chi   +4\frac{1}{\chi}\bar{\slpa} W_{N}\int\bound \frac{1}{\chi}\slpa W_{N}d\chi   +\notag\\
& -4 \frac{1}{ \chi^2}\bar{\slpa}\slpa  W_{N}\int\bound d\chi' W_N  +8 \frac{1}{ \chi^2}\bar{\slpa}\slpa  W_{N}\int\bound d\chi' \left(\chi-\chi'\right) W_{N,0} \notag\\
&-4\left(\dchi W_N\right)^2+8 W_{N,0}\dchi W_N-4 W_{N,0}^2    +2   \bar{\slpa}\slpa W_N^2  \frac{1}{\chi}  \notag\\
&+\frac{1}{\chi^2}\slpa^2 W_{N}\int\bound d\chi'\frac{\chi - \chi'}{ \chi'} \bar{\slpa}^2 W_{N}+ \frac{1}{\chi^2}\bar{\slpa}^2 W_{N}\int\bound d\chi'\frac{\chi - \chi'}{ \chi'}  \slpa^2 W_{N} +\notag\\
&+   2\frac{1}{\chi^2}\bar{\slpa}\slpa W_{N} \int\bound d\chi'\frac{\chi - \chi'}{\chi'}\bar{\slpa}\slpa W_{N}       - 4\frac{1}{\chi^2}\bar{\slpa}\slpa W_{N}  \int\bound d\chi' W_N      +\notag\\
& -\frac{4}{\chi^2} W_N  \int\bound d\chi' \frac{\chi - \chi'}{\chi'}\bar{\slpa}\slpa W_{N} +\frac{4}{\chi} W_N  \int \bound d\chi' \frac{1}{\chi}\bar{\slpa}\slpa W_{N}+\notag\\
&  +\frac{4}{\chi} W_{N,0}\int\bound d\chi' \frac{\chi - \chi'}{\chi'}\bar{\slpa}\slpa W_{N} +\frac{4}{\chi}\slpa \bar{\slpa}W_{N,0}\int\bound d\chi' W_N + \notag  \\
&+ \frac{2}{\chi^2}\slpa \bar{\slpa}\slpa W_N       \int\bound \frac{\chi-\chi'}{\chi'}\bar{\slpa} W_{N}d\chi' + \label{eq:0D4WW}\\
&\left.\left.+  \frac{2}{\chi^2}\slpa \bar{\slpa}^2 W_N      \int\bound\frac{\chi-\chi'}{\chi'}\slpa W_{N}d\chi'\notag\right]\right\}
\end{align}
and 
\begin{align}
\tDm{h}=&\frac{1}{z_S +1} \cc{4}\left\{-\frac{\chi_S}{2} h_{rr}\left(\chi_S\right)+\frac{1}{2}\ichi \frac{\chi_S}{\chi}\left(\slpa {}_{-1}h_{r}+\bar{\slpa} {}_{1}h_{r} -h_{rr}\right)\right.+\notag\\
&+\intchi \frac{1}{2\chi}\left(\slpa \bar{\slpa} h_{rr}-\chi h_{rr,0}\right).\label{eq:D4h}
\end{align}
The contributions to ${}_{0}\tDD^{(4)}$ from the redshift perturbations are divided into $\tDm{\delta z}$, $\tDm{v}$, $\tDm{v^2}$, and $\tDm{v^4}$, where $\tDm{\delta z}$ denotes the perturbations independent of the peculiar velocity and $\tDm{v}$, $\tDm{v^2}$, and $\tDm{v^4}$ refer to the terms involving the peculiar velocity at different powers:
\begin{align}
    \tDm{\delta z}=&
\frac{2\chi_S}{(1+z_S)}\cc{4}\Bigg\{
 \left[ -  \frac{1}{\chi_S}\int^{\chi_{S}}_{0}d\chi \left( 2W_{N}-\left(\chi_{S}-\chi-\frac{c}{\HH_s}\right)  \frac{1}{\chi}\bar{\slpa}\slpa W_N \right) \right.+\notag\\
%
            &\left.+\frac{c }{\chi_S\HH_S} \left(-\chi_{S}\frac{dV_{NS}}{d\chi_S}+ 2 W_{NS}\right)\right]\left( U_{NS}-2\int^{\chi_S}_0 d\chi W_{N,0} \right)+\notag\\
         +&   \left(1-\frac{\HH'_S c^2}{2\chi_S\HH^3_S}-\frac{3c}{2\HH_S \chi_S}\right)\left( U_{NS}-2\int^{\chi_S}_0 d\chi W_{N,0}\right)^2+\notag\\
&+\left(1-\frac{c}{\HH_S \chi_S}\right)\left\{2U_{PS}-4\int^{\chi_S}_{0}d\chi W_{P,0}-\frac{5}{2}U_{NS}^2   -\frac{1}{2}\int^{\chi_S}_{0}d\chi h_{rr,0}+\right.\Bigg.\notag\\
       & -2 \left(\int^{\chi_S}_{0}d\chi W_{N,0}\right)^2+ 4\left(U_{NS}\int^{\chi_S}_{0} d\chi W_{NS}\right)_{,0}+ 4W_{NS,0}\ichi W_N+\notag\\
       &+2 \frac{1}{\chi_S}\bar{\slpa} U_{NS} \int_{0}^{\chi_S}d\chi\left(\chi_S-\chi\right)\frac{1}{\chi}\slpa W_N    +2 \frac{1}{\chi_S}\slpa U_{NS} \int_{0}^{\chi_S}d\chi\left(\chi_S-\chi\right)\frac{1}{\chi}\bar{\slpa} W_N   +   \notag\\
		&-4\frac{d}{\chi_S} U_{NS}  \int^{\chi_S}_{0}d\chi''W_{N} +V_{NS}\left(U_{NS}-2\ichi W_{N,0}\right)+\notag\\
&+\frac{c}{\HH_S}\left(  \frac{d U_{NS}}{d\chi_S}U_{NS}-2\frac{d U_{NS}}{d\chi_S}\int^{\chi_S}_{0}d\chi W_{N,0} -2W_{NS,0} U_{NS}    +2W_{NS,0}2\int^{\chi_S}_{0}d\chi W_{N,0} \right)\notag\\  
	   &+\int^{\chi_S}_{0} d\chi\left(  4 W_N \dchi U_{N}   - 4U_{N,0} W_{N} -4   W_{N,0} W_{N} - 4  W_{N,00}\int^{\chi}_{0}d\chi'W_{N} +\right.\notag\\
        %
        &-2 \frac{1}{\chi}\slpa U_{N}  \int_{0}^{\chi}d\chi' \frac{1}{\chi'}\bar{\slpa}W_N    -2   \frac{1}{\chi}\bar{\slpa} U_{N}  \int_{0}^{\chi}d\chi'\frac{1}{\chi'}\slpa W_N  +   \notag\\
	&- 2\frac{1}{\chi}\slpa W_{N,0}\int_{0}^{\chi}d\chi\frac{\chi - \chi'}{\chi}\bar{\slpa} W_N   -2\frac{1}{\chi^2}\slpa W_N   \int_{0}^{\chi}d\chi\frac{\chi - \chi'}{\chi}\bar{\slpa} W_N  +\notag\\
	&\left. \left.   -2\frac{1}{\chi}\bar{\slpa} W_{N,0} \int_{0}^{\chi}d\chi'\frac{\chi - \chi'}{\chi}\slpa W_N         -2\frac{1}{\chi^2}\bar{\slpa} W_N    \int_{0}^{\chi}d\chi'\frac{\chi - \chi'}{\chi}\slpa W_N     \right)\right\}\Bigg\},
\end{align}
\begin{align}
    \tDm{v}=&\frac{2\chi_S}{(1+z_S)}\cc{4}\Bigg\{
\frac{\vp}{\chi_S}\left[ \int_{0}^{\chi_{S}}d\chi \frac{\chi_{S}-\chi-\frac{c}{\HH_S}}{\chi}\frac{1}{2}\left(\frac{d}{d\chi}\left(\chi \slpa {}_{1}B_N\right)+\slpa^2B_{Nr}-2\chi B_{Nr,0}\right)\right.+\notag\\
&\left.+\frac{c}{\HH_S}B_{NSr} - \int^{\chi_S}_0 d\chi  \left( B_{Nr} +\left(2\chi_S-\frac{\HH'_S c^2}{\HH^3_S} -\frac{3c}{\HH_S }\right)B_{Nr,0}\right) \right]+\notag\\
&+\left(1-\frac{c}{\HH_S \chi_S}\right)\left\{
	 -\vpl\left[      \frac{1}{2}{}_{-1}B_{NS}   +\frac{1}{2}\int^{\chi_S}_{0}d\chi \frac{1}{\chi}\left(\bar{\slpa} B_{Nr} +{}_{-1}B_{N} \right)      + \right]\right.+\notag\\
	&-\vm\left[    \frac{1}{2}{}_{1}B_{NS}   +\frac{1}{2}\int^{\chi_S}_{0}d\chi  \frac{1}{\chi}  \left(\slpa B_{Nr}   +{}_{1}B_{N} \right)\right]  +\notag\\
	&\left.+\vp \vp'  \frac{c}{\HH_S}\left[  V_{NS}      -\int^{\chi_S}_{0}d\chi  W_{N,0}    -\left( 3+\frac{\HH'_S c}{\HH^2_S} \right) \left( U_{NS}-2\int^{\chi_S}_0 d\chi W_{N,0}\right) \right]\right\}\Bigg\},
\end{align}
\begin{align}
    \tDm{v^2}=&\frac{2\chi_S}{(1+z_S)}\cc{4}\Bigg\{\left(1-\frac{c}{\HH_S \chi_S}\right)\left\{\vs^2\left( \frac{3}{2}V_{NS} -\frac{1}{2}U_{NS}\right.\right. +\notag\\
    &\left. -2\int^{\chi_S}_{0} d\chi W_{N,0}+\frac{1}{2}\frac{c}{\HH_S}\frac{d U_{NS}}{d\chi}   -\frac{c}{\HH_S}W_{NS,0} \right)+\vp^2\Bigg[3U_{NS} -V_{NS}     +\notag\\
    &-6\int^{\chi_S}_{0} d\chi W_{N,0}-3\frac{c}{\HH_S}\frac{d W_{NS}}{d\chi}  +3\frac{c}{\HH_S}W_{NS,0}+\notag\\
    &+\left(5\frac{c}{\HH_S}+\frac{\HH'_Sc^2}{\HH^3_S} - \frac{1}{2}\frac{c^2}{\HH_S^2} \frac{d}{d\chi_S}   \right)\left(\frac{1}{2}\frac{d U_{NS}}{d\chi_S}-W_{NS,0} \right)  \Bigg]+\notag\\
	&+\vp \vp'  \frac{c}{\HH_S}\left[  V_{NS}      -\int^{\chi_S}_{0}d\chi  W_{N,0}    -\left( 3+\frac{\HH'_S c}{\HH^2_S} \right) \left( U_{NS}-2\int^{\chi_S}_0 d\chi W_{N,0}\right) \right]\notag\\
	&+\vp^{\prime 2}\left( 2\frac{c^2}{\HH_S^2 }\int^{\chi_S}_{0}d\chi W_{N,0}  -\frac{c^2}{\HH_S^2 } U_{NS} \right)+ \vs \vs'\frac{c}{\HH_S}\left( U_{NS}-2\int^{\chi_S}_{0}d\chi W_{N,0}\right)   +\notag\\
%
%
&    -\frac{c}{\HH_S }\vp'\vm\frac{1}{2}\int^{\chi_S}_{0}d\chi' \frac{1}{\chi'}\slpa W_{N}    -\frac{c}{\HH_S }\vp'\vpl\frac{1}{2}\int^{\chi_S}_{0}d\chi' \frac{1}{\chi'}\bar{\slpa} W_{N}  +  \notag\\
        %
	&  +\vp\frac{c}{\HH_S}\vm'\frac{1}{2}\int^{\chi_S}_{0}d\chi' \frac{1}{\chi'}\slpa W_{N}   +\vp\frac{c}{\HH_S}\vpl'\frac{1}{2}\int^{\chi_S}_{0}d\chi' \frac{1}{\chi'}\bar{\slpa} W_{N}  +  \notag\\
	& +\vp\vm\frac{1}{2} \frac{1}{\chi_S}\slpa W_{NS}    +\vp\vpl \frac{c}{2\HH_S\chi_S}\bar{\slpa} W_{NS}  +\notag\\
	&\left. +\frac{\vp c^2}{\HH_S^2}\vp'' \left( U_{NS}-2\int^{\chi_S}_0 d\chi W_{N,0}\right) \right\}+\notag\\
+&\left[ -  \frac{1}{\chi_S}\int^{\chi_{S}}_{0}d\chi \left( 2W_{N}-\left(\chi_{S}-\chi-\frac{c}{\HH_s}\right)  \frac{1}{\chi}\bar{\slpa}\slpa W_N \right) \right.+\notag\\
%
            &\left.+\frac{c }{\chi_S\HH_S} \left(-\chi_{S}\frac{dV_{NS}}{d\chi_S}+ 2 W_{NS}\right)\right]\left( \frac{1}{2}\vs^2 -\vp^2 +\frac{c}{\HH_S }\vp' \vp \right)+\notag\\
%
&+\left(2-\frac{\HH'_S c^2}{\chi_S\HH^3_S}-\frac{3c}{\HH_S \chi_S}\right)\left( U_{NS}-2\int^{\chi_S}_0 d\chi W_{N,0}\right)\left(\frac{1}{2}\vs^2  +\frac{c}{\HH_S }\vp' \vp \right)+\notag\\
&-\left(2-\frac{\HH'_S c^2}{\chi_S\HH^3_S} -\frac{3c}{\HH_S \chi_S}\right)\frac{c}{\HH_S}\left[-\vp' \vp U_{NS}    +2\vp' \vp\int^{\chi_S}_{0} d\chi W_{N,0}  \right]+\notag\\
%
&+\left\{4V_{NS} - 2\frac{1}{\chi_S} \int^{\chi_{S}}_{0}d\chi \Big[W_{NS}-\left(\chi_{S}-\chi\right)  \frac{1}{2\chi}\bar{\slpa}\slpa W_N \Big]+\right.\notag\\
    &-\left(V_{NS}+\chi_S\frac{d}{d\chi_S}V_{NS}  +4  W_{NS}   +\int^{\chi_{S}}_{0}d\chi  \frac{1}{\chi}\bar{\slpa}\slpa W_N  \right)\frac{c}{2\HH_S \chi_S}\left(   3+\frac{\HH'c}{\HH^2}     \right)+\notag\\
%
%
&+\left(\frac{13c}{2\chi_S  \HH_S}    +\frac{\HH''_S c^3}{2\chi_S  \HH^4_S}-\frac{3 \HH'^2 c^3}{2\chi_S  \HH^5_S}-2    +\frac{\HH'_S c^2}{\chi_S\HH^3_S}  \right) U_{NS}+\notag\\
&+\left(4  -\frac{\HH''_S c^3}{\chi_S  \HH^4_S}+\frac{3 \HH'^2 c^3}{\chi_S  \HH^5_S}+\frac{ \HH'_S c^2}{\chi_S  \HH_S^3}      -\frac{4c}{\HH_S \chi_S}  \right)  \int^{\chi_S}_{0} d\chi W_{N,0}  +\notag\\
    &\left.+\frac{c^2}{\HH^2_S \chi_S}\left(\frac{d}{d\chi_S}V_{NS}+\frac{1}{2}\chi_S\frac{d^2}{d\chi_S^2}V_{NS}  -  \frac{d}{d\chi_S}W_{NS}   -\frac{1}{2}\bar{\slpa}\slpa W_{NS}\right)\right\}\vp^2+\notag\\
&  -\left(2-\frac{\HH'_S c^2}{\chi_S\HH^3_S} -\frac{3c}{\HH_S \chi_S}\right)\vp  \left( \vm\frac{1}{2}\int^{\chi_S}_{0}d\chi' \frac{1}{\chi'}\slpa W_{N}+\right. \notag\\
&\left.+ \vpl\frac{1}{2}\int^{\chi_S}_{0}d\chi' \frac{1}{\chi'}\bar{\slpa} W_{N}\right)  \Bigg\},
\end{align}
and 
\begin{align}
    \tDm{v^4}=&\frac{2\chi_S}{(1+z_S)}\cc{4}\Bigg\{\left(1-\frac{c}{\HH_S \chi_S}\right)\Big\{
      \frac{3}{2}\vp^2\vs^2-\vp^4     +\frac{1}{8}\vs^4     +\frac{1}{2}\vp\frac{c}{\HH_S} \vp\vs \vs'   \notag\\
&    +\frac{\HH'_Sc}{\HH^2_S} \frac{1}{2}\frac{\vp^2c}{\HH_S}\vs \vs'  +\left(2+3\frac{\HH'_S c}{\HH^2_S} -\frac{\HH^{\prime 2}_S c^2}{\HH^4_S}-\frac{\HH''_S c^2}{\HH^4_S}   \right)\frac{\vp^3}{6}\frac{c}{\HH_S}\vp'     -\frac{1}{2}\frac{\vp^2c^2}{\HH_S^2}\vs^{\prime 2}     +\notag\\
 	&    -\frac{\vp c}{\HH_S}\vp'\left(3+ \frac{\HH'_S c}{\HH^2_S} \right)  \frac{1}{2}\vs^2    +   \frac{c}{\HH_S}\vs \vs'\frac{1}{2}\vs^2    -\frac{1}{2}\frac{\vp^2c^2}{\HH_S^2}\vs \vs''    +    \notag\\
	 &    -\frac{c^2}{\HH_S^2 }\vp^{\prime 2}\frac{1}{2}\vs^2  -\frac{c^2}{\HH_S^2 }\vp^{\prime 3}\frac{1}{\HH_S }\vp    -3\left(1+\frac{\HH'_S c}{\HH^2_S}  \right)\frac{\vp^3}{6}\frac{c^2}{\HH_S^2}\vp''  +\notag\\
        &   +\frac{\vp^3}{6}\frac{c^2}{\HH_S^2}\frac{\vp''' c}{\HH_S}      +\frac{\vp^2 c}{\HH_S}\vp^{\prime 2}\left( \frac{1}{2}-\frac{3}{2}\frac{\HH'_S c}{\HH^2_S} \right) \frac{c}{\HH_S }  +\notag\\
	&   +\vp\frac{c^2}{\HH_S^2}\vp'' \frac{1}{2}\vs^2   +\frac{3c^2}{2\HH_S^2}\frac{c}{\HH_S}\vp^2\vp'\vp''+ +\vp^2\frac{c}{\HH_S}\vp^{\prime 2} \int^{\chi_S}_{0}d\chi W_{N,0}  \Big\}+\notag\\
%
    &  +\left(     1-\frac{\HH'_S c^2}{2\chi_S\HH^3_S}-\frac{3c}{2\HH_S \chi_S}\right)\left[    \frac{1}{4}\vs^4       +\frac{c^2}{\HH_S^2 }\vp^{\prime 2} \vp^2          +\frac{c}{\HH_S }\vp' \vp \vs^2   \right]+\notag\\
& -\left(    -\frac{11c}{2\chi_S  \HH_S}+3+\frac{\HH''_S c^3}{2\chi_S  \HH^4_S}-\frac{3 \HH'^2 c^3}{2\chi_S  \HH^5_S}-\frac{3 \HH'_S c^2}{\chi_S  \HH_S^3}         \right)\frac{1}{2}\vs^2 \vp^2+\notag\\
        &-\left(2-\frac{\HH'_S c^2}{\chi_S\HH^3_S} -\frac{3c}{\HH_S \chi_S}\right)\frac{c}{\HH_S}\left[    -\vp' \vp\frac{1}{2}\vs^2 -\vp^{\prime 2} \vp^2\frac{1}{\HH_S }    \right.+\notag\\
        &  + \vp^2\frac{d U_{NS}}{d\chi} - \vp^22W_{NS,0} + \vp^2 \vs \vs'   + \left.  -\vp^3\frac{c}{\HH_S}\vp''  \right]+\notag\\
%
&+\left(-\frac{5c}{2\chi_S  \HH_S}+1+\frac{\HH''_S c^3}{2\chi_S  \HH^4_S}     -\frac{ \HH'^2 c^3}{2\chi_S  \HH^5_S}    +\frac{ \HH'_S c^2}{\chi_S  \HH_S^3}     -2\frac{\HH'_Sc}{\HH^2_S}              \right)\frac{c}{\HH_S }\vp' \vp^3 \notag\\
& +\left(1+\frac{\HH'_S c^2}{24\chi_S\HH^3_S}    -\frac{13c}{12\chi_S  \HH_S}     -\frac{c^3 \HH''_S}{12\chi_S  \HH^4}       +\frac{1 \HH'^2 c^3}{4\chi_S  \HH^5_S}       -\frac{\HH_S^{\prime \prime \prime}c^4}{24\chi_S  \HH_S^5}         +\right.\notag\\
&\left.-\frac{15c^4 \HH^{\prime 3}_S}{24\chi_S  \HH_S^7}       +\frac{5 c^4\HH'_S \HH''_S}{12\chi_S  \HH^6}      \right)\vp^4\Bigg\}
\end{align}
Now we compute the imaginary part ${}_{0}\tDD_{\rm I}$ of ${}_{0}\tDD$. Note that ${}_{0}\tDD_{\rm I}$ comprises only terms off the diagonal of the Jacobi mapping $\tDD_{ab}$ and therefore only consists of terms involving $\bar{n}^{i}_a \bar{n}^{j}_b$. At both order $\Oo{2}$ and order $\Oo{3}$ the Jacobi mapping $\DD_{ab}$ is symmetric in the indices $a$ and $b$. Consequently, using \eqref{eq:C22} and \eqref{eq:B30} the rotation $\omega$ vanishes at these orders.
%
%
%
At order $\Oo{4}$ the only term that contributes to ${}_{0}\tDD_{\rm I}^{(4)}$ stems from the product of the second order contracted Riemann with the second order Jacobi mapping  $\RR^{(2)b}_{\hspace{2mm}a}\DD^{(2)}_{\; bc}$. Note that all contributions from the redshift perturbations in $\tDD_{ab}^{(4)}$ in \eqref{eq:Dz44} are symmetric and consequently do not contribute to the rotation $\omega$.
\begin{align}
{}_{0}\tDD_{\rm I}^{(4)}=&\cc{4}\frac{1}{z_S+1}\intchi \frac{1}{2}\left(\eminus{a}\eplus{b}-\eplus{a}\eminus{b}\right)4\bar{n}^{s}_c \bar{n}^{cr}W_{N,as}\int^{\chi}_0\int^{\chi'}_0 d\chi'd\chi'' \chi''W_{N,rb}\notag\\
=&\frac{1}{z_S+1}\ichi \frac{\chi_S - \chi}{\chi^2} \left(\bar{\slpa}^2W_N\int^{\chi}_0\ d\chi' \frac{\chi-\chi'}{\chi'} \slpa^2 W_N +\right.\notag\\
&\left.  -\slpa^2W_N\int^{\chi}_0 d\chi'\frac{\chi-\chi'}{\chi'} \bar{\slpa}^2 W_N\right).\label{eq:tDD0I4}
\end{align}

\section{Spin Operators}
\subsection{Real and imaginary contributions using spherical spin operators}
\label{App:C1}
In \eqref{eq:kappa} and \eqref{eq:gamma}, the convergence $\kappa$, the shear $\gamma$, and the rotation $\omega$ have been defined via the real and imaginary part of ${}_{0}\DD$ and ${}_{2}\DD$. In this section, we examine the derivatives $\slpa$ and $\bar{\slpa}$, and its application on scalars, vectors, and tensors in order to understand which contributions of ${}_{0}\DD$ and ${}_{2}\DD$ are real or complex. We start with various combinations of the derivatives applied on the scalar functions $X$ and $Y$:
\begin{align}
\bar{\slpa}\slpa X=&\left(\slpabb\right)\left(\slpae\right)X=\left(\pt^2+\cot\theta+\frac{1}{\sin^2\theta}\pp^2\right)X\label{eq:C1}\\
=&\Delta_{\theta\phi} X \in \mathbb{R}\label{eq:C2}\\[5pt]
\slpa X\bar{\slpa}X=&\left(\slpae\right)X\left(\slpab\right)X=\pt X\pt X+\frac{1}{\sin^2\theta}\pp X\pp X \in  \mathbb{R}\label{eq:C3}\\[5pt]
\slpa X\bar{\slpa}Y+&\slpa Y\bar{\slpa}X=2\left(\pt X\pt Y+\frac{1}{\sin^2\theta}\pp X\pp Y\right)\in \mathbb{R}\label{eq:C4}\\[5pt]
\slpa\slpa X=&\left(\slpaee\right)\left(\slpae\right)X\label{eq:C5}\\
=&\left[\sin\theta\pt \left(\frac{1}{\sin\theta}\pt\right)-\frac{1}{\sin^2\theta}\pp\pp+i2\frac{1}{\sin\theta}\pp\pt\right]X \in\mathbb{C}\label{eq:C6}\\
\bar{\slpa}\bar{\slpa} X=&\left(\slpabb\right)\left(\slpab\right)X \\
=&\left[\frac{1}{\sin\theta}\pt \left(\sin\theta\pt\right)-\frac{1}{\sin^2\theta}\pp\pp  -i2\frac{1}{\sin\theta}\pp\pt\right]X \in\mathbb{C}\\
\slpa X \slpa Y=&\pt X\pt Y-\frac{1}{\sin^2\theta}\pp X\pp Y+\frac{i}{\sin\theta}\left(\pt X\pp Y+\pt Y\pp X\right) \in \mathbb{C}\label{eq:C7}
\end{align}
We will find combinations like \eqref{eq:C1} - \eqref{eq:C4} in the expression for $\tDm{2}$ and $\tDm{4}$ in \eqref{eq:tDD2s} and \eqref{eq:tDDsplit}, respectively. In sections \ref{sec:sec3} we discuss the physical interpretations of the real and the imaginary part of ${}_{0}\tDD$ and ${}_{2}\tDD$. While $\textrm{Re}\big[{}_{2}\tDD\big]$ and $\textrm{Im}\big[{}_{2}\tDD\big]$ are both related to the shear \eqref{eq:gamma}, $\textrm{Re}\big[{}_{0}\tDD\big]$ and $\textrm{Im}\big[{}_{0}\tDD\big]$ are proportional to the convergence $\kappa$ and rotation $\omega$ \eqref{eq:kappa}, respectively. 

Next we discuss the slashed derivatives applied to vectors. In ${}_{0}\tDD^{(3)}$ \eqref{eq:tDD3s} and ${}_{2}\tDD^{(3)}$ \eqref{eq:2tDD3s} we find the following combinations:
\begin{align}
\bar{\slpa}{}_{1}B=&-\left(\slpabb\right)\left(B_{\theta}+iB_{\phi}\right)\label{eq:C8}\\
=&-\left[\frac{1}{\sin\theta}\pt\left(\sin\theta B_{\theta}\right)+\frac{1}{\sin\theta}\pp B_{\phi}\right]+i\left[-\frac{1}{\sin\theta}\pt\left(\sin\theta B_{\phi}\right)+\frac{1}{\sin\theta}\pp B_{\theta}\right]\in \mathbb{C}\\[5pt]
\slpa{}_{-1}B=&-\left(\partial_{\theta}+\frac{i}{\sin\theta}\partial_{\phi}+\cot\theta\right)\left(B_{\theta}-iB_{\phi}\right)\\
=&-\left[\frac{1}{\sin\theta}\pt\left(\sin\theta B_{\theta}\right)+\frac{1}{\sin\theta}\pp B_{\phi}\right]+i\left[\frac{1}{\sin\theta}\pt\left(\sin\theta B_{\phi}\right)-\frac{1}{\sin\theta}\pp B_{\theta}\right]\in \mathbb{C}\label{eq:C11}\\[5pt]
\bar{\slpa}{}_{1}B+&\slpa{}_{-1}B=-2\left(\pt B_{\theta}+\frac{1}{\sin\theta}\pp B_{\phi}\right)\in \mathbb{R}\label{eq:C12}\\[5pt]
\slpa{}_{1}B=&-\left(\slpaee\right)\left(B_{\theta}+iB_{\phi}\right)\\
=&-\left[\sin\theta\pt\left(\frac{1}{\sin\theta} B_{\theta}\right)-\frac{1}{\sin\theta}\pp B_{\phi}\right]-i\left[\sin\theta\pt\left(\frac{1}{\sin\theta} B_{\phi}\right)+\frac{1}{\sin\theta}\pp B_{\theta}\right]\in \mathbb{C}\label{eq:C13}
\end{align}

There are slashed derivatives of the tensor potential in both ${}_{0}\tDD$ \eqref{eq:tDDsplit} and ${}_{2}\tDD$ \eqref{eq:DZ44}. However, the combinations only involve $h_{rr}$, which is a scalar function of spin-0 like $X$ and $Y$ in \eqref{eq:C1} - \eqref{eq:C7}, and ${}_{\pm1}h_r$\footnote{${}_{\pm1}h_r$ expressed in terms of spherical coordinates yields ${}_{\pm1}h_r=h_{r\theta}\pm i h_{r\theta}$ analogously to ${}_{\pm1}B=B_{\theta}\pm iB_{\phi}$.}, which is a spin-$\pm1$ function such as ${}_{\pm1}B$ in \eqref{eq:C8} - \eqref{eq:C13}. Thus, we can use the above relations to compute the real and imaginary part.

\subsection{Useful relations}
\label{app:stuff}
In this subsection, we list useful relations using the spin-weighted formalism. First we look at derivatives of the basis vectors $\eplus{i}$, $\eminus{i}$, and $e_{r}^{i}$.
\begin{align}
\eplus{i}\bk{j}_{,i}=&\frac{1}{\chi}\eplus{j}\\
\eminus{i}\bk{j}_{,i}=&\frac{1}{\chi}\eminus{j}\\
\eplus{i}e_{+,i}^{j}=& \frac{1}{\chi}\cot\theta \eplus{j}\\
\eminus{i}e_{-,i}^{j}=&  \frac{1}{\chi}\cot\theta \eminus{j} \\
e^{i}_{\pm}e_{\mp,i}^{j}=&-\frac{1}{\chi}\cot\theta e_{\mp}^{j}-\frac{2}{\chi}\bk{j}\\
\eplus{i}\eplus{j}e^{m}_{r,ij}=&\frac{1}{\chi^2}\cot\theta \eplus{m}
\end{align}
For the scalar function $X$ we find the following relations useful:
\begin{align}
\eplus{i}\eplus{j}X_{,ij}=&\frac{1}{\chi^2}\slpa^2 X\\
\eplus{i}\bk{j}X_{,ij}=& -\frac{1}{\chi}\slpa X_{,r}+\frac{1}{\chi^2}\slpa X   \\
\eminus{i}\bk{j}X_{,ij}=& -\frac{1}{\chi}\bar{\slpa} X_{,r}+\frac{1}{\chi^2}\bar{\slpa} X   \\
\eminus{i}\eminus{j}X_{,ij}=&\frac{1}{\chi^2}\bar{\slpa}^2 X\\
\eplus{i}\eminus{j}X_{,ij}=&\eminus{i}\eplus{j}X=\frac{1}{\chi^2}\bar{\slpa}\slpa X+\frac{2}{\chi}X_{,r}\label{eq:C22}\\
\eplus{i}\eplus{j}\bk{m}X_{,ijm}=&\frac{1}{\chi^2}\slpa^2 X_{,r}  -\frac{2}{\chi^3}\slpa^2X \label{eq:C23}\\
\eplus{i}\eminus{j}\eplus{m}X_{,ijm}=&  \eplus{i}\eplus{j}\eminus{m}X_{,ijm}    =-\frac{1}{\chi^3}\slpa \bar{\slpa}\slpa X      -\frac{4}{\chi^2}\slpa X_{,r}   +\frac{2}{\chi^3}\slpa X   \label{eq:C24}\\
\eminus{i}\eplus{j}\eminus{m}X_{,ijm}=&      \eminus{i}\eminus{j}\eplus{m}X_{,ijm}        =-\frac{1}{\chi^3}\bar{\slpa} \slpa\bar{\slpa} X     -\frac{4}{\chi^2}\bar{\slpa} X_{,r}     +\frac{2}{\chi^3}\bar{\slpa} X   \label{eq:C25}\\
\eplus{i}\eminus{j}\bk{m}X_{,ijm}=&\eminus{i}\eplus{j}\bk{m}X_{,ijm}  =   \frac{1}{\chi^2}\slpa\bar{\slpa} X_{,r}-\frac{2}{\chi^3}\slpa\bar{\slpa} X-2\frac{1}{\chi^2}X_{,r}+\frac{2}{\chi}X_{,rr}\\
\eplus{i}\eminus{j}\eminus{m}X_{,ijm}=&-\frac{1}{\chi^3}\slpa \bar{\slpa}^2 X-\frac{4}{\chi^2}\bar{\slpa}X_{,r}+\frac{4}{\chi^3}\bar{\slpa} X\\
\eminus{i}\eplus{j}\eplus{m}X_{,ijm}=&-\frac{1}{\chi^3}\bar{\slpa} \slpa^2 X-\frac{4}{\chi^2}\slpa X_{,r}+\frac{4}{\chi^3}\slpa X\\
X^{,i}X_{,i}=&\left(\bk{i}\bk{j}+\frac{1}{2}\eplus{i}\eminus{j}+\frac{1}{2}\eminus{i}\eplus{j}\right)X_{,i} X_{,j}\notag\\
=&X_{,r}X_{,r}+\frac{1}{\chi^2}\slpa X\bar{\slpa} X.\label{eq:C26}
\end{align}
Let $Y^{i}$ be a vector field. We can express $Y^{i}$ in terms of the basis $\left\{\bk{i},\eplus{i},\eminus{i}\right\}$ as $Y^{i}=Y_r \bk{i}+\frac{1}{2}{}_{-1}Y\eplus{i}+\frac{1}{2}{}_{1}Y\eminus{i}$. Then, the following relations can be found:
\begin{align}
\eplus{i}\eminus{j}\bk{m}Y_{m,ij}=&\frac{1}{\chi^2}\left(\bar{\slpa}\slpa Y_r+2\chi Y_{r,r}+\bar{\slpa}{}_1 Y+\slpa{}_{-1} Y  -2Y_r\right)\label{eq:B30}\\
\eminus{i}\eplus{j}Y_{i,j}=&-\frac{1}{\chi}\bar{\slpa}{}_{1}Y+\frac{1}{\chi}2Y_r\\
\eplus{i}\eminus{j}Y_{i,j}=&-\frac{1}{\chi}\slpa {}_{-1}Y+\frac{1}{\chi}2Y_r\\
\eplus{i}\eplus{j}Y_{i,j}=&-\frac{1}{\chi}\slpa {}_{1}Y\\
\eplus{i}\eplus{j}\bk{m}Y_{m,ij}=&\frac{1}{\chi^2}\slpa^2 Y_r\\
Y^{i}Y_{i}=&\left(\bk{i}\bk{j}+\frac{1}{2}\eplus{i}\eminus{j}+\frac{1}{2}\eminus{i}\eplus{j}\right)Y_i Y_j\notag\\
=&Y_{r}Y_{r}+{}_{-1}Y{}_{1}Y.
\end{align}
For a tensor field $Z_{ij}$, which can be expressed as $Z^{ij}=Z_{rr}\left[\bk{i}\bk{j}-\frac{1}{4}\left(\eplus{i}\eminus{j}+\eminus{i}\eplus{j}\right)\right]+{}_{-1}Z_r \frac{1}{2}\left(\eplus{i}\bk{j}+\bk{i}\eplus{j}\right)+  {}_{1}Z_r \frac{1}{2}\left(\eminus{i}\bk{j}+\bk{i}\eminus{j}\right)+ \frac{1}{4}{}_{-2}Z\eplus{i}\eplus{j} +\frac{1}{4}{}_{2}Z\eminus{i}\eminus{j}$    the following relations hold:
\begin{align}
\eplus{i}\eplus{j}Z_{ij}=&{}_{2}Z\\
\eplus{i}\eminus{j}Z_{ij}=&-Z_{rr}\\
\eplus{i}\eplus{j}\bk{m}Z_{im,j}=&-\frac{1}{\chi}\slpa {}_{1}Z_r-\frac{1}{\chi}{}_{2}Z\\
\eplus{i}\eminus{j}\bk{m}Z_{im,j}=&-\frac{1}{\chi}\bar{\slpa}{}_{1}Z_r+\frac{3}{\chi}Z_{rr}\\
\eminus{i}\eplus{j}\bk{m}Z_{im,j}=&-\frac{1}{\chi}\slpa {}_{1}Z_r+\frac{3}{\chi}Z_{rr}\\
\eplus{i}\eplus{j}\bk{m}\bk{n}Z_{mn,ij}=&\frac{1}{\chi^2}\slpa^2 Z_{rr}+4\frac{1}{\chi^2}\slpa {}_{1}Z_r+\frac{2}{\chi^2}{}_{2}Z\\
\eplus{i}\eminus{j}\bk{m}\bk{n}Z_{mn,ij}=&\frac{2}{\chi^2}\left(\frac{1}{2}\slpa \bar{\slpa}-\chi \partial_r -3\right)Z_{rr}+\frac{2}{\chi^2}\left(\slpa {}_{-1}Z_r+\bar{\slpa} {}_{1}Z_r\right).
\end{align}

Let ${}_{s}T$ be a function of spin $s$. The operators $\slpa$ and $\bar{\slpa}$ obey the following commutation rule
\begin{align}
\left(\bar{\slpa}\slpa-\slpa\bar{\slpa}\right)_{s}T=2s {}_{s}T.\label{eq:STS}
\end{align}

\bibliographystyle{unsrt}
\bibliography{WL_PF_Paper}




\end{document}